\documentclass[11pt]{article}
\usepackage[margin=1in]{geometry}
\usepackage{microtype,amsmath,amssymb,amsthm}
\usepackage{graphicx,booktabs,natbib,algorithm,algorithmic}
\usepackage[hidelinks]{hyperref}
\newif\ifsubmissionlayout\submissionlayouttrue
\newtheorem{theorem}{Theorem}
\newtheorem{proposition}[theorem]{Proposition}
\newtheorem{corollary}[theorem]{Corollary}
\theoremstyle{remark}\newtheorem{remark}[theorem]{Remark}
\newcommand{\KL}{\mathrm{KL}}
\newcommand{\A}{\mathcal A}\newcommand{\Y}{\mathcal Y}\newcommand{\M}{\mathcal M}
\title{Adaptive Experimentation Can Hide Model Error:\\Detection Limits and Falsification Budgets}
\author{Ramnarayan Choudhary\\MBZUAI\\\texttt{ramnarayan.ramniwas@mbzuai.ac.ae}}
\date{}
\hypersetup{pdftitle={Adaptive Experimentation Can Hide Model Error: Detection Limits and Falsification Budgets},pdfauthor={Ramnarayan Choudhary},pdfsubject={Research preprint on adaptive experimental design and model adequacy}}
\begin{document}
\maketitle
\begin{abstract}
An experiment can be informative about the parameters of a working model and
uninformative about whether that model is wrong. Adaptive design can reinforce
this mismatch by repeatedly selecting settings where an incorrect model fits.
We bound the power of any valid adequacy test in terms of the experimenter's own
sampling distribution, then distinguish information absent from the selected
data from information lost by the detector. A computable diagnostic measures
unresolved error relative to a declared expansion of the working model. It
motivates a training-free rule that reserves part of the design budget for
falsification. The empirical record is mixed: the independently specified primary
study fails detection eligibility, and a later numerator diagnostic fails its
improvement criteria. After the frozen rule fails on fresh continuous-null
panels, a disclosed secondary horizon-calibrated revision
detects 35.8\% of misspecified campaigns versus 31.7\% for a class-separation
control at $T=240$, but has no resolved advantage at $T=480$. In a corrected
289-unit measured replay it improves on random design and trails
class-discrimination EIG. A conditional futility bound explains why remaining
time matters without proving the gate constant or its power. These results
support separating parameter learning, model checking and detector performance;
they do not establish a universally superior design policy or prospective
scientific discovery.
\end{abstract}

\section{Introduction}
\label{conf:intro}
Experimental design determines which features of a scientific system become
observable. A common objective, expected information gain (EIG), chooses
measurements that reduce uncertainty about parameters within a working model.
That objective is useful when the model is adequate. When it is not, the same
measurements may give precise estimates while revealing little about the missing
mechanism. Improving inference or optimizing EIG more accurately does not remove
this mismatch between parameter learning and model checking.

A simple example makes the distinction precise. A working straight-line response
can interpolate observations at two endpoints. Suppose the true response agrees
there but has an interior bump. An experimenter confined to those endpoints
cannot distinguish truth from the interpolating line, however many measurements
it collects or how capable its inference engine becomes. The failure concerns
selected observations, not reasoning quality. The same issue arises when each
setting is separately compatible with some working-model member but only a
combination of settings rules out every member. Detection therefore depends on
joint design coverage, not simply on visiting an unusual setting.

This concern has deep roots in design for inadequacy and model discrimination
\citep{box1959basis,atkinson1972planning,tommasi2009bayesian,catanach2023metrics}.
Recent work on learning to outgrow a theory already combines information blind
spots, complementary experiments, exploration and anytime-valid class rejection
\citep{ma2026outgrow}. We do not claim these ideas as new. Our contribution is an
own-design power analysis, a distinction between acquisition and detector
bottlenecks, an explicitly qualified computable diagnostic, and an auditable
cross-setting evaluation of a training-free intervention.

\begin{figure}[t]
\centering
\includegraphics[width=\linewidth]{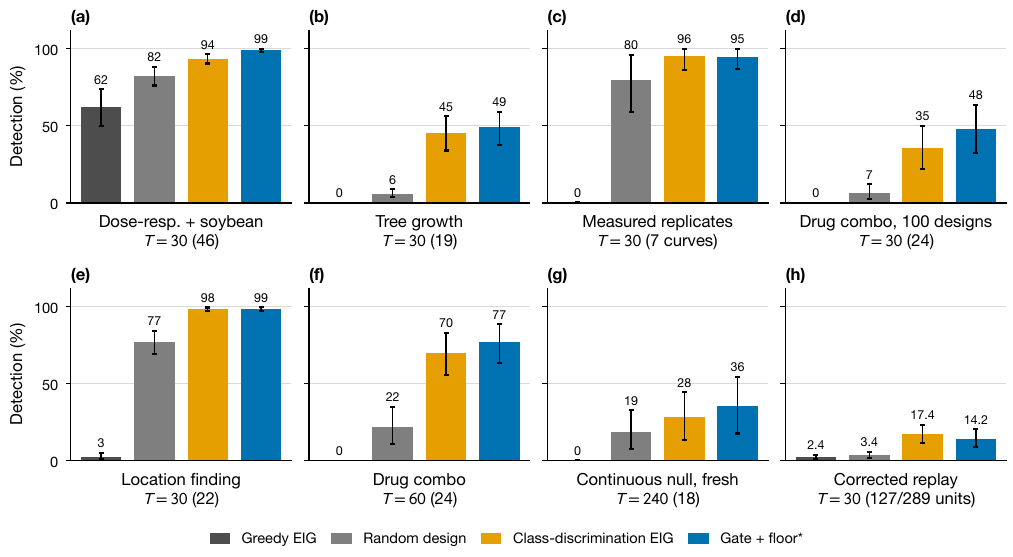}
\caption{\textbf{Detection recovery varies across settings and comparators.}
Panels (a--e) use the frozen $.1$ gate at $T=30$; (f--h) use the revised
$\log(20)/T$ gate (asterisk in legend). Descriptive 95\% intervals use 5,000
world/curve resamples in (a--g), seed 0; (h) resamples 20 cell lines with unit
weighting, seed 20261026. Campaigns per world/unit: 50 in (g,h), 200 otherwise.
Counts label worlds/curves; (h) shows 127 low-$p$ units of 289. Dose/soybean and
location use earlier random-detectable pools; both measured replays use full-data
fit labels. Class EIG exceeds the rule in (h). These heterogeneous evaluations
are neither pooled nor eight confirmations of one unchanged rule. The technical
record retains separation, all-world results and registered failures.}
\label{conf:fig:evidence}
\end{figure}

Figure~\ref{conf:fig:evidence} summarizes the intervention's promise and its
limits. In several evaluated settings greedy EIG rarely detects a planted error,
while targeted discrimination recovers substantial detection. Recovery can
coexist with small relative prediction-risk premiums on correctly specified
controls. This does not mean that the rule saves experimental observations:
every arm receives the same horizon. Nor does it establish that more capable
agents generally become blind. The frontier-model observations use selected
worlds, unequal completion and very small campaign counts.

We follow the information from experiment selection to rejection, separating
limits imposed by the sampling policy from evidence lost through the detector's
predictive numerator. This separation guides the intervention and its evaluation:
reserving measurements for falsification can repair one bottleneck while leaving
the other intact. The technical record contains the proofs, full experimental
detail and every previously registered failure.

\paragraph{Evidence boundary.}
This version reports the studies evaluated in the technical record, including
negative results. A later original-prior, outward-certified development study has
finished collection and passed an independent index/summary reconciliation.
Its full retained-history verification is pending, so its numerical outcomes
are excluded from the scientific comparisons here. Completion and index agreement
alone do not establish scientific acceptance or fresh-world confirmation.

\section{Related work}
\label{conf:related}
Design for inadequacy and Bayesian discrimination already address model checking
and its trade-off with estimation. Expected discriminatory information is close
to our KL targeting score \citep{catanach2023metrics}: with common one-step
posterior predictives $q_t,p_t$, their Eq.~41 gives
$D_t(a)=\KL(q_t(\cdot\mid a)\|p_t(\cdot\mid a))$; memberwise ED adds
within-expansion information $U_t(a)$. This standard decomposition can change
action rankings, without establishing superiority over their trajectory-design
method. Our contribution separates power available in a policy's selected data,
evidence captured by a proper detector, and prediction risk incurred by
intervention. These quantities can disagree within one environment.

Modern Bayesian design includes amortized policies learned before deployment
\citep{rainforth2023modern,foster2021dad}, and robustness methods address
misspecification through generalized updates or robust objectives
\citep{barlas2025gboed,abdulsamad2026maximin}. A bias-against design criterion
controls misleading evidence under a joint Bayesian model
\citep{chen2026evidential}; rejecting a scientific family under a fixed off-model
truth is a different question. Learning to outgrow a theory is the closest
structural-discovery predecessor \citep{ma2026outgrow}. It already identifies
within-class information blind spots and uses complementary class discrimination,
exploration and anytime-valid rejection alongside learned structural revision.
Our own-design ceiling applies to any experimenter, while its best-allocation
separation addresses what good allocation could achieve. The distinction is
between a policy's observable discrepancy and the best available discrepancy,
not between having and lacking model-class exploration.

Joint family/parameter EIG with LLM-proposed families and residual feedback
is also studied by \citet{yuh2026bmd}; \citet{yang2026openended} address
closed- and open-world adequacy with possibilistic reasoning. Our certificate
is relative to represented alternatives, not unrestricted open-world knowledge.

Our Ma-style controls isolate parts of that predecessor. Completed secondary
comparisons use one expanded MAP member and a finite shared-parameter acquisition
fit. The separate original-prior certified development uses edit-specific
nuisance-marginal predictions and continuous shared-null projection. Acquisition
fits are distinct from scientific detector nulls. Neither control reproduces
the learned joint proposer, complexity/calibration/cost ranking, original fixed
exploration or structural revision. Neither wins nor close detection rates
support superiority to the complete method; that comparison requires a faithful
implementation on common classes, with training and deployment resources stated.

Evidence-driven allocation is established in multi-armed testing:
\citet{sandoval2026multiarmed} give oracle growth and high-confidence delay
guarantees, while \citet{lin2026sampleefficient} analyze adaptive sampling for
multiple testing. Their conditional-increment premises require care for our
shared-parameter null: ratios of successive profile e-process values need not
be valid one-step e-values (Appendix~\ref{app:deadlineoracle}). Neither method
is evaluated here as a complete baseline.

Anytime-valid inference supplies false-rejection control under adaptive sampling
\citep{ramdas2023game}. It does not make the experiment informative, guarantee a
well-fitting numerator or eliminate numerical approximation. Sequential
controlled sensing already uses forced exploration under its own separation
assumptions \citep{deshmukh2019controlled}. Likewise, finite-horizon e-value
curtailment and horizon-aware betting precede our futility rationale
\citep{baas2026curtailment,taga2026horizon,clerico2026timesensitive}. The new role here is to audit a
particular scientific-design intervention against those distinct premises,
including cases where its promised recovery fails.

\section{Power limits under adaptive design}
\label{conf:ceiling}
At step $s$, a predictable, possibly randomized policy selects $a_s\in\A$ and
observes $Y_s\in\Y$ from a stationary truth $P(\cdot\mid a_s)$. We take finite
action and outcome spaces. The scientific null is a working family
$\M_0=\{p_\theta:\theta\in\Theta\}$, which may have a continuous parameter
space. All comparisons use the same policy mechanism under truth and null;
policy factors cancel from path likelihood ratios. Define
\[
\delta_\theta(a)=\KL(P(\cdot\mid a)\|p_\theta(\cdot\mid a)),\quad
\nu_T(a)=T^{-1}\mathbb E_P\sum_{s=1}^T\mathbf1\{a_s=a\},
\]
and $D(\nu)=\inf_\theta\sum_a\nu(a)\delta_\theta(a)$. Let $A_T$ be any
history-measurable rejection event with probability at most $\alpha$ under
every working-model parameter. It can be the event of an anytime crossing by $T$.
Write $\mathrm{kl}$ for binary relative entropy.

\begin{theorem}[Own-design power ceiling]
\label{conf:thm:ceiling}
For any such test and policy,
\[
P(A_T)\le\mathrm{kl}^{-1}_\alpha\{T D(\nu_T)\},\qquad
\mathrm{kl}^{-1}_\alpha(K)=\max\{p\in[\alpha,1]:\mathrm{kl}(p\|\alpha)\le K\}.
\]
\end{theorem}
\noindent
The proof is the pathwise KL chain rule followed by data processing onto
$A_T$: $\KL(P^\pi\|Q^\pi_\theta)=T\sum_a\nu_T(a)\delta_\theta(a)$.
Minimize over $\theta$. The bound uses no alternative library, numerator or
assumption about the agent's reasoning (full proof: Theorem~\ref{thm:ceiling}).
If one null member agrees with truth on every design the policy ever visits,
then rejection probability is at most $\alpha$ at every horizon. Repeatedly
measuring those settings cannot fix the problem.

The ceiling also yields a necessary resource condition: power at least
$1-\beta>\alpha$ requires $TD(\nu_T)\ge\mathrm{kl}(1-\beta\|\alpha)$.
For an endpoint-saturated policy, this constrains how much sampling must leave
the saturated support. It does not say that any chosen exploration budget is
sufficient. In particular, a policy-independent ceiling based on
$D^\star=\max_\nu D(\nu)$ is still an upper bound. A value near one does not
mean that a practical test, or even an optimal test, attains that power.

\ifsubmissionlayout
\paragraph{A concrete coverage example.}
Consider two Bernoulli settings, $A$ and $B$. Two working hypotheses and a
truth have success probabilities
\[
\begin{array}{c|cc}
 & A & B\\\hline
\theta_1 & .2 & .8\\
\theta_2 & .8 & .2\\
P & .2 & .2
\end{array}
\]
A policy using only $A$ is indistinguishable from $\theta_1$; one using only
$B$ is indistinguishable from $\theta_2$. Either has $D(\nu)=0$, and any
valid test has power at most $\alpha$. Balanced occupancy instead gives
$D(\nu)=\tfrac12\KL(\mathrm{Ber}(.2)\|\mathrm{Ber}(.8))=.3\log4>0$.
Each setting is individually matched by a working hypothesis, yet their
combination exposes the class error. This example illustrates the information
condition; it does not assert which settings EIG will choose or guarantee that
a particular detector captures all available separation.

\else
Appendix~\ref{app:submissionclarity} gives a two-setting coverage example.
\fi

\paragraph{Acquisition loss and numerator loss are different.}
Our detector is universal inference,
\[
 E_t=\frac{Q_t}{\sup_\theta L_{\theta,t}},\qquad
 Q_t=\prod_{s\le t}q_s(Y_s\mid a_s,H_{s-1}),\quad
 \tau=\inf\{t:E_t\ge1/\alpha\}.
\]
Predictable proper or sub-probability $q_s$ makes $E_t$ dominated, at each true
null parameter, by a nonnegative test supermartingale. Ville's inequality then
controls any-time false rejection \citep{ramdas2023game}. The supremum must cover
the scientific null; a finite grid denominator alone does not validate a
continuous null. Along each history,
\[
\log E_t=\underbrace{\log L_{P,t}-\sup_\theta\log L_{\theta,t}}_{\text{realized truth/null signal}}
 -\underbrace{(\log L_{P,t}-\log Q_t)}_{\text{numerator regret}}.
\]
Neither term is necessarily nonnegative on an individual history. Inserting the
simulator truth as numerator is a retrospective diagnostic, never a deployable
procedure. A poor alternative can miss signal that good acquisition collected;
an ideal numerator cannot reveal an error observationally absent from a history.
Appendix~\ref{app:partitionrepair} gives a constructed continuous-null example
where a proper local-sharing numerator repairs this loss at a fixed budget;
its registered reused-history diagnostic fails the improvement and world-loss
criteria, and fresh-panel performance remains untested.

\paragraph{What suffices, under additional assumptions.}
For finite $\Theta$ and an expansion $\M_1$ containing the truth with a uniform
prior, define $d_\theta(a)=-2\log\sum_y\sqrt{P(y\mid a)p_\theta(y\mid a)}$.
With probability at least $1-\beta$, simultaneously in time,
\[
\log E_t\ge\min_\theta\sum_{s\le t}d_\theta(a_s)
                 -2\log(|\Theta|/\beta)-\log|\M_1|.
\]
Square-root likelihood-ratio martingales and a union bound prove this sufficient
condition (Theorem~\ref{thm:achieve}). The minimum follows accumulation over
settings: complementary observations can reject a class even if every setting
is individually matched by some parameter. This guarantee is conditional on
cumulative separation and truth membership; the acquisition rule below is not
proved to enforce its premises. It is not a matching minimax rate for the ceiling.

\section{A posterior diagnostic for represented model error}
\label{conf:certificate}
Own-design discrepancy involves the unknown truth. A practical agent instead
has a working predictive $p_t$, an expanded predictive $q_t$, and an expansion
posterior $w_t(h)$. For a declared protected-target weighting $\mu$, compute
\[
 B_t=\sum_x\mu(x)\mathbb E_{h\sim w_t}\KL(p_h(\cdot\mid x)\|p_t(\cdot\mid x)).
\]
This is exactly $B_t=U_t+D_t$, where
\[
 U_t=\sum_x\mu(x)\mathbb E_h\KL(p_h(\cdot\mid x)\|q_t(\cdot\mid x)),\quad
 D_t=\sum_x\mu(x)\KL(q_t(\cdot\mid x)\|p_t(\cdot\mid x)).
\]
The identity follows by inserting $q_t$ into each log ratio and using
$\mathbb E_h p_h=q_t$. $U_t$ measures disagreement among represented hypotheses;
$D_t$ measures their average predictive disagreement with the working model.
Under a truth drawn from the expansion's actual prior, $B_t$ is posterior
expected working-model KL error. Under that Bayesian joint, $U_t$ decreases in
conditional expectation by the information the next observation carries about a
fresh protected-target outcome. It can increase on an individual trajectory.

These statements identify both what the agent can compute and what it cannot
claim. Small $U_t$ alone leaves $D_t$ uncontrolled. A wrong singleton expansion
can have $B_t=U_t=0$ despite positive true error. Fixed-truth calibration for a
finite correctly represented truth is possible, but its conservative factor
$1/(\beta\pi(P))$ can be large. Off-library truths and continuous quadrature do
not inherit that finite-model guarantee. The diagnostic is a certificate relative
to declared alternatives, not a universal certificate that the model is adequate
(Propositions~\ref{prop:cert}--\ref{prop:ignorance}).

Two audited constructions guide the acquisition rule. First, opposite-sign
alternatives can cancel in the expansion's predictive average, causing
predictive-KL targeting to choose an irrelevant design. Averaging each
hypothesis's discrepancy before maximization avoids that particular cancellation.
Second, even the correct target score can be disabled too early by an
ignorance-only gate: a two-model example takes one useful observation and then
never rejects on any response path. Replacing the score cannot repair that gate
(Proposition~\ref{prop:triggerfailure}). These are mechanism-specific results,
not proofs that one replacement is uniformly optimal.

\section{Reserving a falsification budget}
\label{conf:rule}
Define expected discrepancy at a candidate setting by
$\mathrm{ED}_t(a)=\mathbb E_{h\sim w_t}\KL(p_h(\cdot\mid a)\|p_t(\cdot\mid a))$.
Before rejection, the rule takes an ED-maximizing action when $B_t\ge b^\star$
or a predictable floor coin succeeds; otherwise it follows the base designer.
The floor probability is $0.5/\sqrt{t+1}$. Some benchmarks switch to expanded
inference after rejection; the continuous-null comparisons hold prediction in
the working class. This distinction matters when interpreting prediction risk.
The method needs a specified expansion and numerical inference, but no policy
training or language-model calls on overridden steps.

\ifsubmissionlayout
\ifsubmissionlayout
\begin{algorithm}[t]
\else
\begin{algorithm}[H]
\fi
\caption{Revised falsification rule with fixed-class prediction}
\label{alg:revisedrule}
\begin{algorithmic}[1]
\REQUIRE Base policy, working family $\M_0$, expansion, target $\mu$, $T$, $\alpha$
\STATE Set $b=\log(1/\alpha)/T$ and $\mathrm{rejected}=\mathrm{false}$.
\FOR{$t=0,\ldots,T-1$}
\STATE From $H_t$, compute $p_t$, expansion posterior, $B_t$, and $\mathrm{ED}_t$.
\STATE Draw an independent $Z_t\sim\mathrm{Ber}(.5/\sqrt{t+1})$.
\IF{not rejected and ($B_t\ge b$ or $Z_t=1$)}
\STATE Choose $a_{t+1}\in\arg\max_a\mathrm{ED}_t(a)$, randomizing ties.
\ELSE
\STATE Choose $a_{t+1}$ from the base policy using $H_t$.
\ENDIF
\STATE Observe $Y_{t+1}$ and update inference and the proper path numerator $Q_{t+1}$.
\STATE Evaluate $E_{t+1}=Q_{t+1}/\overline L_{0,t+1}$ using a valid null upper bound.
\STATE Record rejection when its conservative comparison reaches $1/\alpha$.
\ENDFOR
\end{algorithmic}
\end{algorithm}

Algorithm~\ref{alg:revisedrule} makes the time ordering explicit. The target
weights sum to one, so $B_t$ and ED are measured in nats per observation.
Planning posteriors need not equal the detector numerator; proper predictable
sequence likelihoods determine validity. Physical observations continue through
$T$, including after rejection, and this version keeps prediction in $\M_0$.
The frozen $.1$ gate and expansion-after-rejection variants remain distinct.
Trigger activity, target separation and detector capture are separate conditions;
the algorithm alone proves none of their empirical success criteria.

\else
The complete pre-outcome action order is in Algorithm~\ref{alg:revisedrule}.
\fi

\paragraph{A revised gate, not retroactive confirmation.}
The original frozen gate is $b^\star=0.1$. It succeeds in several $T=30$
evaluations and fails on fresh continuous-null panels at longer horizons. A
development pass on earlier worlds motivated the revision
$b^\star_T=\log(1/\alpha)/T$, declared as a secondary arm before fresh-panel
outcomes. At $T=30$ it is close to 0.1, not identical; the earlier results remain
results for the frozen rule. An earlier use of the same calibration for an
ignorance trigger failed and stays a failure. Distinguishing revisions from
confirmations prevents an appealing final formula from rewriting the evidence.

A conditional futility calculation supplies a limited rationale. Fix a history
$H_t$, a working parameter $\theta$ selected at that history and frozen thereafter,
and $\lambda>0$. Put
\[
 R_t(\theta)=Q_t/L_{\theta,t},\qquad
 \bar D_\lambda(h,\theta)=\max_a D_{1+\lambda}(p_h(\cdot\mid a)\|p_\theta(\cdot\mid a)),
\]
and $c=\log(1/\alpha)-\log R_t(\theta)-(T-t)\bar D_\lambda$. Remaining
crossing probability under truth $h$ and any continuation is bounded by
\[
 \phi_\lambda(c)=\min\{1,(1+\lambda^{-1})\lambda^{1/(1+\lambda)}
                       e^{-\lambda c/(1+\lambda)}\}\quad(c>0),
\]
with bound one otherwise. The proof controls numerator/truth by one Ville bound
and a normalized R\'enyi exponential martingale by another, then optimizes their
union bound (Proposition~\ref{prop:futility}). Infinite divergence is vacuous.
Posterior averaging gives an agent-computable $F_t$ only under a correctly
specified expansion prior. It is not a fixed-truth certificate off the library.

The inequality relates remaining time, evidence deficit and maximum R\'enyi
separation. Replacing them by total $T$, zero initial log evidence and mean ED
is a heuristic. It proves neither the constant of $b^\star_T$ nor a nontrivial
lower power bound. Also $R_t(\theta)\ge E_t$, so silently substituting $E_t$
can invalidate the upper bound. Finite-horizon e-value curtailment is already
studied by \citet{baas2026curtailment}; horizon-aware betting and hopeless regions
by \citet{taga2026horizon}. Our conditional composite-null bound uses standard
martingale tools; we claim no first futility principle.

\paragraph{The constant needs more than mean ED.}
An independent registered analysis supplies a counterexample even with correct
representation and the same prior for planning and detection: null Bernoulli
$1/2$, equal prior on truths $1/100$ and $99/100$, $\alpha=0.05$, $T=5$.
ED is $0.63715>\log(20)/5$ at every history, but no possible prefix statistic
reaches 20. Thus gate activity alone does not imply any positive power for this
detector. A randomized deadline oracle has exactly $76.08\%$ power on the same
five observations; this established Neyman--Pearson/Doob construction isolates
detector choice in the example (Appendix~\ref{app:deadlineoracle}).
For a finite null, a conditional sufficient margin is instead
\[
 \min_\theta\{\log(Q_t/L_{\theta,t})+K_\theta\}
 \ge b+r+\varepsilon+\sqrt{\tfrac V2\log(|\Theta|/\beta)},
\]
where $b=\log(1/\alpha)$, $K_\theta$ lower-bounds remaining cumulative true/null
KL, $r$ bounds conditional numerator regret, $\varepsilon$ numerical loss,
and $V$ bounds the sum of conditional log-ratio range widths squared.
If these premises hold with probability at least $1-\eta$, Hoeffding and a
finite union give remaining power at least $1-\eta-\beta$.
Appendix~\ref{app:gateachievability} states the stronger parameter-specific bound,
proof, coverage assumptions linking ED to $K_\theta$, and exact rational checks.
The inequalities use standard concentration tools; the contribution is the
explicit gate counterexample and accounting of the missing premises. They do
not certify the continuous-null production rule.

A separate information-budget argument applies when the represented truth and
planning and numerator priors match, updates are exact, and there is no floor.
The implemented hedged continuous-null detector with off-library alternatives
does not satisfy those premises. A floor prevents permanent cessation of
probing, but does not by itself establish that the chosen settings separate
every false class. Guarantee statements and empirical gate comparisons must
therefore remain separate.

\section{Experiments}
\label{conf:evidence}
The studies test different parts of the argument. Controlled worlds isolate
acquisition and detector failures; fitted and measured-response studies test
transfer beyond those constructions; continuous-null panels test the revised
gate under a larger scientific null. Development, frozen-rule evaluations and
revised-rule evaluations remain separate. Unless stated otherwise, uncertainty
summaries resample worlds or curves, since campaigns within a world are not
independent scientific units. Passing a registered point threshold does not
establish equivalence or a population guarantee, and we do not pool these
heterogeneous studies into one effect.

\begin{table}[t]
\centering\small
\caption{\textbf{The revision helps, but does not erase frozen failures.}
Detection percentages; fresh continuous-null panels each have 18 non-null worlds
and 50 campaigns/world/arm. Drug transfer reuses 24 earlier worlds, 200 campaigns
per world. Sep.\ is the partial class-separation adaptation.}
\label{conf:tab:horizons}
\begin{tabular}{lrrrr}
\toprule
Panel & Frozen 0.1 & Revised & Class EIG & Sep.\\
\midrule
Fresh continuous null, $T=240$ & 2.78 & 35.78 & 28.44 & 31.67\\
Drug transfer, $T=60$ & 59.31 & 76.79 & 70.17 & 69.77\\
Fresh continuous null, $T=480$ & 10.22 & 52.00 & 46.89 & 51.44\\
\bottomrule
\end{tabular}
\end{table}

\paragraph{Controlled mechanisms and the negative primary study.}
The generic and mechanistic evaluations demonstrate severe detection collapse
in several families, but accuracy effects vary by environment. Four of seven
generic families remain largely undetectable at longer budgets, and two
mechanistic environments show partial or absent collapse. A registered predictor
of the accuracy effect fails. Constructive oracle and crossed comparisons show
both acquisition-limited and numerator-limited behavior; no single diagnosis
explains the entire record.

The independently specified primary comparison is particularly informative.
It uses a continuous logistic null, priors fixed without truth-index access,
12 policies, 16 response replicates and two horizons, completing all 9,228
planned histories. The strongest capped control detects $4/288=1.39\%$ at
$H=60$ and $14/288=4.86\%$ at $H=120$, below the registered 20\% eligibility
minimum. No confirmation horizon is selected. The fixed-budget candidate is
not the later gated rule. Its H120 truth-numerator diagnostic crosses on
68/288 histories versus 13 actual rejections; full cycling gives 112 versus 30.
These unavailable-truth diagnostics establish missed realized signal on some
histories, not a deployable repair. At H120 no fixed convex reweighting of the
two existing proper numerator components can add candidate rejections on the
saved histories. At H60 the broad component does add two, so the conclusion is
horizon-specific. The mean any-test ceiling of 0.679 at H120 is an upper bound,
not evidence that 67.9\% power is attainable (Section~\ref{sec:independentprimary}).

\paragraph{Audit and reporting contract.}
Acquisitions precede outcomes; proper path likelihoods determine crossings
at $\alpha=.05$. Caps count intervention opportunities even when override
and base actions coincide. Prediction uses the full horizon after rejection;
expansion-after-rejection comparisons are identified separately. Risk premiums
use the stated control baseline. Registered criteria, descriptive contrasts
and retrospective diagnostics remain distinct. Reusing worlds or fitted
observations does not create independent validation.

\paragraph{Fitted curves, measured replicates and location finding.}
Frozen-rule evaluations include dose response, soybean and tree growth,
a 100-design drug-combination simulator, measured-replicate replay and
location finding. Figure~\ref{conf:fig:evidence} shows the registered pools.
Dose/soybean and location pool only worlds whose earlier random-design detection
is at least 20\%; all-world results and reversals are retained. The location
benchmark is a modified DAD-style environment, not a reproduction of the original
trained policy. The class-separation control uses a single MAP alternative and a
shared finite-grid null parameter for acquisition, not the complete system of
\citet{ma2026outgrow}. Its all-world location detection exceeds the floor rule
at both horizons, despite positive selected-pool results for the rule.

Measured-replicate replay is stronger evidence than drawing outcomes from fitted
curves: the experiment chooses among recorded replicate outcomes. However,
labels, transformations and preprocessing use full-data fits. It is not an
untouched-unit evaluation or prospective lab experiment. The seven-curve
misspecified subset gives 94.79\% detection for the frozen rule and 95.50\%
for class EIG. Available curves contain only 12--28 observations, so these
$T=30$ campaigns do not establish without-replacement experimental validation.
Prediction gains after rejection may also reflect class expansion, which some
comparisons disallow. We report discovery and prediction separately.

A later 300-unit replay of measured combination wells
\citep{jaaks2022combinations} is withheld as supporting evidence: its
normalization treats solvent annotations on active-drug wells as negative
controls. The original results and G1 failure remain documented, with the
data audit and required corrected rerun in Appendix~\ref{app:largereplay}.
The smaller measured-replicate study above uses a different dataset and pipeline.

A registered correction fixes control roles and metadata exclusions before new
fitting, retaining 289 distinct-drug units. Among 127 low-$p$ fit-labelled units,
the revised rule detects 14.16\%, versus 3.42\% for random, 17.37\% for class
EIG and 16.02\% for separation. Its paired cell-line-bootstrap contrast is
$+10.74$ points $[6.46,16.24]$ against random and $-3.21$ points
$[-5.70,-.68]$ against class EIG (20 cell lines). The registered collapse
criterion again fails; restoration, five-point comparator margin and high-$p$
alarm/risk criteria pass their descriptive point thresholds. These passes do
not establish inferential noninferiority or a population ten-point gain.
Separate known-grid-null simulations record zero revised-rule alarms in 5,780
campaigns, without making high-$p$ empirical units true nulls. Prediction endpoints
include class expansion after rejection. Full-data labels, reused units and
with-replacement replay remain limits (Appendix~\ref{app:correctedreplay}).

\paragraph{Fresh continuous-null panels and transfer of the revision.}
Table~\ref{conf:tab:horizons} keeps the frozen failures visible. The $T=240$ panel
uses 24 fresh worlds (six null), 50 campaigns per world/arm, a broader uniform
bump library, and a half structural/half per-action Beta sequence hedge. Its
continuous denominator uses a concavity upper bound. The formula is valid in
exact arithmetic, but implementation uses float64 plus a $10^{-9}$ pad, rather
than the primary pipeline's outward-certified arithmetic. Independent outward
checks on 131 fixed count fixtures through $T=480$ give positive slack; historical
prefixes are not universally certified. The generator probabilities were
rechecked bit-exact against the original at all three world seeds. Priors and
libraries differ, so this is not an isolated horizon extension of the primary.

At $T=240$, frozen restoration/comparator criteria fail. The secondary revised
gate passes its three point criteria; its descriptive paired-world contrast
against separation is $+4.1$ points $[1.4,7.2]$. At $T=480$, a second fresh
panel gives $+0.6$ points $[-1.8,2.7]$: no resolved advantage, not a statistical
equivalence result. Near-null worlds remain hard, and all arms miss two matched
bumps in the $T=240$ panel. Large upper ceilings do not isolate the cause.
Zero null alarms in 300 campaigns per arm do not prove calibration beyond the
mathematical detector premises.

Later fixed-design diagnostics find large truth-numerator gaps on four difficult
bumps, but do not partition losses on actual adaptive histories. A three-expert
numerator fails its registered improvement criterion: 52.0\% for both versions
on the reused $T=480$ panel. Neither a universal coding-cost floor nor
irreparability follows from that negative result (Appendix~\ref{app:oct26audit}).

At $T=60$ the revision is applied unchanged to earlier drug, location,
dose/soybean and tree worlds. All four point criteria pass; drug detection
improves from 59.31\% to 76.79\%, above class EIG's 70.17\%. Shared saved
columns reproduce the prior drug run exactly, including every one of its 32
world/control rows. This is gate transfer on previously evaluated worlds, not
fresh-world confirmation. Control premiums range from $+0.015\%$ to
$+0.112\%$ for the revision. At $T=480$ they are $+0.014\%$ for the revision
and $+0.365\%$ for separation. These are relative total prediction-risk
premiums; small values can mask relative differences in reducible error.

\section{Language-model experimenters}
\label{conf:llm}
Selected frontier-model pilots concentrate designs and detect much less often
than random sampling. This association is consistent with the own-design
ceiling but does not identify capability as a cause. Model, prompt, provider and
selection vary together; campaigns are few and sometimes incomplete. A monitor
prompt can improve detection, but completed ablations do not isolate a causal
contribution of numerical certificate values from inadequacy framing or random
reference information. Setting-specific pointing gains remain unresolved.

The stronger practical intervention is a harness that directly reserves actions
for falsification. With the frozen $T=30$ gate, wrapped Argon detects 26/32
misspecified campaigns (81.25\%); all 48 planned campaigns complete, including
8 wrapper and 8 plain controls. Wrapper controls have no alarms and a paired
risk premium of $+0.226\%$. Plain/monitor reference detections are 12.5\% and
71.9\% from separate interleaved reruns. These are external references, not a
concurrent randomized wrapper comparison.

Wrapped Sol completes 24 of 32 planned non-null campaigns before the cap and
detects 19 (79.17\% campaign-weighted; 81.25\% equally weighted over the 16
represented worlds). A separate control-completion run brings the combined
control allocation to eight pairs, zero wrapper alarms and $+0.201\%$ premium.
Initial W3 is preserved as FAIL/unassessable; completion W3 passes. The eight
non-null missing campaigns stay missing: full planned detection could be
59.38--84.38\% without assumptions. Available-case W1--W2 pass against external
7.7\% plain and 61.5\% monitor references. Neither missingness nor these small
controls permits a general safety claim.

An API-free control wraps greedy and random designers on the same worlds with
200 seeds. Their detection becomes 76.3\% and 76.9\%, versus 0.1\% and 57.3\%
plain. Registered point criteria pass; a two-seed preview was included. These
similar rates support robustness on the selected panel, not agent independence
or equivalence to frontier models. Because the harness takes about half the
actions, improvement need not represent improved model reasoning.

\section{Discussion}
The main distinction is between learning parameters and making a model's errors
observable. An own-design power ceiling can show that a sampling policy leaves
little power for any valid adequacy test. When useful
signal is present, however, a poorly specified predictive numerator can still
lose it. Reporting detection alone cannot identify which failure occurred.

The posterior diagnostic and the budget rule make this distinction operational
within a declared model expansion. They do not remove dependence on that
expansion. A small certificate can reflect missing alternatives, while a large
certificate need not imply enough cumulative evidence to reject. The futility
bound explains a dependence on remaining time; the counterexample shows why
mean discrepancy and a threshold constant alone cannot supply achievability.

The empirical findings support a qualified intervention claim. Reserving actions
for falsification repairs substantial detection loss in several evaluated
settings. It also fails registered criteria, loses to class EIG on the corrected
measured replay, and has no resolved advantage over separation on the longer
fresh continuous-null panel. The findings support a setting-dependent
intervention claim; the revised rule remains distinct from the frozen hypothesis.
Prediction, detection and computation should remain separate endpoints: the
rule does not reduce the number of observations, and a small total prediction
premium can conceal a larger change in reducible error.

Use beyond these studies requires care about statistical validity and the
source of the evidence. Mathematical false-rejection control depends on the
scientific null, predictable numerator and numerical enclosure; a small control
panel with zero alarms cannot establish those premises. Replay with full-data
labels and replacement does not provide prospective validation on new scientific
units. The selected language-model pilots also leave the causes of detection
collapse unresolved, including any role of capability or numerical monitoring. Faithful learned-design comparisons and independent-prior confirmation
would strengthen a method claim; prospective experiments would be needed for a
discovery or deployment claim. The present evidence instead supports treating
model checking as a distinct design objective and auditing both the information
acquired and the evidence a detector can retain.
\label{conf:mainend}
\clearpage

\clearpage
\appendix
\section*{Reproducibility and study status}
The technical record retains registered failures, source/run identities,
assumptions, numerical qualifications and retrospective diagnostics. Private
review materials contain the manuscript source and deterministic checks;
separate study-specific packets preserve frozen code, fixed inputs, retained
outputs and independent audits for the corrected replay and the failed numerator
diagnostic. They do not collectively reproduce every empirical study or
independently rederive every saved numerical certificate. No public code/data
archive is linked in this version. The later certified development study has
completed collection; its independent index/summary audit is separate from the
pending full verification of retained histories. Its numerical outcomes remain
outside this version's scientific evidence. Any later confirmation would require
its own prior registration and fresh-world evaluation. This preprint makes no
claim of general policy superiority or prospective scientific validation.

\section{Introduction}
\label{sec:intro}

A scientific agent that chooses its own experiments is graded on data it selected
itself. Bayesian experimental design makes this choice principled: pick the design that
maximizes expected information gain about the parameters of the working model
\citep{rainforth2023modern,foster2021dad}. Language-model scientists make the same choice by
reasoning \citep{gandhi2025boxinggym,huang2025popper}. Both are evaluated by how fast they
learn \emph{inside} the model they hold. Real scientific models are wrong somewhere, so a
second question matters as much: when the model is wrong, does the agent's own
experimentation leave it any chance of noticing?

Figure~\ref{fig:capability} gives the answer for ten experimenters on the same planted
errors. Random experimentation catches the error in 58\% of campaigns. Small open language
models do worse or about as well, and get better with size. Then the trend breaks: two
frontier reasoning models from two providers catch the error in 6--9\% of campaigns, close to
the 0\% of the information-maximizing optimizer, and their experiments are nearly as
concentrated as its; across four frontier models, detection falls roughly as design
concentration rises. These observations motivate a conditional claim: when estimation-oriented designs leave
little discrepancy visible, they also limit detection power. We formalize that limit and
measure the detection--accuracy trade-off of reserving effort for model discrimination.
The pilot does not establish that increasing capability generally causes blindness.

\begin{figure}[t]
  \centering
  \includegraphics[width=\linewidth]{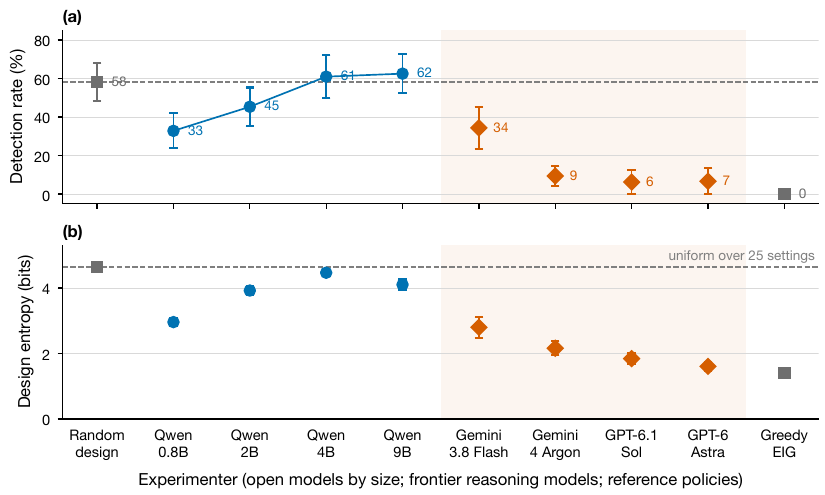}
  \caption{\textbf{Several tested estimation-oriented experimenters retain little detection power.}
  (a) Rate at which a valid adequacy test detects a planted model error within 30
  experiments, and (b) entropy of the experimenter's design distribution, on 16 misspecified
  worlds in four families (15 for GPT-6 Astra). Open Qwen3.5 models (blue) keep detection
  power by scanning or repeating settings; several frontier reasoning models
  (red, two providers) concentrate their designs like the information-maximizing
  optimizer (greedy EIG). The association between concentration and detection in
  these selected worlds is not a causal capability comparison. Whiskers are
  $\pm1$ s.e.\ across worlds; open models use 4 campaigns per world, frontier models 1--2
  (pilot), random design and greedy EIG 200. The dashed lines mark random design's detection
  rate and the entropy of a uniform design.}
  \label{fig:capability}
\end{figure}

The question is old in classical design. \citet{box1959basis} asked for designs under
which ``there is a high chance that inadequacy of $f(\xi)$ to represent $g(\xi)$ will be
detected,'' and \citet{atkinson1972planning} constructed designs that serve estimation
and an adequacy test against a nested extended model at the same time. We study the setting in which designs are now chosen sequentially and adaptively by an agent, the
agent's model is updated after every outcome, and the decision to trust or revise the
model has to be made at a data-dependent time. Static lack-of-fit arguments do not
transfer directly, because the design distribution is itself a function of the
(misspecified) posterior \citep{esponda2016berk,fudenberg2021limit}.

\paragraph{Contributions.}
\begin{enumerate}
\item \textbf{A power ceiling for adaptive designs} (Section~\ref{sec:theory}). For any
  policy --- greedy EIG, an amortized design network, or a language model --- the
  probability that an anytime-valid adequacy test ever fires within $T$ steps is at most
  the largest $p$ with $\mathrm{kl}(p\,\|\,\alpha) \le T\,D(\nu_T)$, where $D(\nu_T)$ is the
  misfit of the best working model on the policy's own design occupancy $\nu_T$. Policies that concentrate on a
  design set where one working-model member agrees with the truth have power at most
  $\alpha$ at every horizon. A sufficient condition, proved with a test supermartingale that needs no bound on
  likelihood ratios, uses cumulative separation, including errors identifiable only by combining several
  individually inconclusive designs. It is a sufficient guarantee, not a matching minimax rate.
\item \textbf{An expansion-relative uncertainty diagnostic} (Section~\ref{sec:certificate}). An agent-computable
  quantity that splits exactly into predictive uncertainty and model discrepancy, equals the posterior expected error of
  the working model when the truth is drawn from the expansion's prior, and whose ignorance part is
  a supermartingale that falls in conditional expectation by the predictive information each experiment carries about the
  target, so the agent can condition additional discrimination effort on expansion-relative uncertainty.
  A finite-model fixed-truth bound gives a conservative calibration, while exact
  counterexamples separate failure of action ranking from premature return to the base policy.
  Its interpretation does not extend to truths outside the proposed expansion.
\item \textbf{Generic-family detection collapse; environment-dependent accuracy}
  (Section~\ref{sec:experiments}). Across seven families and randomly drawn
  misspecifications, greedy EIG removes essentially all detection power, while its effect
  on accuracy changes sign across families. We report a pre-registered predictor of the
  accuracy effect that failed.
\item \textbf{Design and detector trade-offs} (Sections~\ref{sec:budget} and
  \ref{sec:crossed}). Budgeted discrimination recovers much of random-design detection
  in the initial generic benchmark, while fitted-curve stress tests expose fixed-budget
  recovery failures. Crossed comparisons separate acquisition from detector scope.
  A frozen Bernoulli comparison retains a gain over the tested hybrid under a
  continuous-null denominator; Gaussian endpoint gains are negligible.
  The broader categorical numerator has an explicit finite-sample coding cost.
  The completed independent-prior exact-cap comparison yields no eligible
  confirmation horizon, limiting broader claims for the proposed acquisition.
\item \textbf{Language-model experimenters} (Section~\ref{sec:llm}). A harness that lets
  an LLM choose experiments while detection is scored exactly. Small open models keep their
  detection power by scanning the design list; frontier reasoning models from two providers
  design like optimizers, leave little misfit visible on their own designs, and lose it.
  A tested monitor prompt raises measured detection in two models even without targeted
  setting suggestions. The completed component ablation does not isolate numerical-score
  use, and setting-specific gains remain unresolved at this pilot sample size.
\end{enumerate}

\section{Related work}
\label{sec:related}

\paragraph{Design for model inadequacy.} \citet{box1959basis} trade the bias of a
fitted response surface against the chance of detecting its inadequacy;
\citet{atkinson1972planning} designs experiments for an original model and a nested
extension simultaneously; model-discrimination criteria such as Bayesian KL-optimality
\citep{tommasi2009bayesian} choose designs where rival models disagree. \citet{catanach2023metrics} explicitly define
expected discriminatory information through predictive KL divergence and study its
trade-off with information gain in dynamical systems. This is a close predecessor of
our discrepancy criterion. Under common one-step posterior predictives their
Eq.~41 gives $D_t(a)=\KL(q_t(\cdot\mid a)\|p_t(\cdot\mid a))$, while
memberwise ED adds the within-expansion information $U_t(a)$. This standard
identity can change action rankings, without proving superiority over their
trajectory-design method; our focus is adaptive design's effect on testing power. The
de la Garza phenomenon \citep{delagarza1954spacing} explains why estimation-optimal designs
use minimal support, which is exactly the regime in which our ceiling bites. Our
falsification budget combines established discrimination ideas with an anytime-valid
test; we do not claim the first sequential model-discrimination method or the idea of
reserving design effort for adequacy checking.

\paragraph{Adaptive design under misspecification.} Modern Bayesian experimental design
is reviewed by \citet{rainforth2023modern}; amortized policies include
\citet{foster2021dad}. Robustness to misspecification has been pursued through
generalized Bayesian updates \citep{barlas2025gboed} and max--min objectives
\citep{abdulsamad2026maximin}. A distinct evidence-aware objective,
\citet{chen2026evidential}'s bias-against criterion, controls misleading evidence
under the joint Bayesian model through a variance-penalized EIG bound; that
objective is different from rejecting a composite scientific model class.
We study the latter detection question. \citet{bach2007active}
analyses active learning for misspecified generalized linear models. In economics,
\citet{esponda2016berk} and \citet{fudenberg2021limit} characterize where misspecified
agents who choose their own data end up (Berk--Nash equilibria); our own-design misfit
$D(\nu)$ is the residual at such a fixed point.

\paragraph{Model-class revision and a close contemporary predecessor.}
\citet{ma2026outgrow} already combine within-class information blind spots,
class-level discrimination across complementary experiments, exploration, and anytime-valid
class rejection with structural revision. Our work must be evaluated against this close
predecessor, and we do not claim to introduce those concepts. Their blind spot is a
two-design Gaussian example in which one-step information gain never observes the class
error; their separation quantity $J^\star$ is the \emph{best achievable} class separation over
design allocations, and the scaling of rejection cost with $\log(1/\alpha)/J^\star$ is reported
empirically; and their discovery policy is learned, pairing structural edits with experiments,
from 2,000 training environments. The contributions here are complementary. (i) The ceiling
of Theorem~\ref{thm:ceiling} bounds the power of \emph{any} level-$\alpha$ test through the
misfit visible on the policy's \emph{own} design occupancy, and the sufficient condition gives the
sufficient direction under additional premises; neither establishes a matching minimax rate.
(ii) We measure how often this happens across generic and mechanistic families, on measured
replicate data, in a 100-design screen and in frontier language-model agents. (iii) The
blindness certificate is computable from the agent's own posteriors and comes with
Bayesian and fixed-truth calibration and an information budget. (iv) The final rule needs no
training: it gates probing on the certificate, uses a target score that avoids the predictive
cancellation constructed in Appendix~\ref{app:predictivecancellation}, keeps an exploration floor so that
probing never stops forever, and its cost on a correct model is measured.

\paragraph{Evidence-driven allocation and finite deadlines.}
Multi-armed testing already optimizes evidence allocation:
\citet{sandoval2026multiarmed} establish oracle growth and high-confidence
rejection-time guarantees, and \citet{lin2026sampleefficient} analyze adaptive
sampling for multiple testing. Their conditional-increment premises are not
implied by the anytime validity of our profile-ratio process; its successive
value ratios can fail that premise (Appendix~\ref{app:deadlineoracle}).
\citet{clerico2026timesensitive} derive deadline-optimal testing from a
Neyman--Pearson event and its null Doob martingale. We use that established
construction to distinguish detector-specific zero power from unavailable
information in our existing finite-horizon example, not as a new general
optimality principle or a complete empirical comparison.

\paragraph{Anytime-valid testing.} We use e-processes and Ville's inequality
\citep{ramdas2023game}, with a universal-inference denominator so the test is valid at
every parameter of the working model under any adaptive design.

\paragraph{Controlled sensing, coverage and acquisition cost.}
Forced exploration is standard in sequential controlled sensing;
\citet{deshmukh2019controlled} ensure order-$\sqrt T$ sampling of each control
under their composite-testing assumptions. \citet{vershinin2025cost} distinguish
sample count from expected physical acquisition cost. Appendix~\ref{app:cycliccoverage}
combines explicit cyclic coverage with a universal alternative to remove the
pointwise informative-target assumption in a finite stationary setting.
Appendix~\ref{app:compactcoverage} extends the calculation to compact
continuous nulls with explicit covering and numerical-error penalties.
This distinct cyclic variant remains an unlaunched prototype; neither result
establishes practical efficiency or automatically covers the fixed-budget
policies in our completed independent-prior benchmark. \citet{yang2026openended} also study identifiability
and reliable scientific discovery, with stated generator-coverage and
finite-state assumptions for their recovery guarantees; neither their result
nor ours establishes unrestricted discovery.

\paragraph{Scientific agents.} \citet{dubova2026against} show in a multi-agent
simulation that theory-motivated experiment choice produces an ``illusion of epistemic
success'' and that random choice yields better theories; we give the detection side of
that observation a formal ceiling, test it with the acquisition function machine-learning
systems actually use, and evaluate targeted budgets as possible alternatives to random sampling.
BoxingGym \citep{gandhi2025boxinggym} benchmarks LLM experimental design;
POPPER \citep{huang2025popper} uses sequential e-values to validate a given hypothesis;
\citet{jhaveri2026failing} find confirmation bias in LLM rule discovery; LLM-IDEA
\citep{shetty2026llmidea} diagnoses non-identifiability. Our policy-level power analysis complements these studies. The ceiling follows standard
relative-entropy chain rules and data processing; its role here is to expose the
information bottleneck induced by the experimenter's design occupancy.

\section{Setup}
\label{sec:setup}

An agent runs $T$ experiments. At step $t$ it chooses a design $a_t \in \A$ (finite) and
observes $y_t \in \Y$ (finite) drawn from the unknown truth $P(\cdot\mid a_t)$. Its
working model is $\M_0 = \{p_\theta(\cdot\mid a) : \theta \in \Theta\}$; an expansion
$\M_1 \supseteq \M_0$ is available. A \emph{policy} $\pi$ maps the history
$H_{t-1} = (a_1, y_1, \ldots, a_{t-1}, y_{t-1})$, plus any internal randomness, to a
distribution over $\A$. Greedy EIG, an amortized design network and a prompted language
model are all policies. We write $P^\pi$ for the law of $H_T$ when the truth is $P$, and
$Q^\pi_\theta$ when the truth is $p_\theta$; the same policy acts in both worlds.

\paragraph{Adequacy e-process.} Let $q_s(\cdot\mid a)$ be any predictive distribution
fixed before $y_s$ is observed (we use the $\M_1$ posterior predictive). Define
\begin{equation}
\label{eq:eprocess}
E_t \;=\; \frac{\prod_{s \le t} q_s(y_s \mid a_s)}{\sup_{\theta\in\Theta} \prod_{s\le t} p_\theta(y_s\mid a_s)},
\qquad \tau = \inf\{t : E_t \ge 1/\alpha\}.
\end{equation}
The agent \emph{detects} misspecification at $\tau$. Without expansion it keeps its policy;
with expansion it switches to $\M_1$ at $\tau$ and continues with greedy EIG under $\M_1$.

\paragraph{Own-design misfit.} Let $\delta_\theta(a) = \KL(P(\cdot\mid a)\,\|\,p_\theta(\cdot\mid a))$
be the misfit of $\theta$ visible at design $a$, $\nu_T^\pi(a) = T^{-1}\,\mathbb{E}_{P^\pi}
\sum_{t\le T} \mathbf{1}\{a_t = a\}$ the policy's expected design occupancy, and
\begin{equation}
\label{eq:misfit}
D(\nu) \;=\; \min_{\theta\in\Theta} \sum_{a\in\A} \nu(a)\,\delta_\theta(a)
\end{equation}
the misfit of the best working model on the designs the policy actually uses.
Under uniform random design $D$ is the misfit of the projection onto $\M_0$ under the
uniform design measure; under a policy that concentrates its designs, $D$ can be zero
even though $\M_0$ is wrong elsewhere.

\section{A power ceiling for adaptive designs}
\label{sec:theory}

We assume $p_\theta(y\mid a) > 0$ for all $\theta, a, y$.

\begin{proposition}[Validity under adaptive design]
\label{prop:validity}
If the truth is $p_{\theta_0}$ for some $\theta_0\in\Theta$, then for every policy $\pi$,
$Q^\pi_{\theta_0}(\tau < \infty) \le \alpha$.
\end{proposition}

\begin{proof}
$E_t \le M_t := \prod_{s\le t} q_s(y_s\mid a_s)/p_{\theta_0}(y_s\mid a_s)$. Conditional on
$H_{s-1}$ and $a_s$, $\mathbb{E}[q_s(y_s\mid a_s)/p_{\theta_0}(y_s\mid a_s)] = \sum_y q_s(y\mid a_s) = 1$,
so $M$ is a nonnegative $Q^\pi_{\theta_0}$-martingale with $M_0 = 1$, whatever rule chose
$a_s$. Ville's inequality gives $Q^\pi_{\theta_0}(\sup_t M_t \ge 1/\alpha) \le \alpha$.
\end{proof}

Write $\mathrm{kl}(p\,\|\,q) = p\log\frac{p}{q} + (1-p)\log\frac{1-p}{1-q}$ for the binary
relative entropy and $\mathrm{kl}^{-1}_\alpha(K) = \max\{p\in[\alpha,1] : \mathrm{kl}(p\,\|\,\alpha)\le K\}$.

\begin{theorem}[Power ceiling]
\label{thm:ceiling}
For every policy $\pi$, every $\theta\in\Theta$ and every horizon $T$, let
$K_\theta = T\sum_a \nu_T^\pi(a)\,\delta_\theta(a)$, with $\nu^\pi_T$ the expected occupancy under the
truth. Then
\begin{equation}
\label{eq:ceiling}
P^\pi(\tau \le T) \;\le\; \mathrm{kl}^{-1}_\alpha(K_\theta) \;\le\; \alpha + \sqrt{K_\theta/2},
\qquad\text{hence}\qquad
P^\pi(\tau \le T) \;\le\; \mathrm{kl}^{-1}_\alpha\big(T\,D(\nu^\pi_T)\big).
\end{equation}
\end{theorem}

\begin{proof}
The event $A = \{\tau\le T\}$ is determined by $H_T$ and the policy's internal randomness.
The two path laws $P^\pi$ and $Q^\pi_\theta$ share every policy factor, so by the chain rule
$\KL(P^\pi\|Q^\pi_\theta) = \mathbb{E}_{P^\pi}\sum_{t\le T}\delta_\theta(a_t) = K_\theta$. By data
processing, $\mathrm{kl}(P^\pi(A)\,\|\,Q^\pi_\theta(A)) \le K_\theta$. By
Proposition~\ref{prop:validity}, $Q^\pi_\theta(A)\le\alpha$; if $P^\pi(A) > \alpha$, then because
$q\mapsto\mathrm{kl}(p\,\|\,q)$ is nonincreasing on $[0,p]$,
$\mathrm{kl}(P^\pi(A)\,\|\,\alpha)\le K_\theta$, which is the first inequality. Pinsker's
inequality $\mathrm{kl}(p\,\|\,\alpha)\ge 2(p-\alpha)^2$ gives the second. Minimizing over
$\theta$ gives the bound in terms of $D$.
\end{proof}

The proof uses only that $\{\tau\le T\}$ is an event of the history with probability at most $\alpha$
under $Q^\pi_\theta$. The same bound therefore holds for \emph{any} test, sequential or fixed-horizon,
with level $\alpha$ at $\theta$, including the most powerful test of $p_\theta$ against the true $P$,
which no agent can run because it requires knowing $P$. Section~\ref{sec:detector} uses this oracle
to separate information loss from detector weakness.

\begin{remark}
The ceiling holds for any numerator $q_s$, any expansion and any agent. It depends on the
policy only through where it looks, $\nu_T^\pi$, and not through how it reasons. A
language-model scientist is subject to it exactly as greedy EIG is.
\end{remark}

\begin{corollary}[Saturating designs never detect]
\label{cor:saturation}
Suppose $\nu_T^\pi$ is supported on $S\subseteq\A$ for all $T$ and $\M_0$ is saturated on
$S$: some $\theta_S$ has $p_{\theta_S}(\cdot\mid a) = P(\cdot\mid a)$ for all $a\in S$.
Then $P^\pi(\tau < \infty)\le\alpha$, as if the model were correct.
\end{corollary}

Estimation-optimal designs for a $k$-parameter model typically use about $k$ support
points \citep{delagarza1954spacing}, and a $k$-parameter family can often reproduce the
truth's response at $k$ points. EIG-greedy policies that concentrate near such designs
therefore drive $D(\nu_T)$ toward zero. Blindness means exactly this: some working model nearly
matches the truth on the occupied designs. That happens when the discrepancy lies off the support,
and also when it lies on the support but a working model can absorb it there.

\begin{corollary}[A necessary falsification budget]
\label{cor:budget}
If $P^\pi(\tau\le T) \ge 1-\beta > \alpha$, then $T\sum_a\nu_T^\pi(a)\,\delta_\theta(a) \ge
\mathrm{kl}(1-\beta\,\|\,\alpha)$ for every $\theta$ (about $2.65$ nats for $\alpha=\beta=0.05$).
In particular, if $\M_0$ is saturated on $S$ by $\theta_S$ and the policy places a fraction
$\gamma$ of its designs outside $S$, then
$\gamma \ge \mathrm{kl}(1-\beta\,\|\,\alpha) / (T \max_a \delta_{\theta_S}(a))$.
\end{corollary}

\begin{proof}
$\mathrm{kl}^{-1}_\alpha$ is increasing, so $\mathrm{kl}^{-1}_\alpha(K_\theta)\ge 1-\beta$ forces
$K_\theta\ge\mathrm{kl}(1-\beta\,\|\,\alpha)$. For the second claim, $\delta_{\theta_S}=0$ on $S$, so
$\sum_a \nu(a)\delta_{\theta_S}(a)\le\gamma\max_a\delta_{\theta_S}(a)$.
\end{proof}

Corollary~\ref{cor:budget} is necessary. A sufficient guarantee holds when the finite
expansion contains the truth, but it must allow several designs to discriminate jointly.
Define the R\'enyi divergence of order $\tfrac12$ and its cumulative separation by
\begin{equation}
\label{eq:cumulative-separation}
d_\theta(a)=-2\log\sum_y\sqrt{P(y\mid a)p_\theta(y\mid a)},\qquad
R_t=\min_{\theta\in\Theta}\sum_{s\le t}d_\theta(a_s).
\end{equation}
The minimum is taken \emph{after} accumulating the evidence over designs: different
parameters may fit each setting separately, while no single parameter fits them together.

\begin{theorem}[A sufficient cumulative falsification budget]
\label{thm:achieve}
Let $\Theta$ and $\M_1$ be finite, let $q_s$ be the posterior predictive of $\M_1$
under a uniform prior, and let $P\in\M_1$. For every adaptive policy, with probability
at least $1-\beta$, simultaneously for all $t$,
\begin{equation}
\label{eq:achieve}
\log E_t\ge R_t-2\log(|\Theta|/\beta)-\log|\M_1|.
\end{equation}
Consequently detection occurs by the first time
$R_t\ge\log(1/\alpha)+\log|\M_1|+2\log(|\Theta|/\beta)$, if such a time occurs.
\end{theorem}

\begin{proof}
The mixture numerator is at least $|\M_1|^{-1}\prod_{s\le t}P(y_s\mid a_s)$.
For each $\theta$, define
\[
\rho_\theta(a)=\sum_y\sqrt{P(y\mid a)p_\theta(y\mid a)}.
\]
Conditional on the past and $a_s$, the mean under $P$ of the square-root
likelihood ratio is $\rho_\theta(a_s)$.
Thus the process
\[
M_t^\theta=\prod_{s\le t}\frac{\sqrt{p_\theta(y_s\mid a_s)/P(y_s\mid a_s)}}{\rho_\theta(a_s)}
\]
is a nonnegative martingale. Ville's inequality and a union bound over $\Theta$
give, with probability at least $1-\beta$, for every $t$ and $\theta$,
\[
\sum_{s\le t}\log\frac{P(y_s\mid a_s)}{p_\theta(y_s\mid a_s)}
\ge\sum_{s\le t}d_\theta(a_s)-2\log(|\Theta|/\beta).
\]
Taking the minimum over $\theta$ and applying the mixture lower bound proves the result.
\end{proof}

No likelihood-ratio bound is required. If $N_t$ observations use designs satisfying
$\min_\theta d_\theta(a)\ge\Delta$, then $R_t\ge N_t\Delta$, recovering a sufficient
condition in terms of the \emph{realized} number of such probes. For independent budget
coins with probability $\gamma$, $N_T$ is random: replacing it by its mean $\gamma T$
is not a probability guarantee. If every selected probe satisfies this separation,
$\Pr(N_T<\gamma T/2)\le\exp(-\gamma T/8)$; requiring
$\gamma T\Delta/2\ge\log(|\M_1|/\alpha)+2\log(|\Theta|/\beta)$ and
$\gamma T\ge8\log(1/\eta)$ therefore gives power at least $1-\beta-\eta$.
These are sufficient conditions, not a guarantee that our predictive-KL rule finds
such settings or an optimal budget frontier. The necessary KL and sufficient
R\'enyi conditions need not match without further comparability assumptions.

\paragraph{Why cumulative separation matters.}
Consider two Bernoulli settings, null probability vectors $(0.2,0.8)$ and $(0.8,0.2)$,
and truth $(0.2,0.2)$. Each setting alone is matched by one null parameter, so
$\min_\theta d_\theta(a)=0$ everywhere. Alternating the settings instead gives
$R_T=(T/2)[-2\log(0.8)]$ for even $T$. Our independent constructive audit verifies
this distinction: at $T=60$ the exact final-time rejection probability is $0.9982$,
whereas repetition of one setting produced zero rejections in 2,000 campaigns.
Exact enumeration through $T=12$ gives false-alarm probabilities below $0.008$
under both null parameters for fixed and outcome-adaptive policies. This validates
an implementation and illustrates the theorem's scope; it is not external validation.

\section{Detection when the alternative is inadequate}
\label{sec:alternative-scope}
The ceiling concerns the information in the selected designs. A particular detector can
also fail because its alternative predicts those outcomes poorly. Along every history,
\begin{equation}
\label{eq:two-bottlenecks}
\log E_t=\left[\log L_{P,t}-\max_\theta\log L_{\theta,t}\right]
-\left[\log L_{P,t}-\log L_{1,t}\right],
\end{equation}
where $L_{P,t}$ is the unavailable true likelihood and $L_{1,t}$ is the finite
alternative mixture. This is an exact diagnostic decomposition, not an agent-computable
certificate; neither bracket must be nonnegative on an individual path.

\paragraph{Alternative prediction regret.}
Suppose both candidate classes are finite and positive, and the realized action proportions
converge almost surely to $\nu$, possibly random. A bounded martingale strong law for
candidate log likelihoods relative to $P$, and the constant log-sum-exp overhead of the
finite mixture, give
\[
\frac{\log E_t}{t}\longrightarrow D_0(\nu)-D_1(\nu),\qquad
D_i(\nu)=\min_{h\in\M_i}\sum_a\nu(a)\KL(P_a\,\|\,p_{h,a}).
\]
This is a finite-class specialization of established misspecified-likelihood asymptotics
\citep{chatterjee2018bayes}. When $D_0>0$ but $D_1=D_0$, the limiting linear evidence
growth vanishes despite informative data. Zero linear growth does not imply zero finite-time
power or rule out every sublinear crossing.

\paragraph{A broad categorical alternative.}
Unknown-alternative sequential tests using universal coding are established
\citep{sreedharan2013universal}. Here assign an independent uniform Dirichlet prior over
the $K$ outcome probabilities at each of the finitely many designs. With counts $n_{a,y}$
and $N_a=\sum_y n_{a,y}$, its predictable density is
$q_{\mathrm{sat},t}(y\mid a)=(n_{a,y}+1)/(N_a+K)$ and its sequence likelihood is
\[
L_{\mathrm{sat},t}=\prod_a\frac{(K-1)!\prod_y n_{a,y}!}{(N_a+K-1)!}.
\]
For every stationary categorical truth and every realized history,
\begin{equation}
\label{eq:broad-regret}
\log L_{\mathrm{sat},t}\ge\log L_{P,t}-C_t,\qquad
C_t=\sum_a\log\binom{N_a+K-1}{K-1}.
\end{equation}
Indeed the multinomial count probability is at most one, and the Dirichlet marginal equals
the inverse multinomial coefficient divided by $\binom{N_a+K-1}{K-1}$. This gives a
transparent conservative bound, not an optimal universal-coding constant.

\begin{proposition}[Broad-alternative cumulative separation]
\label{prop:broad}
For the finite positive null of Theorem~\ref{thm:achieve}, any common adaptive policy,
and any stationary categorical truth on the fixed finite design set, with probability
at least $1-\beta$, simultaneously for all $t$,
\[
\log\frac{L_{\mathrm{sat},t}}{\max_\theta L_{\theta,t}}
\ge R_t-C_t-2\log(|\Theta|/\beta).
\]
The true probabilities need not be members of the proposed finite expansion.
\end{proposition}

\begin{proof}
Apply the square-root likelihood-ratio martingale event in the proof of
Theorem~\ref{thm:achieve} to every null parameter and time, then combine its
true-likelihood lower bound with \eqref{eq:broad-regret}.
\end{proof}

Independent uniform exploration with a fixed positive probability samples every design
at a positive asymptotic rate. If $P$ differs from each finite null member somewhere,
$R_t$ then grows linearly almost surely, while $C_t=O(|\A|(K-1)\log t)$, giving eventual
rejection almost surely. The complexity penalty can nevertheless defeat power at practical
budgets. This argument does not cover unrestricted physical dynamics, a growing design set,
nonstationary responses, or a continuous null approximated by a finite grid.
For example, the new fitted-curve setup has 50 designs and 61 outcome categories
(60 finite bin edges plus both tails). At $T=30$, the broad Dirichlet regret bound
ranges from $54.87$ nats if every observation is at one design to $123.33$ nats if
all are at distinct designs. With 625 null members and $\alpha=\beta=0.05$,
the sufficient rejection certificate requires $R_t>76.73$ to $145.19$ nats.
This is a deterministic cost calculation, not a measured failure of a broad
detector on those datasets; a negative lower bound does not imply low actual power.

\paragraph{A fixed likelihood hedge.}
For proper sequence likelihoods $L_{j,t}$ and fixed positive weights summing to one,
$L_{\mathrm{hedge},t}=\sum_j\lambda_jL_{j,t}$ retains null validity and obeys
$\log E_{\mathrm{hedge},t}\ge\log E_{j,t}+\log\lambda_j$ for every expert and path.
We test a three-way mixture of a structured expansion, the full categorical alternative,
and a one-design-contamination alternative that integrates uniform categorical probabilities
at one design and uses a shared null member elsewhere. This is an established mixture
construction, not a new testing principle. It adds flexibility with an explicit likelihood
penalty and does not promise equal detection times.

\section{Falsification budgets}
\label{sec:budget}

Corollary~\ref{cor:budget} says detection must be paid for with design mass where the
working model is wrong. The agent does not know where that is, but it can ask where the
expansion and the working model disagree. For the Bayes-factor process $\prod_s q_s/\prod_s p_s$, the
expected one-step log-growth at design $a$ under the expansion's predictive $q_t$ is exactly
$\KL(q_t(\cdot\mid a)\,\|\,p_t(\cdot\mid a))$, where $p_t$ is the working model's predictive. For the
valid e-process \eqref{eq:eprocess}, whose denominator is a running maximum likelihood, it is at most
$\KL(q_t(\cdot\mid a)\,\|\,p_{\hat\theta_{t-1}}(\cdot\mid a))$ with $\hat\theta_{t-1}$ the running maximum-likelihood
point, and we use the posterior predictive as a smoothed proxy (Section~\ref{sec:growth} reports the completed comparison
with designs aimed at the e-process itself). A \emph{falsification budget} $\gamma$
runs the base policy, but at each step, with probability $\gamma$, plays
\begin{equation}
\label{eq:disc}
a_t \in \arg\max_{a\in\A}\; \KL\big(q_t(\cdot\mid a)\,\big\|\,p_t(\cdot\mid a)\big).
\end{equation}
This is the sequential analogue of a compound estimation-and-discrimination design
\citep{atkinson1972planning,tommasi2009bayesian,catanach2023metrics}; the predictive-KL
criterion itself is established rather than a new acquisition function. The natural control spends the same
budget on a uniformly random design (\emph{random mixing}); uniform random design is the
case $\gamma = 1$ of random mixing.

\subsection{Knowing when you are blind}
\label{sec:certificate}

Theorem~\ref{thm:ceiling} says an agent cannot tell whether its model is wrong on designs it
never visits. It can quantify unresolved predictive uncertainty \emph{within an expansion}:
whether hypotheses in that expansion still disagree on the protected target. This
is not a certificate that all possible errors could have been revealed. Let $w^1_t$ be the posterior over the expansion $\M_1$, $q_t$ its predictive
and $p_t$ the working model's predictive, and let $\mu$ weight the protected target. Define
the \emph{blindness certificate}
\begin{equation}
\label{eq:cert}
B_t = \sum_x \mu(x)\, \mathbb{E}_{h\sim w^1_t}\KL\big(p_h(\cdot\mid x)\,\big\|\,p_t(\cdot\mid x)\big).
\end{equation}

\begin{proposition}[Certificate]
\label{prop:cert}
(i) $B_t = U_t + D_t$, with \emph{ignorance}
$U_t = \sum_x\mu(x)\,\mathbb{E}_{h\sim w^1_t}\KL(p_h(\cdot\mid x)\,\|\,q_t(\cdot\mid x))$ and
\emph{evidence} $D_t = \sum_x\mu(x)\,\KL(q_t(\cdot\mid x)\,\|\,p_t(\cdot\mid x))$.
(ii) If the truth is drawn from the prior of $\M_1$, then
$\mathbb{E}\big[\sum_x\mu(x)\KL(P(\cdot\mid x)\,\|\,p_t(\cdot\mid x))\,\big|\,H_t\big] = B_t$ and
$\mathbb{E}\big[\sum_x\mu(x)\KL(P(\cdot\mid x)\,\|\,q_t(\cdot\mid x))\,\big|\,H_t\big] = U_t$.
\end{proposition}

\begin{proof}
(i) For each $x$, $\mathbb{E}_h\sum_y p_h\log\frac{p_h}{p_t} = \mathbb{E}_h\sum_y p_h\log\frac{p_h}{q_t}
+ \sum_y q_t\log\frac{q_t}{p_t}$, because $\mathbb{E}_h p_h = q_t$. (ii) Under the Bayesian joint, and
because the policy's randomness is independent of the truth, the truth given $H_t$ is distributed as
$w^1_t$ (policy factors cancel), and both functionals are evaluated at quantities fixed by $H_t$.
\end{proof}

The identities in (ii) are Bayesian: they average over truths drawn from the expansion's prior.
A looser fixed-truth bound is available for a fixed finite expansion; neither statement
covers truths outside $\M_1$.

\begin{proposition}[Fixed-truth calibration in a finite expansion]
\label{prop:fixedtruth}
Let $\M_1$ be finite with strictly positive outcome probabilities and a positive
prior $\pi$ fixed before the data. Fix $P\in\M_1$. Under every predictable policy,
with $P$-probability at least $1-\beta$, simultaneously for all $t$,
\[
w^1_t(P)>\beta\pi(P),\qquad
\mathrm{err}^1_t\le\frac{U_t}{\beta\pi(P)},\qquad
\mathrm{err}^0_t\le\frac{B_t}{\beta\pi(P)},
\]
where $\mathrm{err}^1_t=\sum_x\mu(x)\KL(P(\cdot\mid x)\|q_t(\cdot\mid x))$
and $\mathrm{err}^0_t$ uses $p_t$ instead of $q_t$.
The computable set $C_t=\{h:w^1_t(h)>\beta\pi(h)\}$ contains $P$ at every time
on that event, so $\max_{h\in C_t}\sum_x\mu(x)\KL(p_h(\cdot\mid x)\|p_t(\cdot\mid x))$
also upper-bounds $\mathrm{err}^0_t$.
\end{proposition}

\begin{proof}
Writing $Q_t=\sum_h\pi(h)L_{h,t}$, the ratio $Q_t/L_{P,t}$ is a nonnegative
mean-one martingale under $P$. Ville's inequality gives
$Q_t/L_{P,t}<1/\beta$ simultaneously in time with probability at least $1-\beta$.
Since $w^1_t(P)=\pi(P)L_{P,t}/Q_t$, the posterior bound follows. Keep the
nonnegative $P$ summand in the definitions of $U_t$ and $B_t$ to obtain the
error bounds; the last claim follows from $P\in C_t$.
\end{proof}

For a uniform prior the factor is $m/\beta$, with $m=|\M_1|$.
Thus small $U_t$ controls uncertainty in the expansion but leaves $D_t$ in the
working-predictive error bound. A sufficient conservative gate for working error
at most $\varepsilon$ is $B_t\le\beta\pi_{\min}\varepsilon$; the confidence-set
maximum can avoid that worst-prior-mass factor. These are represented-truth
statements for one fixed expansion and prior, not guarantees under arbitrary
class revision or continuous-truth quadrature. Appendix~\ref{app:continuousriskgate}
gives a standard likelihood-confidence-set extension over a declared continuous
family, with a protected-risk gate and an implemented outward optimization
reference connected to a proper sequence numerator and causal campaign recorder.
A small registered feasibility study finds no intervention savings and substantial
computational overhead (Appendix~\ref{app:riskgatefeasibility}); broader benefits remain unestablished;
a finite grid maximum is not a certified continuous bound.
Section~\ref{sec:wrongexp}
reports the completed test of that limitation. $U_t$ is computable from the agent's own models and data, with no access
to the truth. It has a second reading that explains how it falls.

\begin{proposition}[Expected ignorance decreases by target-predictive information]
\label{prop:ignorance}
For each target $x$, $U_t(x) = \mathbb{E}_{h\sim w^1_t}\KL(p_h(\cdot\mid x)\,\|\,q_t(\cdot\mid x)) =
I(h;Y_x\mid H_t)$, the mutual information between the expansion's hypothesis and a fresh outcome at
$x$. If the truth is drawn from the prior of $\M_1$, then for every policy
\[
\mathbb{E}\big[U_{t+1}\,\big|\,H_t, a_{t+1}\big] \;=\; U_t \;-\; \sum_x\mu(x)\,I\big(Y_x;\,y_{t+1}\,\big|\,H_t, a_{t+1}\big)\;\le\;U_t .
\]
\end{proposition}

\begin{proof}
The identity is the definition of mutual information with $q_t = \mathbb{E}_{h\sim w^1_t}p_h$. Given
$h$, the next outcome $y_{t+1}$ and a fresh $Y_x$ are independent, so
$I(h;Y_x,y_{t+1}) = I(h;y_{t+1}) + I(h;Y_x\mid y_{t+1})$ and
$I(y_{t+1};h,Y_x) = I(y_{t+1};h)$ give $I(h;Y_x\mid y_{t+1}) = I(h;Y_x) - I(Y_x;y_{t+1})$, all
conditional on $(H_t,a_{t+1})$; averaging over $y_{t+1}$ and weighting by $\mu$ gives the claim.
\end{proof}

So $U_t$ is a nonnegative supermartingale under the Bayesian joint: its conditional mean
decreases by target-predictive information. It need not decrease on an individual trajectory.
In the constructive audit, one observation raises $U$ from $0.0799$ to $0.3267$ nats,
while its conditional mean falls to $0.0544$. A restricted singleton expansion can
have $B=U=0$ while true protected KL error is $0.4159$ nats. Low ignorance therefore
cannot establish adequacy when the expansion is wrong. The
decrease is the expected predictive information gain of \citet{bickfordsmith2023prediction}, computed
under $\M_1$ rather than $\M_0$; a working-model EIG step that the expansion deems uninformative about
the target leaves the agent exactly as blind as before. $U_t$ is large when members of the expansion
that agree on the designs used so far disagree on the target, which is when concentrated designs
leave the own-design misfit of Theorem~\ref{thm:ceiling} small; this link is heuristic, since the
ceiling depends on the unknown truth. An agent that computes $U_t$ therefore knows when it is blind
\emph{relative to its expansion}, even though it cannot know whether it is wrong. This suggests an
\emph{adaptive falsification budget}: play the targeted design \eqref{eq:disc} while $U_t\ge u^\star$
and the base policy otherwise, so that targeted effort depends on unresolved uncertainty within the expansion. Under the Bayesian joint, the nonnegative supermartingale converges but need not
fall below $u^\star$; before expansion, the rule then continues to select targeted designs.

\begin{corollary}[Interpretation of the pre-expansion uncertainty threshold]
\label{cor:adaptive}
Suppose the truth is drawn from the prior of $\M_1$. Before adequacy-triggered expansion, at every history $H_t$ at which the adaptive rule
selects its base policy for step $t+1$,
\begin{gather*}
\mathbb{E}\Big[\sum_x\mu(x)\KL\big(P(\cdot\mid x)\,\|\,q_t(\cdot\mid x)\big)\,\Big|\,H_t\Big] < u^\star
\quad\text{and}\\
\mathbb{E}\Big[\sum_x\mu(x)\KL\big(P(\cdot\mid x)\,\|\,p_t(\cdot\mid x)\big)\,\Big|\,H_t\Big] < u^\star + D_t.
\end{gather*}
\end{corollary}

\begin{proof}
Before expansion, the rule selects its base policy at step $t+1$ only when $U_t<u^\star$; apply
Proposition~\ref{prop:cert}(ii) and $B_t = U_t + D_t$.
\end{proof}

In words: whenever the pre-expansion rule selects its base policy, its posterior expected error about the
target, under the expansion's prior, is below the tolerance $u^\star$ plus the discrepancy $D_t$ it
has \emph{already} seen. This is a posterior-mean statement under the expansion prior, not a frequentist
high-probability guarantee for a fixed truth. It also does not imply the adequacy
test has rejected: predictive discrepancy $D_t$ and rejection are distinct.

\paragraph{Action ranking and intervention triggering have different failure modes.}
Predictive averages can cancel: Appendix~\ref{app:predictivecancellation} gives
a finite positive construction in which predictive-KL targeting keeps selecting
an uninformative design with probability at least $3/4-\alpha$.
Changing the score does not by itself repair the trigger.
Proposition~\ref{prop:triggerfailure} gives a two-model example containing the truth
where both predictive-KL and expected-discrepancy targeting choose the useful
design, but $U$ falls below $0.1$ after its first response on either branch.
The policy then returns to an uninformative base forever, the rejection statistic
stays at $7/5$ or $3/5$, and working target error remains
$\operatorname{kl}(0.9\|0.5)$. The same failure occurs at $u=\log(20)/30$.
Appendix~\ref{app:adaptiveaudit} audits a conditional detect-or-resolve statement;
its strong informativeness assumption is not proved for either greedy rule.
A whole-certificate gate changes the error interpretation and probing effort,
but using the same numeric threshold on $B$ is not a universal repair.

\section{Experiments}
\label{sec:experiments}

\paragraph{Families.} Seven grid-exact families with 25 designs each: dose--response
(logistic, Bernoulli), linear regression (Gaussian), Michaelis--Menten kinetics (Gaussian),
pharmacokinetic decay (Gaussian), a Poisson log-linear rate, hyperbolic temporal
discounting with a two-dimensional design (Bernoulli), and a two-dimensional logistic
boundary (Bernoulli). Gaussian outcomes are binned into 16 classes, so inference on the
parameter grid (576--625 points) is exact. The expansion $\M_1$ adds one local Gaussian bump
in link space (5 centres in 1-D, 9 in 2-D, 4 amplitudes), and contains $\M_0$.
Misspecified truths are $\M_0$ members plus a random local bump (random centre, width and
sign) or a random smooth deviation; well-specified controls are $\M_0$ members.
The protected target is predictive accuracy over all 25 designs (expected Brier score
against the truth). Horizon $T = 60$, $\alpha = 0.05$, 200 seeds per world and policy.
The simulator reproduces the original pilot study to five decimals. Pass/fail criteria for
every experiment were written down before it ran.

\paragraph{Detectability.} In four families --- Poisson, dose--response and both
two-dimensional families --- even random design almost never detected the planted errors
within 60 experiments, and extending the horizon to 200 did not change that
(pre-registered; random design then detected in 3.3\%, 11.2\%, 0\% and 0\% of campaigns,
with no false alarms). That is a limit of the expansion class and the horizon, not of the
design rule, and we report those families rather than drop them. Even there, greedy EIG
detected nothing in any of the 12 Poisson worlds while random design reached 67.5\% in one.
Our primary descriptive population includes all misspecified worlds. We additionally
call a world \emph{detectable} if random design detects in at least 20\% of seeds; 21 of 84 worlds
qualify (12 PK, 5 Michaelis--Menten, 4 linear). The all-world audit is post hoc. This subgroup is selected using
the same random-design outcomes reported in the comparison, so subgroup estimates
are descriptive and potentially selection-biased. A confirmatory evaluation needs
a separate screening seed set and independently generated worlds.

\begin{figure}[t]
  \centering
  \includegraphics[width=\linewidth]{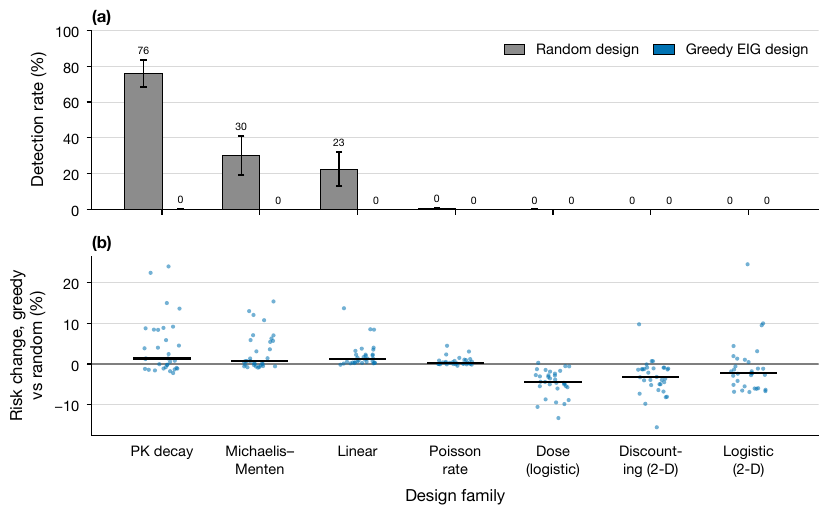}
  \caption{\textbf{Greedy EIG yields low detection in the generic families; its accuracy cost
  changes sign.} (a) Rate at which the valid adequacy test \eqref{eq:eprocess} fires within
  60 experiments, mean over 12 misspecified worlds per family (200 seeds each); whiskers
  are $\pm 1$ s.e.\ across worlds. (b) Change in protected risk of greedy EIG relative to
  random design, fixed working model, one dot per misspecified world (32 per family,
  200 seeds each); bars are medians.}
  \label{fig:collapse}
\end{figure}

\subsection{Detection collapses under EIG design}
\label{sec:collapse}

Figure~\ref{fig:collapse}a shows the main result. Over all 84 misspecified worlds the
adequacy test fired in 18.4\% of random-design campaigns and in 0.03\% of greedy-EIG
campaigns; on the 21 detectable worlds, 71.2\% versus 0.1\%. Per family, random versus
greedy detection was 75.8\% versus 0.2\% (PK), 30.2\% versus 0\% (Michaelis--Menten) and
22.6\% versus 0\% (linear). The test never fired in any of the 33{,}600 well-specified
campaigns (28 controls $\times$ 200 seeds $\times$ 6 policies), consistent with
Proposition~\ref{prop:validity}. All four pre-registered criteria of this experiment passed.

\paragraph{Own-design misfit tracks the collapse.} For every world and policy we computed the
policy's design occupancy $\nu_T$, its own-design misfit $D(\nu_T)$ exactly on the parameter
grid, and the ceiling of Theorem~\ref{thm:ceiling}. Across 336 (world, policy) cells the
observed detection rate never exceeded the ceiling (pre-registered; zero violations, a check of the
implementation and of Monte Carlo error rather than evidence for the theorem). The ceiling is loose: On the
21 detectable worlds greedy EIG used a median of 4 designs with own-design misfit
$0.0059$ nats per step, giving a median ceiling of $0.31$ (Pinsker: $0.47$) against an
observed detection rate of $0.001$; random design spread over all 25 designs with misfit
$0.30$. The evidence for the mechanism is therefore not the bound but the ranking: own-design misfit
ranks detection across all misspecified cells (Spearman $\rho = 0.77$). The targeted budget of Section~\ref{sec:budget} restores own-design misfit to
the random-design level ($0.30$) while using a median of 9 designs.

\paragraph{Is it the detector?}
\label{sec:detector}
The universal-inference test is conservative, so collapse could reflect a weak detector rather than
uninformative data. We applied four detectors to the same campaigns ($T=30$, 200 campaigns per world
and policy, the three detectable generic families and both mechanistic environments): the
universal-inference test; a posterior predictive check (discrepancy $-\sum_s\log p_\theta(y_s\mid a_s)$);
a fixed-horizon Bayes factor of $\M_1$ against $\M_0$; and an \emph{oracle}, the most powerful test of
the own-design projection $p_{\theta^\ast}$ against the true $P$, calibrated to level $0.05$ by running the
same policy on data from $p_{\theta^\ast}$. No agent can run the oracle, but by the remark after
Theorem~\ref{thm:ceiling} its population power bounds that of any test with level $\alpha$ at $\theta^\ast$.
The reported oracle uses a quantile estimated from 200 simulated null campaigns,
so its empirical power is an estimate rather than a certified upper bound; independent
calibration and calibration uncertainty are needed for a confirmatory comparison. On the 12
detectable generic worlds, greedy EIG's data defeated every detector: the oracle detected in 31.5\% of
campaigns against 99.6\% for random design, the predictive check in 10.5\% against 81.8\%, and the Bayes
factor in 4.8\% against 84.8\%. The collapse there is a loss of information. Two-compartment
pharmacokinetics is different: the oracle detected in 87.3\% of greedy campaigns, the Bayes factor in
46.5\% and the valid test in only 15.8\%, so most of that environment's partial collapse is the price
of a valid test that does not know the alternative. With substrate inhibition no detector showed a
collapse. Our pre-registered pooled criteria, which required greedy's oracle and practical detection
to fall below half of random design's over all 41 detectable worlds, failed, because those worlds are
dominated by the 20 substrate-inhibition worlds; the breakdown by environment is post hoc. The
ceiling held in all 228 oracle cells, and on the well-specified controls the predictive check raised
false alarms at about its nominal 4\%, the Bayes factor at most 0.5\% and the valid test none. The
valid test is also far from the oracle under random design (64\% against 99.6\%), which leaves room
for more powerful valid detectors.

\subsection{The accuracy effect is family-dependent}
\label{sec:accuracy}

The popular story that optimal experimentation under a wrong model also produces a
\emph{worse} model does not survive randomization over misspecifications
(Figure~\ref{fig:collapse}b). Across 224 misspecified worlds with a fixed working model,
greedy EIG was significantly worse than random design ($z > 3$) in 39.7\% of worlds and
significantly better in 41.1\%. Median relative changes ranged from $-4.4\%$
(dose--response) to $+1.3\%$ (PK); the largest losses reached $+24.5\%$. In the
dose--response family that motivated the pilot study, greedy design was never
significantly worse. A pre-registered predictor, which held that the loss occurs when the
error lies where greedy design does not look, failed with the wrong sign (AUROC 0.36):
losses occur where greedy design \emph{over}-samples the error. This matches
Corollary~\ref{cor:saturation}: the working model absorbs the discrepancy into its
parameters on a small support. It is detection, not accuracy, that adaptive design
reliably destroys.

\begin{figure}[t]
  \centering
  \includegraphics[width=\linewidth]{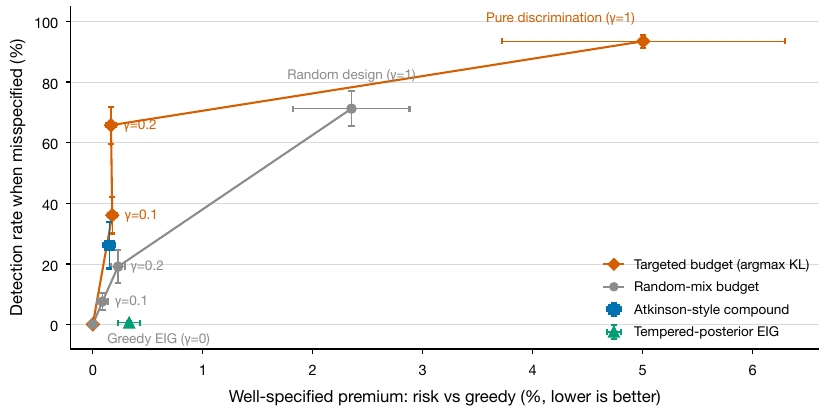}
  \caption{\textbf{A targeted falsification budget buys most of randomization's detection
  at a fraction of its cost.} Horizontal axis: mean relative increase in protected risk
  over greedy EIG on the 28 well-specified controls. Vertical axis: detection rate on the 21
  detectable misspecified worlds. Whiskers are $\pm 1$ s.e.\ across worlds; 200 seeds per
  world and policy; the working model is expanded on detection. Lines connect the measured
  budget settings and are guides only; pure discrimination is the targeted budget at
  $\gamma=1$. Separate markers show two baselines: an Atkinson-style compound criterion and
  greedy EIG under a tempered (generalized-Bayes) posterior.}
  \label{fig:budget}
\end{figure}

\subsection{Falsification budgets}
\label{sec:budget-results}

Figure~\ref{fig:budget} compares targeted budgets \eqref{eq:disc} with equal-budget random
mixing. Random design detects in 71.2\% of campaigns on detectable worlds but costs
$+2.35\%$ ($\pm 0.53$) risk in well-specified worlds. A targeted budget of $\gamma=0.2$
detects in 65.8\% (92\% of random design) at $+0.17\%$ ($\pm 0.04$); $\gamma=0.1$ detects in
36.1\% at $+0.18\%$. Random mixing at the same budgets detects in only 19.2\% and 7.6\%.
Targeting beat random mixing in all 21 detectable worlds at both budgets (sign test
$p = 4.8\times 10^{-7}$). Because the agent expands its model when the test fires, detection
also buys accuracy: on detectable worlds, protected risk relative to greedy EIG fell by
14.4\% under random design, 11.5\% under $\gamma=0.2$ and 6.8\% under $\gamma=0.1$.
A post hoc audit retaining all 84 misspecified worlds gives detection rates of
$17.68\%$ for the targeted 20\% budget, $18.44\%$ for random design, and
$4.84\%$ for equal-budget random mixing. The targeted budget's mean difference
from greedy EIG is $+17.65$ percentage points, with a descriptive 95\% bootstrap
interval $[13.49,22.15]$ obtained by resampling worlds within each family.
This interval reflects variation across the existing worlds and does not separately
model within-world campaign uncertainty; it is not an independent confirmation.
In an earlier run, using the pilot's Bayes-ratio detector (valid only for the
Bayes-marginal null), the pre-registered target of 80\% of random design's detection was
missed at $\gamma=0.1$ (65\%); $\gamma=0.2$ was then pre-registered for the valid-detector
run reported here.

\paragraph{Baselines.} We compared three alternatives on the same worlds and seeds, with
pre-registered criteria, all of which held. \emph{This tempered-posterior baseline does not restore detection:} greedy EIG under a tempered posterior (power $0.5$), as a simple robustness-inspired baseline, detected in 0.6\% of campaigns and cost
$+0.33\%$ in well-specified worlds. This baseline uses ordinary predictive EIG with tempered weights; it is not an
implementation of the Gibbs EIG acquisition of \citet{barlas2025gboed}, and does
not establish that GBOED or robust design more broadly fails to detect error. \emph{The tested compound setting is weaker than the tested stochastic budget:} maximizing normalized EIG plus half the normalized discrepancy, a compound-design-inspired heuristic rather than a reproduction of
\citet{atkinson1972planning}, detected in 26.2\% at
$+0.15\%$; the $\gamma=0.2$ budget detected 2.5 times as often at the same cost
($+0.17\%$). We conjecture, but have not tested, that a criterion applied at every step
keeps returning to designs whose discrepancy the current fit already absorbs, while
occasional full-weight probes do not. \emph{Pure discrimination over-pays:} spending every
experiment on \eqref{eq:disc} ($\gamma = 1$) detected most often (93.5\%) and reduced risk
most on detectable worlds ($-15.1\%$), but cost $+5.00\%$ when the model was right, twice the
premium of random design. Neither the compound coefficient nor the tempering power was optimized on a separate
development set, so these comparisons do not establish superiority to the respective
method families. Between greedy design and pure discrimination, the budget
$\gamma$ traces the trade-off between staying blind and paying for insurance; small budgets
sit near its efficient end (Figure~\ref{fig:budget}).

\paragraph{Sensitivity.} We varied one setting at a time around the defaults on the three
families where errors are detectable (pre-registered; all criteria held, and no setting
produced a false alarm). Greedy EIG never detected in more than 0.9\% of campaigns, while
random design detected in 64--82\%, across test levels $\alpha\in\{0.01,0.05,0.1\}$, horizons
$T\in\{30,60,120\}$ and expansion widths $\{0.06,0.12,0.24\}$. The targeted budget's detection
rose smoothly with its size: 14\%, 36\%, 66\%, 79\% and 88\% for $\gamma=0.05$, 0.1, 0.2, 0.3 and
0.5, with a premium that stayed below 40\% of random design's in every setting. A narrower
expansion (width 0.06) made the budget slightly more sensitive than random design (70.4\%
versus 69.4\%); a broader one made it less (56.7\% versus 73.4\%), plausibly because a broader
expansion class describes the localized planted errors less well.

\subsection{Mechanistic environments}
\label{sec:realistic}

The families above plant generic deviations. We also built two environments with physically
real errors and the expansion a scientist would actually try next, each with 200 log-spaced
designs instead of 25 (pre-registered; 20 misspecified and 4 well-specified worlds each,
200 campaigns per world and policy). In \emph{enzyme kinetics} the working model is
Michaelis--Menten and the truth, like the expansion, adds substrate inhibition
($K_i$ between 0.5 and 20 times $K_m$). In \emph{pharmacokinetics} the working model is
one-compartment elimination and the truth, like the expansion, is two-compartment (a fast
distribution phase holding 20--80\% of the dose, 3--30 times faster).

Detection collapse is environment-dependent here, and the pre-registered collapse criterion
failed in both. With substrate inhibition greedy EIG still detected in 94.6\% of campaigns
(random design 98.9\%): estimating $V_{\max}$ requires measurements at high substrate
concentration, which is exactly where inhibition shows, so the error lies on the optimizer's
own designs and Theorem~\ref{thm:ceiling} imposes no ceiling. With two-compartment
kinetics the collapse was partial: greedy EIG detected in 29.3\% against 80.9\% for random
design on the 11 worlds where any policy could detect (in the other 9, the fast phase was too
fast to be seen on the design grid by any policy). The falsification budget was robust in
both: at $\gamma=0.2$ it detected in 99.9\% and 87.4\% of campaigns, matching or exceeding random
design, at a well-specified premium of $+0.08\%$ and $+0.18\%$ against $+1.14\%$ and $+1.22\%$. In
pharmacokinetics, where the error mattered, it also cut protected risk by 27.3\% relative
to greedy EIG, the same as random design. The lesson is the one the theory states: an
optimizer goes blind when some working model can match the truth on the designs it needs, which is
common but not universal, and a small targeted budget is cheap insurance in either case.

\subsection{Expansion-relative uncertainty in practice}
\label{sec:cert-results}

\begin{figure}[t]
  \centering
  \includegraphics[width=\linewidth]{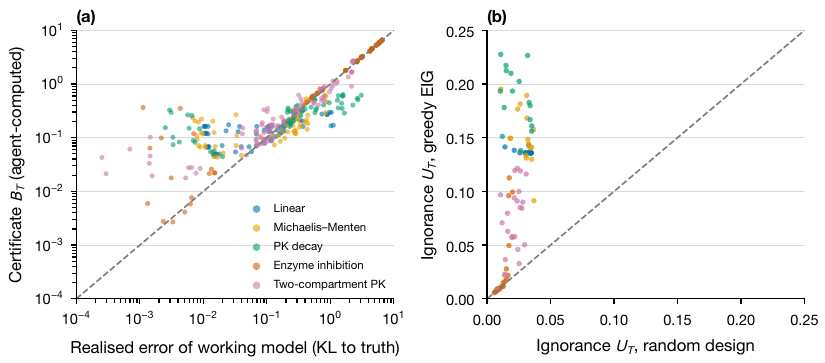}
  \caption{\textbf{Expansion-relative diagnostics correlate with realized error in the tested worlds.}
  (a) The blindness certificate $B_T$ of Proposition~\ref{prop:cert}, computed by the agent from
  its own models and data, against the realized protected error of its working model, for every
  (world, policy) cell; the dashed line is equality. (b) The ignorance part $U_T$ under greedy
  EIG against random design, one point per world; the dashed line is equality. 96 worlds in five
  environments (76 misspecified, 20 well-specified), 200 campaigns each, $T=60$.}
  \label{fig:certificate}
\end{figure}

We computed the certificate of Section~\ref{sec:certificate} in 96 worlds: the three generic
families where errors are detectable and the two mechanistic environments (pre-registered,
with one amendment recorded before any full run: after a one-world pilot showed that $B_t$ also
grows when an error is \emph{detected}, the blindness criterion and the adaptive trigger were
switched from $B_t$ to its ignorance part $U_t$). The diagnostic is associated with realized error in these worlds
(Figure~\ref{fig:certificate}a): across all cells the rank correlation between $B_T$ and the
realized error was 0.91, and within each environment 0.63--0.99, with the median ratio of
certificate to realized error between 0.90 and 1.30. It errs on the cautious side where the
error is small, because it also counts what the data could not rule out. The diagnostic also distinguishes unresolved uncertainty under the two policies (Figure~\ref{fig:certificate}b): greedy EIG ended with more ignorance than
random design in 100\% of worlds in four environments, about five times as much on average
(0.11 versus 0.02). The exception is informative: with substrate inhibition, where
Section~\ref{sec:realistic} found that greedy EIG is \emph{not} blind, its ignorance was higher in
only 62\% of worlds and close to random design's. This association is supportive in the tested environments; it is not proof of
calibration outside the expansion prior or a universally reliable blindness detector.

The adaptive budget, which spends targeted designs only while $U_t \ge u^\star$, used fewer of them
when the model was right (2.7--6.6 per campaign at $u^\star\ge 0.2$, against 12 for a fixed 20\%
budget) and at $u^\star = 0.1$ matched random design's detection on detectable worlds (84.0\% against
83.9\%) at a premium of $+0.08\%$ against $+0.89\%$. Against the frontier traced by fixed budgets
it detected 0.10--0.20 more at equal premium for $u^\star\in\{0.1,0.2,0.3\}$, meeting the
pre-registered criterion, but the 95\% bootstrap intervals over worlds include zero (for
$u^\star=0.2$, $[-0.01, 0.33]$), because the premiums being compared are a few hundredths of a
percent. We therefore claim that adaptive budgeting is a cheap operating point, not that it
dominates fixed budgets.

\subsection{Crossing acquisition and detector alternatives}
\label{sec:crossed}
A registered mechanism pilot uses 24 newly generated Bernoulli worlds in three shape groups
(single-cell, opposite two-cell and smooth cubic discrepancies), plus nine exact null controls.
There are nine designs, a nine-member linear-probability working class, and local and quadratic
finite expansions. Nine acquisition policies each generate 256 campaigns; seven numerator
experts score the \emph{same} history with a shared maximum-null denominator. Rejection changes
neither design nor prediction, which remains under $\M_0$.

At $T=600$, on identical random-design histories, the quadratic/full/contamination hedge
increases detection over the quadratic-only numerator by $39.99$, $80.86$ and $90.48$ percentage
points in the three shape groups. World-bootstrap intervals are $[33.10,47.56]$,
$[70.02,90.63]$ and $[80.27,97.66]$. This is a detector-scope effect; broad numerators
leave greedy endpoint histories almost entirely unrejected. Their complexity also matters:
the fully saturated alternative alone is weaker than the local finite alternative for some
localized errors.

The proposed quadratic hybrid, allocating 10\% of observations to predictive-KL targeting
and 10\% to random probing, missed its registered 10-point improvement criterion on pooled
two-cell/cubic worlds: the observed gain with a broad hedge was $9.40$ points
($[1.29,22.10]$), so that criterion fails. Its control excess-Brier premium is
$0.00002724$ at $T=600$, below the declared absolute limit but about 33\% of the small
remaining greedy excess error. Absolute and relative prediction costs tell different stories.

A symmetric Bernoulli expansion can have exactly the same initial mixture predictive as the
null despite substantial alternative uncertainty. If only zero-deviation endpoints are observed,
this cancellation can persist. Writing $q_1=\sum_h w_1(h)P_h$ for the expanded
posterior predictive, the standard information decomposition gives
\[
\mathbb E_{h\sim w_1}\operatorname{KL}(P_h\Vert q_0)
=\operatorname{KL}(q_1\Vert q_0)+I_{w_1}(h;Y),
\]
at each candidate design and observed history. The expected-KL score therefore
also rewards parameter information within the expansion. Its advantage over
mean-predictive KL cannot alone identify a distinct class-discrimination
mechanism; alternative-parameter EIG and combined-latent EIG remain required
controls. Class-label information with nuisance integration is already an
established design target \citep{drovandi2014modeluncertainty,hainy2022classification}.
We included a stronger posterior-expected-KL acquisition
control at the same 20\% budget; its good development performance motivated a frozen independent
comparison against the stronger local hybrid on new shapes and two observation modalities.
The frozen comparison is complete: on all 24 new Bernoulli worlds at $T=600$, the
expected-discrepancy control detects in $85.29\%$ against $61.60\%$ for the local hybrid,
a gain of $23.68$ points ($[20.07,27.10]$), meeting its registered ten-point target.
Fully random acquisition with the same hedge detects in $85.34\%$ on these worlds.
Thus this operating point approximately matches its detection while retaining an 80\%
estimation allocation; the result does not establish better detection than random design.
On the discretized Gaussian observations both methods are near full detection, with
only a $0.033$-point difference at this horizon. There is no substantive Gaussian gain
there. The initial mixture distributions coincide under the Bernoulli observation model
but differ under the Gaussian categories, confirming the algebraic modality distinction.
At the earlier predeclared Gaussian horizon $T=200$, expected discrepancy detects in
$96.35\%$, the hybrid in $85.48\%$, and the mean-predictive-KL control in $96.92\%$.
Thus the control is useful but does not establish superiority over the best tested acquisition
in that modality. The modalities also differ in signal/noise and coding dimension, so performance differences
do not isolate a pure causal effect of cancellation. A subsequent robustness replay uses
a conservative continuous-linear-null likelihood upper bound to test whether the Bernoulli
advantage survives removal of the coarse parameter-grid denominator. In the boundary-safe
confirmation, expected-discrepancy detection is $63.20\%$ versus $41.41\%$ for the hybrid,
a $21.79$-point difference ($[18.02,25.81]$), meeting the unchanged ten-point target.
The replay includes nine additional controls with continuous coefficients absent from the
inference grid. A supporting tangent globally upper-bounds the concave binomial likelihood
over endpoint probabilities in $[0,1]^2$, so its conservatism does not depend on a local
optimizer finding the supremum. The first replay failed its numerical-gap gate on one
greedy control; a child with bounded coordinate rescue clears the gate, with every
finite-detector cell unchanged. Final maximum supporting gap is below $10^{-5}$ and the
maximum sampled continuous-control alarm rate is $0.78\%$. These replay checks do not
supply new-world confirmation or extend the finite-null power theorem to an infinite null.

These are deliberately constructed off-endpoint discrepancies, not a prevalence estimate for
scientific discovery. Intervals resample eight worlds per shape and do not separately model
campaign Monte Carlo uncertainty. A fully calibrated diagnosis of false reassurance, realistic
domains, and faithful close-method comparisons remain necessary.

\paragraph{Exact audit outside the proposed expansion.}
Take truth $(0.2,0.2)$, null Bernoulli vectors $(0.2,0.8)$ and $(0.8,0.2)$, and an expansion
that additionally contains $(0.7,0.7)$. With fixed alternating designs at $T=100$, exact
final rejection probability is $7.44\times10^{-7}$ for that wrong finite mixture, and exceeds
$0.99999$ for the broad Beta alternative and their equal hedge. Here $D_0=D_1=0.4159$
nats, illustrating the alternative-regret bottleneck. Five registered numerical checks
pass, including exhaustive adaptive-path calculations through $T=12$ and 9,656 coding-bound
checks. They corroborate formulas and implementations, not novelty, prevalence or a theorem
proof.

\paragraph{Auditing the nonmyopic comparison.}
An independent finite-control audit checks a proposed two-step planner before larger
comparisons. The null laws have Bernoulli success probabilities $(0.2,0.2)$ and
$(0.8,0.8)$, and the structural alternatives $(0.2,0.8)$ and $(0.8,0.2)$.
The detector uses an equal four-law sequence mixture and the maximum-null denominator.
After histories restricted to the first action, a two-step rejection-probability
objective is identically zero. With fixed-action ties it remains uninformative:
at $T=12$ its power is zero. A class-information fallback gives $57.15\%$ power,
matching simple class information, predictive-KL acquisition and the exact full-horizon
Bayesian planner under the declared prior. All five audit gates pass, including
independent path enumeration and exact null control; this is one constructed example,
not a new acquisition advantage or a general failure of nonmyopic design.
In this setup $E_t\le(1+4^r)/2$ for $r$ measurements at the second action.
At threshold $20$ at least three are necessary. If the base action is fixed first
and only probe steps may use the second, a hard 20\% cap at $T=12$ permits only
two and cannot reject. Probabilistic mixing is a different constraint.
Randomized-tie null EIG is exactly the random control here; both actions are equally
informative under the null, so this case does not establish a prediction-cost trade-off.
Future comparisons require an explicit fallback/reward/horizon and probe-budget
feasibility, as well as strong simple acquisition controls.

\paragraph{Connecting the continuous-logistic detector.}
A separate integration audit uses three settings $x\in\{-1,0,1\}$ and the
scientific null $p(x)=\operatorname{sigmoid}(a+bx)$ for $(a,b)\in[-3,3]^2$.
An outward-decimal supporting-plane bound supplies the denominator; the
numerator is an equal hedge of a fixed U/inverse-U sequence mixture and
per-setting Beta$(1,1)$ marginals. Exhaustive length-12 response paths under
three predictable policies give 39 policy/truth cells. All five registered
checks pass; the largest exact first-crossing risk over 11 declared null laws
and three policies is $1.0254\%$, below $\alpha=0.05$. The finite fixtures are
implementation checks, not empirical validation of every continuous coefficient.
Cycling the three settings detects each declared shape with probability
$58.74\%$; endpoint-only sampling detects neither. At the endpoints each shape
is observationally identical to a valid constant logistic null, so any
level-$\alpha$ test there has power at most $\alpha$ at any horizon.
A simple outcome-adaptive rule gives $68.15\%$ and $9.41\%$ on the two shapes,
illustrating a failure of robustness rather than a new acquisition advantage.
All 12,288 stored paths reconcile in an independent replay. Among 1,281 count
states, eight require coordinate rescue and 19 retain optimizer non-success
flags; nevertheless the maximum independently certified profile gap is below
the registered $10^{-7}$ tolerance, with no unresolved threshold decisions.
The alternatives are deliberately strong and included in the numerator;
prediction costs, independent-prior confirmation and domain validation remain
untested by this audit.

\subsection{Independent-prior development with exact attempt caps}
\label{sec:independentprimary}
A separately registered comparison removes truth-indexed inference priors and
uses the continuous scientific null
$p(x)=\operatorname{sigmoid}(a+bx)$, $(a,b)\in[-3,3]^2$, at nine available actions.
Its inference prior is specified independently of the evaluation worlds,
using $9\times9$ nuisance quadrature, while the test denominator profiles the
full continuous null. The proper numerator is a fixed equal hedge of a
structural mixture and independent per-action Beta$(1,1)$ sequence marginals.
Six null worlds and 18 non-null worlds cover structural bumps, near-null
departures and changes outside the structural prior. Sixteen response-history
replicates, 12 policies and horizons 60 and 120 give 9,216 main campaigns;
12 prespecified $17\times17$ checks complete the allocation of 9,228.
The candidate here is fixed-schedule expected discrepancy, not the later
gated-floor heuristic. Prediction loss averages the nine actions uniformly.

All records pass full runner history replay and independent immutable-source,
raw/index, trace, resource and fixed-class endpoint checks. The recovered
4,928-record prefix is retained unchanged and 4,300 further records complete
remotely. No record is dropped. Capped policies receive exactly 12 or 24
pre-outcome override opportunities, including charges for choosing the base
action. All $H$ observations continue after rejection and prediction remains
in $M_0$, unlike the expansion-after-rejection studies below. The complete
allocation accounts for 830,520 observations and 262,872 charged attempts.

\begin{table}[htbp]
\centering\small
\caption{\textbf{No registered horizon meets the baseline detectability requirement.} Detection percentages over 18 equally weighted non-null worlds, with 16 response-history replicates per world and method. Parentheses give within-world Monte Carlo standard errors on this fixed panel, not population confidence intervals; a zero estimated error does not establish zero uncertainty. All physical observations continue to the horizon. References are ineligible for capped-baseline selection. Run \texttt{2095b9fe}, scientific registration \texttt{22c10ec2}, completed source \texttt{15cd075}.}
\label{tab:primary-independent-prior}
\begin{tabular}{lcrr}
\toprule
Method & Attempts ($60/120$) & $H=60$ & $H=120$\\
\midrule
Expected discrepancy (candidate) & 12/24 & 0.35 (0.35) & 4.51 (1.15)\\
Random overrides & 12/24 & 0.00 (0.00) & 0.69 (0.47)\\
Balanced coverage & 12/24 & 0.00 (0.00) & 0.00 (0.00)\\
Class MI & 12/24 & 1.39 (0.68) & 4.86 (1.24)\\
Alternative-parameter EIG & 12/24 & 0.69 (0.47) & 4.86 (1.21)\\
Combined latent EIG & 12/24 & 0.00 (0.00) & 4.86 (1.18)\\
Predictive KL & 12/24 & 1.39 (0.68) & 4.86 (1.24)\\
Two-step class information & 12/24 & 1.04 (0.59) & 3.12 (0.97)\\
Shared-parameter projection & 12/24 & 0.35 (0.35) & 2.08 (0.75)\\
\midrule
No override (reference) & 0/0 & 0.00 (0.00) & 0.00 (0.00)\\
Full random (reference) & 60/120 & 2.78 (0.95) & 7.29 (1.38)\\
Cycle (reference) & 60/120 & 2.43 (0.86) & 10.42 (1.46)\\
\bottomrule
\end{tabular}
\end{table}

The frozen rule chooses the highest-detection capped control over all 18
non-null worlds, breaking ties by lower null prediction loss and then name.
At $H=60$, class MI wins the tie at $4/288=1.39\%$; at $H=120$,
alternative-parameter EIG wins the tie at $14/288=4.86\%$. The candidate
gives $1/288$ and $13/288$, respectively (Table~\ref{tab:primary-independent-prior}).
Neither control estimate reaches the registered $20\%$--$95\%$ eligibility
range. Consequently \emph{no confirmation horizon or comparator is frozen}.
These are development estimates after control selection, not evidence of
population-best controls, inferential equivalence or universal candidate failure.
Mean null excess log losses are $0.02534$ versus $0.02112$ nats at the shorter
horizon and $0.01029$ versus $0.00994$ at the longer horizon, for the candidate
and respective highest-power controls. No null alarm is recorded in 96 main
campaigns per method and horizon; zero alarms do not establish exact calibration.

The logs retain 224 solver-gap flags in 183 histories. For every flagged
prefix, the stored numerator upper bound divided by the certified feasible-
likelihood lower bound is below 20. Thus even an exact null-profile calculation
cannot create a crossing at those prefixes; no numerical rule or endpoint is
changed. The 12 limited prior-grid pairs all retain nonrejection, but their
action sequences differ by up to 41 of 120 steps and their excess log losses
by up to $0.00709$ nats. These sparse checks do not establish grid-insensitivity
across controls or worlds.

\paragraph{A post hoc check of the observed information bottleneck.}
We additionally insert the simulator's true sequence likelihood into the
recorded denominator bounds, preserving the existing profile-gap acceptance
guard and each policy's saved history. This uses unavailable truth, samples no
outcomes, changes no actions and is not a deployable detector or new power
experiment. At $H=120$, 68 of the candidate's 288 non-null histories cross
under this truth-informed numerator, compared with 13 actual rejections;
55 histories cross only in the truth-informed diagnostic. In contrast, no
no-override history crosses even with that numerator. Full cycling yields
112 truth-informed crossings and 30 actual ones. These checks distinguish
histories that remain unrevealing from histories whose realized signal is
poorly captured by the implemented sequence numerator. They do not identify
an attainable replacement or establish a causal acquisition advantage.
The registered negative decision remains unchanged; fresh superiority
confirmation is not initiated from this study.

\paragraph{Ablating the two proper likelihood components.}
A further post hoc ablation scores the already implemented structural marginal
$L^{\rm str}_t$ and per-action Beta marginal $L^{\rm broad}_t$ separately on every saved history,
retaining the same profile bounds and numerical acceptance guard. All 9,228
original first-rejection endpoints reproduce. On the candidate's 288 non-null
histories at $H=120$, the component crossing counts are 10 and 13, against
13 for their equal hedge; both component event sets are contained in the
hedge's event set, with no unresolved event bounds. Since
$wL^{\rm str}_t+(1-w)L^{\rm broad}_t\le\max(L^{\rm str}_t,L^{\rm broad}_t)$
for every fixed $w\in[0,1]$, changing only
these mixture weights cannot add a rejection on the other saved histories.
At $H=60$ the broad component adds two events, so the conclusion is specific
to the longer-horizon histories. Neither retrospective component selection
nor the unpenalized maximum is a valid combined test at the original level.
No acquisition or prediction policy is rerun; the registered decision stays
unchanged. This narrows a possible repair direction without demonstrating a
useful attainable replacement or new-world improvement.

\subsection{Expansion and data-derived stress tests}
\label{sec:wrongexp}
The wrong-class/width experiment is now complete (96 worlds, 200 campaigns, $T=60$).
On the registered wrong-class random-detectable subgroup, the fixed budget detects in
$83.83\%$ against $84.96\%$ for random design, satisfying its 75\%-recovery criterion.
Within-environment rank correlations of $B_T$ with true working-model KL error range
from $0.396$ to $0.832$, meeting the registered correlation criterion. The expected increase
in false reassurance does \emph{not} occur: the registered conditional proportion is
$11.61\%$ for the wrong class and $13.83\%$ for the right class, so that hypothesis fails.
These conditional denominators and detectable sets differ, and correlation does not
establish out-of-class calibration.

Width mismatch changes the operating point: the narrow expansion's budget gives
$69.31\%$ detection versus $69.38\%$ for random design, while the wide expansion gives
$54.05\%$ versus $70.18\%$. These are expansion-specific selected subgroups; width
variants exist only in the generic environments, so their pooled statistics cannot be
compared directly with a right-class pool that also includes mechanistic environments.

\paragraph{Data-derived simulations.}
We fit richer curves to Indometh, Theoph and DNase data and generate independent outcomes
from those curves (41 fitted/control worlds, 200 campaigns per policy, 50 candidate settings).
This is a fitted-curve simulation, not a prospective experiment or evaluation on held-out
real observations. Working grids are centered on projections of the fitted truth, and
noise is modified by a bin-width floor; these choices limit external-validity claims.

At $T=30$ in the six fitted Indometh worlds, greedy detection is $12.33\%$, random design
$55.58\%$, the fixed 20\% budget $37.17\%$, and the adaptive rule $34.25\%$. The registered
collapse criterion passes, but the budget reaches only $66.9\%$ of random design's detection,
missing the registered $80\%$ target. This robustness criterion fails. At the shorter
$T=12$, the respective rates are $0.92\%$, $20.58\%$, $21.50\%$ and $31.08\%$; this descriptive
short-horizon pattern does not reverse the registered failure at $T=30$.

Exactly well-specified Theoph controls raise no recorded alarms in any arm at $T=30$, with
a fixed-budget total-risk premium of $0.034\%$, meeting the registered control criterion.
The near-well-specified fitted richer Theoph curves are reported separately; non-rejection
of a working model in the original data does not establish adequacy. In the 11 DNase curves,
random detection is $12.00\%$, the fixed budget $5.77\%$ and greedy design $0\%$; the
correlation of fitted discrepancy with random-design detection is $0.917$, descriptively.
Curve-fit uncertainty, fit/evaluation separation, and continuous-parameter grid sensitivity
remain unresolved. Repeated generated campaigns are not additional independent subjects.

\paragraph{Richer expansions on fitted curves.}
A registered follow-up compares the original mechanistic expansion, a finer mechanistic
grid, and its union with local response bumps on the same Indometh/DNase curves.
The union's fixed-budget detection improves by $19.64$ points over the original expansion
on seven commonly detectable worlds, but its recovery ratio is only $0.679$, below the
registered $0.8$ target. Thus X1 fails and X2/X3 pass. The adaptive union arm has
$88.25\%$ detection versus $77.69\%$ for random design on eight selected worlds.
This adaptive finding was exploratory, not the registered fixed-budget hypothesis.
Across all 17 fitted misspecified worlds the corresponding rates are $44.76\%$ and
$38.76\%$. Changing the expansion changes both acquisition and the detector numerator;
this experiment alone does not attribute the improvement to either component.

\paragraph{Confirmation on previously unused fitted datasets.}
The frozen construction is applied to 47 eligible ChickWeight curves and the treated/untreated
Puromycin curves, with 14 exact controls and 200 campaigns per world. These datasets
were unused in development, but their observations are fitted to synthetic response laws;
the inference grids are still centered on projections of the fitted truths.
They are not held-out-observation validation. No eligible fits fail.
On 35 random-detectable worlds at $T=30$, adaptive union detection is $96.47\%$,
random $82.23\%$, and the fixed budget $92.04\%$. C1 passes, but the $4.43$-point
adaptive/fixed gain misses the registered ten-point C2 target. C3 fails because one
of 2,800 adaptive control campaigns rejects, despite a small $0.0182\%$ mean
control total-risk premium. This does not by itself contradict level-$0.05$
validity; the registered requirement was zero observed alarms.
The gain over random remains across all 49 misspecified fitted worlds:
$74.39\%$ versus $59.90\%$, with the fixed budget at $68.87\%$.
These post hoc all-world results remove outcome selection from the descriptive pool.
Figure~\ref{fig:datafollowup} compares the all-world detection differences.
The pool is dominated by 47 chicks; two enzyme-state curves do not establish broad
cross-domain generalization. Adaptive design also uses more targeted probes:
$8.18$ versus $2.30$ before rejection on selected worlds, and $15.18$ versus $6.01$
on controls, out of 30 observations. It is not an equal-intervention-budget comparison,
and post-rejection prediction uses the expanded class. Matched-cost controls and
priors defined independently of fitted truths remain necessary.

\begin{figure}[t]
  \centering
  \begin{minipage}{5.5in}
    \includegraphics[width=\linewidth]{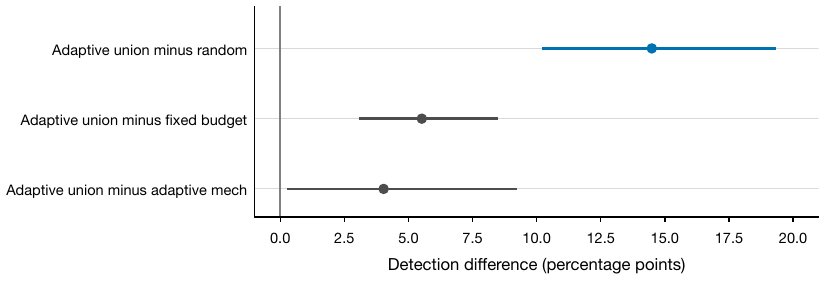}
  \end{minipage}
  \caption{\textbf{The adaptive gain shrinks against stronger controls.}
  All 49 previously unused fitted worlds at $T=30$, with 200 campaigns per world;
  whiskers are descriptive 95\% world-bootstrap intervals, not separate campaign
  uncertainty or population guarantees. Probing budgets differ, and the third
  contrast changes both acquisition and detector expansion. The registered
  selected-world and zero-alarm failures remain failures.}
  \label{fig:datafollowup}
\end{figure}

\paragraph{Second fitted-dataset study: an adaptive non-replication.}
The next registered study uses nine dose-response curves and 48 soybean plots,
plus 21 exact controls, with 200 generated campaigns per world and policy.
These are again fitted-curve simulations with inference grids centered on the
fitted truths, rather than held-out-observation validation. On 46 worlds selected
by observed random-design detection at $T=30$, the original adaptive rule detects
in $71.78\%$, below random design's $82.47\%$, the fixed budget's $91.75\%$,
expanded-model EIG's $86.53\%$, and class-discrimination EIG's $93.61\%$.
Adaptive class-discrimination targeting within the same uncertainty trigger
gives $74.08\%$. The original D1, D4 and D5 gates fail; their criteria are unchanged.
At $T=12$ the selected 26-world adaptive/random comparison reverses
($81.98\%$ versus $64.44\%$, D2 passes), but the heuristic
$u^\star=\log(1/\alpha)/T$ gives $73.13\%$ and fails D6. It is not a proved
calibration of protected risk. The control-cost/false-alarm gate D3 passes;
one of 4,200 adaptive control campaigns rejects at $T=30$.

Post hoc pooling over all 57 misspecified worlds retains the longer-horizon
adaptive deficit: $59.39\%$ versus random's $67.56\%$. The fixed budget gives
$75.48\%$, but this pool is dominated by 48 soybean plots. Its all-world
dose-response detection is $11.72\%$ versus random's $25.50\%$, whereas on
soybean it is $87.44\%$ versus $75.45\%$. Thus neither adaptive nor fixed
targeting is uniformly superior across these families. The original
outcome-selected gates remain the registered analyses; world bootstrap intervals
are descriptive and do not supply independent-subject replication. Shared seed
labels do not align outcome uniforms when policies consume different random
draws, and intervention budgets differ. Aggregate targeted-step counts do not
establish a temporal stopping mechanism; prospective action traces are needed
to test that explanation. These failures strengthen the case for independently
constructed priors, explicit realized-cost accounting and strong classical controls.

\paragraph{Whole-certificate triggering: an improved heuristic on reused worlds.}
A follow-up targets posterior-expected discrepancy and intervenes while
$B_t=U_t+D_t\ge0.1$ and the detector has not fired. On 152 development worlds
(122 misspecified and 30 controls), the 64 random-detectable worlds give
$96.27\%$ detection at $T=30$, versus $92.20\%$ for the original adaptive
rule and $86.95\%$ for expected-discrepancy targeting with the original
$U$ trigger. The original E1, E3 and secondary E4 gates pass. The four-model
construction used to amend the rule before launch is development evidence,
not a new independent test. Control campaigns use a mean 17.74 targeted steps,
versus 15.59 for the original adaptive rule and 6.00 for fixed targeting.

With that score and trigger frozen, a further evaluation reuses exactly the
57 misspecified dose-response/soybean worlds, 21 controls and seeds above.
All 936 shared comparator cells reproduce the previous run exactly; the
score and trigger code remain unchanged. On the selected 46-world pool at
$T=30$, detection is $96.48\%$, versus random's $82.47\%$, fixed targeting's
$91.75\%$ and full class EIG's $93.61\%$; all original F1--F4 point-mean
criteria pass. However, descriptive world-bootstrap intervals for the
differences from fixed targeting and full class EIG include zero:
$4.73$ points $[-0.68,10.19]$ and $2.87$ points $[-1.30,6.89]$.
The rule uses 6.68 targeted steps before detection on these worlds, versus
9.20 for the original adaptive rule and 2.30 for the fixed budget. Its mean total-risk control premium is $0.025\%$;
one of 4,200 control campaigns rejects. These costs are neither matched
physical budgets nor excess-risk premiums.

Across all 57 worlds, detection is $82.24\%$ versus full class EIG's
$79.85\%$. The family weighting matters: on all nine dose-response worlds
the rates are $35.39\%$ versus $39.67\%$, and equal family weighting gives
$63.20\%$ versus $63.53\%$. The positive pooled result is therefore a
promising simulation follow-up, not independent confirmation or uniform
superiority. The experiment also lacks predictive-KL targeting with the same
$B$ trigger, so it is not a full score-by-trigger factorial. Moreover, the
two-model construction in Appendix~\ref{app:adaptiveaudit} gives probability
at least $0.1$ of premature return under a $B$ threshold of $0.1$ itself.
Improving the empirical rule does not supply fixed-truth risk calibration.

\paragraph{New fitted tree-growth families and threshold sensitivity.}
The frozen whole-$B$ rule is also evaluated on two previously unused dataset
families: 14 Loblolly pine trees and five Orange trees, plus 19 exact controls.
Truth is a logistic curve fitted to each tree; the working family is a power
law with an inference grid centered on its fit to that truth curve. No fits
fail. New family and campaign seeds improve independence from the earlier
rule-development examples, but all available tree observations enter these
fits. This remains a fitted-curve simulation, not validation on untouched
observations or a prior specified independently of the truth.

Only one misspecified world has random detection at least $0.2$ at $T=30$,
so the registered fallback evaluates G1--G4 on all 19 worlds; all four pass.
Detection is $49.11\%$ for the new rule, $6.05\%$ for random design,
$7.16\%$ for fixed targeting, $11.08\%$ for the original adaptive rule,
$24.34\%$ for expanded-model EIG and $45.37\%$ for full class EIG.
Our audit reconstructs the fallback pool from the saved world records:
the original printed summary and its $T=30$ bootstrap instead use the single
random-detectable world. Newly computed descriptive world-bootstrap intervals
on all 19 give differences from random and class EIG of $43.05$ points
$[33.63,51.50]$ and $3.74$ points $[1.95,5.45]$, respectively
(5,000 resamples, seed zero). These do not control multiplicity or establish
generalization from two small dataset families. Both family means favor the
new rule over class EIG: $59.50\%$ versus $55.11\%$ for Loblolly and
$20.00\%$ versus $18.10\%$ for Orange. At $T=12$, detection remains low:
$2.39\%$ versus random's $0.18\%$ and class EIG's $1.68\%$.
No control campaign rejects; the mean total-risk premium is $0.0057\%$.
However, control campaigns use 18.63 targeted interventions out of 30,
versus 6.02 for fixed targeting, and the misspecified worlds use 25.06
versus 5.94. A small prediction premium does not mean negligible probing cost.

A separate threshold study reuses 128 development worlds (102 misspecified,
26 controls); 832 unchanged policy/world/horizon cells reproduce the earlier
development run exactly. On 55 random-detectable worlds at $T=30$, thresholds
$0.03$, $0.05$, $0.1$, $0.2$ and $0.3$ give detection
$96.51\%$, $96.51\%$, $95.99\%$, $92.09\%$ and $85.68\%$, respectively.
The original S1--S2 robustness/cost criteria pass, but the synthetic subgroup
falls from $87.88\%$ to $77.83\%$ between thresholds $0.1$ and $0.2$.
Control targeted steps range from 29.59 at $0.03$ to 4.50 at $0.3$;
the respective total-risk premiums are $0.271\%$ and $0.035\%$.
This is a measured trade-off on reused development worlds, not a calibrated
risk threshold or a guarantee of smooth response across families. Neither
study supplies the missing fully crossed, physically matched-cost comparison.

\paragraph{A decaying forced-targeting floor.}
A further variant plays the same expected-discrepancy target with independent
probability $0.5/\sqrt{t+1}$ whenever the whole-$B$ gate is closed and the
test has not rejected (zero-based $t$). Its development study reuses 128 worlds
and seeds; all 1,040 unchanged comparator cells exactly match earlier records.
On the 55 random-detectable worlds at $T=30$, detection is $96.07\%$ versus
$95.99\%$ for the gate alone; all 102 misspecified-world means are $63.73\%$
and $63.69\%$. The original W1--W3 point-mean criteria pass. W1's two-model
construction frequencies---$99.15\%$ versus $76.40\%$ in 2,000 campaigns---were
already observed before registration; their repeated report is not independent
confirmation. The control total-risk premium rises from $0.160\%$ to $0.188\%$,
and targeted steps from 18.52 to 19.78 out of 30; each rule rejects in one of
5,200 control campaigns. These are development tolerance gates, not inferential
non-inferiority tests or physically matched budgets.

We separately calculate finite-history probabilities on that exact two-model
construction, without sampling campaigns or calling the simulation core.
Histories merge by informative-observation and success counts, with exact
rational likelihood/rejection decisions, outward gate bounds and outward
binary64 probability transitions. Independent unmerged branch enumeration
checks short prefixes under both the null and true law. All four registered
numerical checks pass. At $T=200$, rejection under the true law is
$76.7687\%$ for the whole-$B$ gate and $98.7707\%$ with the floor; expected
targeted counts under the null are 2.787 and 16.272. The always-target reference
detects with probability exceeding $99.9999\%$ and uses 196.274 expected probes
under the null. Numerical probability enclosures have widths below
$2.5\times10^{-13}$, conditional on the stated model and arithmetic assumptions;
they are not confidence intervals or new-world evidence.
This mathematical reference does not reconstruct floating-point acquisition
bit for bit; rounding can create score ties in the simulation implementation.
Figure~\ref{fig:forcedcoverage} shows the resulting fixed-policy trade-off.

\begin{figure}[t]
\centering
\includegraphics[width=\linewidth]{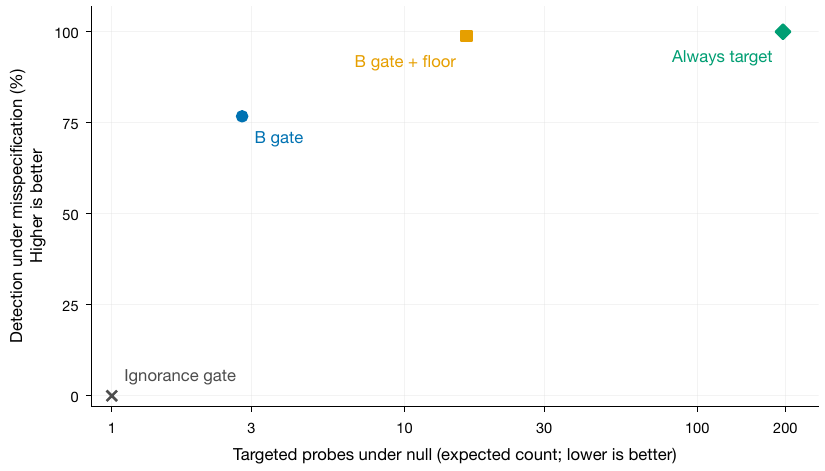}
\caption{\textbf{Preventing permanent cessation has a probing cost.}
Four fixed policies on the two-model construction at $T=200$; the horizontal
axis uses the correct-model law and is logarithmic, while the vertical axis
uses the misspecified law. Deterministic numerical enclosures are smaller
than the markers; no sampled campaigns, swept frontier or interpolation is
shown. Counts stop at rejection. The ignorance-gate point follows analytically.}
\label{fig:forcedcoverage}
\end{figure}

Appendix~\ref{app:forcedcoverage} proves eventual rejection only if forced
targets uniformly separate the represented truth from every null member.
That assumption is not proved for the acquisition rule in general. The
$O(\sqrt T)$ bound covers additional probes when the gate is closed, not all
targeted effort or fixed-truth risk calibration. These qualified results support
testing the variant further, without designating a universally best final rule.

\paragraph{Crossing score, trigger and persistent targeting.}
A separate finite-history audit crosses predictive KL versus expected
discrepancy, $U$ versus $B$ triggers, and the presence of the same forced-target
floor on the two known constructions. It retains 75 system--law--policy cells,
including null laws, both signs of the cancellation departure, fixed-slot
controls and unrestricted coverage references. All six registered numerical
checks pass; unmerged short-history enumeration checks every cell, and the
largest probability-mass enclosure width is $6.34\times10^{-14}$.
Figure~\ref{fig:factorialmechanisms} separates the two mechanisms. In the
two-model example the scores choose the same action at every finite history,
so a score change has no effect. In the four-model example, predictive KL
always chooses $a$ before any $x$ observation: its $x$ predictive is exactly
$1/2$, while its $a$ predictive differs from $1/2$. Both gates and a floor
that repeatedly calls that score therefore retain the same blind design.
At $T=30$ its rejection probability is $0.9122\%$, the null-law probability
on those observations. Expected discrepancy with the whole-$B$ gate and
floor instead rejects with probability $90.5933\%$, but makes 11.607 expected
override attempts versus 29.406 for the blind score; same-as-base attempts
are charged. These are conditional policy effects, not an equal-effort or
new-domain superiority claim. All 30 physical observations continue, with
base actions after rejection; override counts stop at the first crossing.

The fixed-slot control also exposes a reachability limitation. Six potential
overrides at $T=30$ allow at most six informative observations on the two-model
witness, for which
$(1+(9/5)^6)/2=273533/15625<20$; at least seven are required.
Its zero power cannot evaluate score superiority at an adequate cap. This
audit establishes neither a practical matched-budget advantage nor new
independent confirmation; the larger independent-prior comparison remains
separate. A forced-target floor is not the cyclic coverage required by
Appendix~\ref{app:compactcoverage}.

\begin{figure}[t]
\centering
\begin{minipage}{5.5in}
\includegraphics[width=\linewidth]{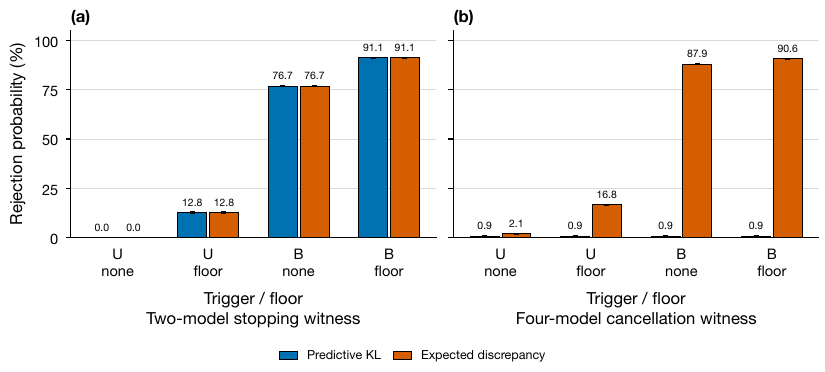}
\end{minipage}
\caption{\textbf{A correct trigger cannot repair a blind target score.}
Two known positive-departure witnesses at $T=30$: scores coincide in (a),
whereas predictive cancellation in (b) survives forced targeting.
Bars show deterministic probability enclosures, smaller than the plotted
marks, not confidence intervals; no sampled seeds. Mirrored departures,
nulls, fixed-slot controls and effort endpoints remain in the complete record.}
\label{fig:factorialmechanisms}
\end{figure}

\paragraph{Frozen floor on previously evaluated fitted worlds.}
The floor variant is next evaluated on the same dose-response, soybean and
tree worlds and campaign seeds used above: 76 misspecified worlds and 40
controls, with 200 campaigns per policy and horizon. All 116 planned world
records are present. Relevant score, trigger, action-selection, coin-stream
and default-coefficient code blocks remain unchanged from the frozen rule;
all 1,160 shared comparator cells exactly reproduce the preceding studies.
The original K1--K3 point-mean tolerance and control criteria pass, without
turning these checks into inferential non-inferiority tests or fresh-data
confirmation.

At $T=30$, the selected 46-world dose-response/soybean pool gives $98.91\%$
detection versus the gate alone's $96.48\%$. The difference is $2.43$ points
with a newly computed descriptive world-bootstrap interval $[0.02,6.17]$
(5,000 resamples, seed zero). Four world means improve and 42 tie; two
soybean plots account for $92.4\%$ of the pooled gain. Across all 19 tree
worlds the rates are $49.18\%$ versus $49.11\%$, a $0.08$-point difference
with interval $[0,0.21]$; two worlds improve and 17 tie. At $T=12$, the
registered pools have identical floor/gate world means. The all-57-world
dose-response/soybean means at $T=30$ are $84.25\%$ versus $82.24\%$, but
the nine dose-response curves still favor full class EIG over the floor
($39.67\%$ versus $36.33\%$). The positive result remains heterogeneous.

On controls at $T=30$, the floor increases mean targeted steps from 11.14
to 13.52 in the dose-response and soybean pool, and from 18.63 to 19.77 in
the tree pool. The floor's mean
total-risk premiums are $0.0279\%$ and $0.0073\%$, respectively; one of
4,200 dose-response/soybean control campaigns and none of 3,800 tree control
campaigns reject. The small prediction premiums do not imply matched or
negligible intervention costs. All observations enter the fitted curves and
truth-centered grids. Bootstrap intervals condition on those worlds, with
no multiplicity correction or propagation of curve-fit uncertainty. Saved
world means do not identify individual campaign histories or temporal mechanisms.

\paragraph{Measured-replicate replay, conditional on the observed datasets.}
A separate study replaces fitted-curve outcome simulation with uniform
resampling from observed replicate bins at each queried positive dose.
There are 20 curves: five spinach curves, two \emph{S.~alba} curves,
ryegrass, lettuce and eleven DNase runs; each has 6--8 queryable doses and
2--4 recorded replicates per positive dose. All 20 planned records, comprising
six policies with 200 campaigns at $T=30$, are present. An independent audit
reconstructs raw-data grouping, bin histograms, noise estimates, nonlinear
fits and the recorded labels from the frozen source and CSV files.

On the seven curves labelled misspecified by the full-data fitted-mean
comparison ($p<0.01$), greedy working-model EIG detects in $0.14\%$ of
campaigns, random design in $79.86\%$, the floor rule in $94.79\%$, and
class-discrimination EIG in $95.50\%$. Greedy design visits 3.24 distinct doses
on average, compared with random design's 7.02. The newly derived paired
world-bootstrap difference of the floor over random is $14.93$ points
with interval $[4.07,28.07]$ (5,000 draws, seed zero); six curves improve and
one ties. Its difference from class EIG is $-0.71$ points,
$[-3.64,1.50]$. These are descriptive intervals conditional on seven curves,
not inferential non-inferiority tests. The original R1--R3 point-mean and
alarm-tolerance criteria reproduce as passing.

Five further curves are labelled ambiguous and excluded from these
label-based primary comparisons. The eight curves labelled adequate
($p>0.2$) have a mean floor-rule alarm
frequency of $2.19\%$, but 33 of their 35 alarms occur on spinach curve~4
($16.5\%$, 33/200). Nonsignificance of a nonlinear fitted-mean comparison
does not establish model adequacy. More fundamentally, each empirical replay
distribution has zero mass in at least 57 of the 61 bins at every queried
action, while every clipped-Gaussian null member assigns positive mass to all
bins. Thus \emph{none} of these replay distributions belongs to the formal
categorical null, including those labelled adequate. Their alarm frequencies
cannot establish null calibration. The same raw observations determine
the parameter grids, control-response scale, bins, noise and labels; none
is reserved as untouched validation. Resampling conditions on the empirical
distribution and adds no new laboratory replicates. Saved results omit
prediction risks, override counts and individual campaign traces, preventing
a cost or predictive-benefit conclusion from this study.

\paragraph{A 100-action constructed interaction screen and a horizon reversal.}
A completed simulation uses ten dose levels for each of two compounds, giving
100 joint actions. The working model is a product of two Hill-response
curves, with known slopes and unknown potencies on a $25\times25$ grid.
There are 24 interaction worlds with off-grid potencies and eight exact-grid
noninteraction controls. A finite alternative adds mechanistic interactions
and local two-dimensional bumps. Outcomes use a binned Gaussian observation
model with known standard deviation $0.05$. This is constructed data, not
a measured drug screen. All 32 world records and 512 policy--horizon cells
are present, with 200 campaign executions per cell. The frozen target,
gate and independent-floor code matches the preceding implementation.

At $T=30$, the registered detectability fallback evaluates all 24 interaction
worlds: greedy detection is zero, random detection is $6.54\%$, the
gate-only rule reaches $45.69\%$, the floor rule $47.69\%$, and
class-discrimination EIG $35.40\%$. All original Z1--Z4 criteria pass.
At $T=60$, however, class EIG reaches $70.17\%$ across all 24 worlds,
against $59.31\%$ for the floor and $22.19\%$ for random. The floor is
worse than class EIG on 23 worlds and ties on one; the new descriptive
world-bootstrap difference is $-10.85$ points, $[-16.19,-6.23]$.
The separate random-selected eight-world pool gives $97.00\%$ for the
floor and $99.38\%$ for class EIG and must not replace the all-world result.
The longer horizon uses the same worlds and seed ranges, not a fresh
independent confirmation.

At $T=30$, the floor raises exact-grid control total Brier risk by $0.0857\%$
relative to greedy and records no alarm in 1,600 control campaigns. Its mean
targeted steps are 12.36 versus 9.98 without the floor; small prediction
premiums do not establish equal or negligible acquisition cost. Grid-null
validity does not extend to a continuous scientific null, and off-grid
potency error contributes alongside interaction error. The result supports
a conditional short-horizon benefit, not a universally best policy.

\paragraph{Post hoc prediction and nonrejection endpoints.}
Reanalysis of the existing saved world means gives mean within-world
relative total Brier-risk changes from greedy of $-1.52\%$, $-2.10\%$ and
$-8.20\%$ for the floor on all 57 dose-response/soybean, 19 tree and 102
development misspecification worlds, respectively. Random design gives
$-3.23\%$, $-2.75\%$ and $-7.59\%$. Thus the floor does not dominate
prediction across these pools. Mean absolute floor-risk differences are
$-0.00753$, $-0.00963$ and $-0.03498$. This score includes irreducible
outcome noise and predictions after possible expansion; the saved means
do not separate excess error or isolate detection's causal benefit.
Across all 57 dose-response/soybean worlds, nonrejection decreases from
$50.01\%$ under greedy to $15.75\%$ under the floor. The smaller
$38.08\%$ to $1.09\%$ comparison applies only to the 46 worlds selected
by random-arm detectability. Nonrejection is not calibrated reassurance
or a direct measure of harmful deployment. These analyses are post hoc,
use previously evaluated worlds and are not fresh confirmatory evidence.

\paragraph{A source-location adaptation.}
We adapt the inverse-square source-location model of \citet{foster2021dad} to
an unexpected second source. This is a modified simulation, not a reproduction
of their neural-design benchmark: we use 100 fixed measurement locations,
a uniform $25\times25$ one-source location grid, 22 binned and probability-floored
outcomes, and signal saturation parameter $m=0.1$ rather than their $10^{-4}$.
The background is $b=0.1$ and log-signal noise standard deviation is $0.5$.
There are 24 two-source worlds and eight exact-grid one-source controls;
the second source has continuous location and strength, while the finite
expansion uses only 25 coarse locations and two extra-source strengths.
The expansion therefore generally omits the exact truth. Six policies are
evaluated with 200 campaigns per world at both declared horizons.

Table~\ref{tab:locationall} retains all two-source worlds. The source-defined
short-horizon selection instead includes 22 worlds whose observed random-arm
detection is at least $20\%$. On that subset the floor exceeds class EIG by
$0.18$ percentage points; on all 24 worlds it is slightly worse. The excluded
worlds are not failures or unusable data. The original four point gates pass,
but their mean-threshold comparisons do not establish inferential noninferiority.
The two horizons reuse worlds and seed ranges, not independent confirmations.

\begin{table}[htbp]
\centering
\small
\caption{Source-location detection (percent), averaging all 24 two-source worlds
equally. These are saved campaign means from run \texttt{fd0610ac}, source
\texttt{7a4ecca}; no random-detectability selection is applied in this table.}
\label{tab:locationall}
\begin{tabular}{rrrrrrr}
\toprule
$T$ & Greedy & Random & Disc 0.2 & Floor & $M_1$ EIG & Class EIG\\
\midrule
30 & 2.40 & 71.83 & 58.19 & 94.25 & 92.58 & 94.65\\
60 & 6.42 & 92.23 & 86.35 & 97.85 & 97.94 & 98.23\\
\bottomrule
\end{tabular}
\end{table}

At $T=30$, mean control expected half-Brier increments over greedy are
$4.24\times10^{-4}$ for the floor and $1.96\times10^{-3}$ for class EIG.
The ratio of their mean within-world relative premiums is $0.217$; their
total-risk ratio is $0.996$, so this is not fourfold lower total risk or
measurement cost. Every policy collects all $T$ observations and can switch
prediction and design to $M_1$ after rejection. The saved targeted-step counter
does not charge class-EIG actions and is not a common intervention budget.
No alarm is recorded in 1,600 control campaigns per policy and horizon, but
this does not establish calibration outside the declared finite null. The
saved output contains 384 world--policy--horizon means; individual campaign
traces are unavailable for an independent detector or acquisition replay.
This study supports allocation effects in the modified finite-grid problem,
not physical source discovery, equal-effort superiority or confirmation of
the independent-prior fixed-class benchmark.

\paragraph{A finite MAP class-separation control.}
A subsequent saved-result comparison implements a plug-in class-separation
score related to \citet{ma2026outgrow}: choose one MAP strictly expanded member,
then maximize its cumulative KL separation from the working class using one
shared finite-grid null parameter across observed action counts plus the next
action. This is a finite one-step adaptation, not the original learned joint
edit--experiment system; it omits candidate-edit search, structural complexity,
calibration/cost ranking, fixed exploration and revision machinery. This finite
acquisition fit does not specify the scientific detector null. All six source-defined point
gates reproduce from 1,320 aggregate cells in 148 reused worlds. The original
short-horizon pools select 22 location and 46 dose/soybean worlds and retain all
19 tree worlds. Across all misspecified worlds, the floor-minus-class-separation
differences at $T=30$ are $+2.42$, $-1.03$ and $+1.50$ percentage points,
respectively; at $T=12$ they are $+0.97$ for dose/soybean and $-0.87$ for tree.
The prediction-premium gates do not measure physical or computational cost,
and the detection gates compare point means rather than inferential
noninferiority. Four unchanged location-policy columns reproduce the earlier
means exactly. Neither those repeated outcomes nor this partial prior-work
adaptation supply fresh confirmation or replace the independent-prior study's
negative horizon decision.

\paragraph{Continuous-null scope and numerical detection.}
Appendix~\ref{app:compactcoverage} extends the finite-null separation argument
to compact log-Lipschitz nulls, combining established covering and likelihood
methods with universal alternative coding, explicit cyclic coverage and
certified numerical error. A deterministic epoch allocation gives a time-uniform
bound with logarithmic model-complexity cost. The result requires the proper
numerator, identifiability, an unbounded coverage schedule and numerical
liveness; it does not certify the evaluated forced-target floor or establish
fast detection at practical budgets.

Two exact calculations illustrate the distinction. For a single Bernoulli
action, a uniform-Beta sequence numerator and threshold 20, substituting the
grid $\{1/4,3/4\}$ for the scientific null $p\in[1/10,9/10]$ gives a
fixed-time rejection probability of
$200205673631/549755813888=36.42\%$ at $T=40$ under the true null $p=1/2$.
The exact continuous profile rejects no count event at that horizon. This is
an exhaustive rational calculation over 41 count events, not an estimated
error rate or a measurement of our production detector. In a separate
two-action complementary history with counts $(20,80)$ and $(80,20)$, the
proper per-action Beta numerator divided by the exact shared-probability
profile exceeds $5\times10^{14}$. The always-valid log-profile upper bound
$U_T=0$, however, yields a statistic at most one on every history and cannot
reject at threshold 20. Conservative computation preserves validity but can
destroy detection; bounded numerical error and actual threshold checks are
separate power premises. Nine deterministic checks verify the fixture
calculations and implementation contracts, not empirical performance.

\paragraph{A continuous working model and a revised gate.}
\label{sec:continuousnull}
The independent-prior development study completed its registered histories and
selected no confirmation horizon: the strongest capped arm detected only
$14/288=4.86\%$ at $H=120$. This result remains a failure of that fixed-budget
candidate. It did not evaluate the later whole-certificate gate with a floor.
A post hoc calculation using Theorem~\ref{thm:ceiling} gives an allocation-uniform
ceiling through
\[
 D^\star=\max_{\nu\in\Delta(\A)}\min_{\theta\in[-3,3]^2}
       \sum_a\nu(a)\KL(p_h(\cdot\mid a)\|p_\theta(\cdot\mid a)).
\]
The mean upper bound on power is 0.611 at $H=60$ and 0.679 at $H=120$
over the 18 non-null worlds. This does not establish attainable power;
a ceiling of one, reached by five of six outside-prior worlds at $H=60$, is vacuous.
Saved-history truth-numerator crossings demonstrate signal on some actual
histories that the implemented numerator misses. They implicate numerator
regret alongside history quality and horizon, without attributing the entire gap
to coding cost. Independent feasible-parameter upper witnesses reproduce the
ceiling means; the original float optimization gap was not an interval certificate.

A subsequent registered study used 24 fresh worlds from the same generator
(seed 2637501), $T=240$, and 50 campaigns per world and arm. Its uniform
41-by-41 working grid and expanded bump library differ from the independent-prior
study; this is a new panel, not a controlled horizon-only extension.
The detector hedges sequence likelihoods from the bump mixture and per-action
Beta(1,1) predictives with weights one half. Its denominator uses the concavity
upper bound over the continuous logistic box $[-3,3]^2$. The formula is valid in
exact arithmetic. The implementation uses float64 and a $10^{-9}$ log-likelihood
pad, not the outward-certified detector of the primary study. An independent
80-digit outward check on 131 fixed count fixtures through $T=480$ found a
minimum slack $9.999\times10^{-10}$; this is numerical assurance on those fixtures,
not certification of all recorded prefixes. The grid pre-filter is safe in exact
arithmetic because a grid maximum is a lower bound on the continuous maximum;
its borderline floating-point decisions are not outward-certified either.

Greedy EIG detects $1/900=0.11\%$, random design $170/900=18.89\%$.
The \emph{frozen} $b^\star=0.1$ rule detects $25/900=2.78\%$ and fails
both restoration/comparator criteria N2--N3; N1 and N4 pass. It takes
falsification actions on about 8\% of steps. After a disclosed development pass
on the earlier worlds, but before fresh-world outcomes, three secondary gates
were declared. The main secondary gate is
\[
 b^\star_T=\log(1/\alpha)/T.
\]
This is a revised rule. At $T=30$, $\log(20)/30=0.0998577\ldots$ is close to
0.1, but not equal; earlier frozen-gate results are not retrospectively relabelled.
The earlier failed calibration of the ignorance trigger (D6) also remains a
failure of a different stopping diagnostic.

At $T=240$ the revised gate detects $322/900=35.78\%$, versus
$285/900=31.67\%$ for class separation and $256/900=28.44\%$ for
class-discrimination EIG. All three secondary point-threshold criteria pass.
Using an independently specified descriptive bootstrap (5,000 paired-world
resamples, seed 0, comparator order class separation, class EIG, random), the
differences are $+4.1$ points $[1.4,7.2]$, $+7.3$ $[2.2,12.9]$,
and $+16.9$ $[7.9,27.2]$. These replace the proposal's unreproduced bootstrap
endpoints; they are not simultaneous intervals or inference across arbitrary
world populations. The null prediction-risk premium is about $+0.03\%$ versus
$+0.43\%$ for class separation; each arm has zero alarms in 300 null campaigns.
Near-null power remains low. All arms miss two matched-bump worlds despite
large upper ceilings; those ceilings alone do not distinguish inadequate
histories from numerator loss.

\paragraph{Transfer to a different horizon on previously evaluated worlds.}
A registered $T=60$ comparison applies the revised gate unchanged to four
previously evaluated families with identical seeds and worlds. This tests transfer
of a gate chosen elsewhere; it is not a fresh-world or untouched-data confirmation.
The drug and tree pools include all misspecified worlds; location and dose/soybean
retain the earlier random-detectable pools. All four registered point criteria pass.
Drug-screen detection is 76.79\% versus 70.17\% for class EIG, 69.77\% for
class separation, and 59.3125\% for the frozen gate. All 32 drug world/control
rows reproduce the saved greedy, random, class-EIG and frozen-gate detection,
risk and action-count aggregates exactly; in particular class EIG is
$0.7016666667$ and frozen is $0.593125$.
Paired-world contrasts against class EIG and class separation are $+6.6$ points
$[4.0,9.5]$ and $+7.0$ $[4.0,10.4]$, respectively. Tree detection is
85.84\% versus 81.21\% and 82.95\%; location and dose/soybean are
99.91\% and 99.97\% for the revised rule. The revised rule's null risk premium
ranges from $+0.015\%$ to $+0.112\%$, below both comparators in each family.
There is one drug control alarm in 1,600 campaigns and one dose/soybean control
alarm in 4,200. Premium means relative excess prediction risk, not fewer
experiments, lower runtime or physical acquisition cost; every arm receives $T$
observations.

\begin{proposition}[Conditional finite-horizon futility]
\label{prop:futility}
Let $E_s=q(H_s)/\sup_{\theta\in\Theta}p_\theta(H_s)$, with predictable
sub-probability predictives $q_u$, and stationary truth $p_h$. Fix $t<T$, an
$H_t$-measurable $\theta\in\Theta$, and $\lambda>0$. Conditional on $H_t$,
freeze $\theta$ throughout the continuation. Write
\[
 R_t(\theta)=q(H_t)/p_\theta(H_t),\qquad
 \bar D_\lambda=\max_a D_{1+\lambda}(p_h(\cdot\mid a)\|p_\theta(\cdot\mid a)),
\]
and $c=\log(1/\alpha)-\log R_t(\theta)-(T-t)\bar D_\lambda$.
For any predictable continuation policy, remaining crossing probability is at most
$\phi_\lambda(c)$, where
\[
 \phi_\lambda(c)=\min\{1,(1+\lambda^{-1})\lambda^{1/(1+\lambda)}
                  e^{-\lambda c/(1+\lambda)}\}\quad(c>0),
\]
and $\phi_\lambda(c)=1$ otherwise. If absolute continuity fails or
$\bar D_\lambda=\infty$, use the vacuous bound one. If the accumulated numerator
is zero, the product numerator stays zero and cannot cross.
\end{proposition}
\begin{proof}
For positive $R_t$, $\log E_s\le\log R_t(\theta)+M_s+S_s$, with
$M_s=\sum_{u=t+1}^s\log(q_u/p_h)(Y_u)$ and
$S_s=\sum_{u=t+1}^s\log(p_h/p_\theta)(Y_u)$.
Under $P_h(\cdot\mid H_t)$, $e^{M_s}$ is a nonnegative supermartingale
started at one: its conditional multiplicative mean is
$\sum_{y:p_h(y\mid a_u)>0}q_u(y)\le1$.
Also $\exp\{\lambda S_s-\lambda\sum_{u=t+1}^sD_{1+\lambda}(a_u)\}$
is a martingale started at one. Predictability of actions and freezing of $\theta$
justify each conditional identity. Two Ville bounds, a union bound, and
$D_{1+\lambda}\ge0$ show that a crossing requires one of events with total
probability at most $\delta+\delta'$, whenever
$\log(1/\delta)+\lambda^{-1}\log(1/\delta')=c$.
Put $\delta=e^{-x}$, $\delta'=e^{-\lambda(c-x)}$. Minimization gives
$x=(\lambda c-\log\lambda)/(1+\lambda)$ and the stated expression.
Bounds above one are trivial, including Ville bounds with $\delta\ge1$.
\end{proof}

For a truth drawn from a correctly specified finite expansion prior, posterior
$\pi_t$, and an $H_t$-measurable working point $\hat\theta_t$, conditioning on
the truth and averaging gives
\[
 F_t=\sum_h\pi_t(h)\inf_{\lambda>0}\phi_\lambda\!\left(
 \log(1/\alpha)-\log R_t(\hat\theta_t)-(T-t)\bar D_\lambda(h,\hat\theta_t)
 \right).
\]
Thus the agent's posterior-predictive probability of a remaining crossing under
any continuation is at most $F_t$. Minimizing over a fixed finite $\lambda$ grid
also gives a valid, weaker bound. This does not give a fixed-truth guarantee for
off-library worlds. The numerator need not equal the planning prior, but
$\pi_t$ must be the actual posterior under the declared generative prior.

The bound relates an evidence deficit to remaining budget times a maximum
R\'enyi divergence. It motivates considering a $1/(T-t)$ scale; replacing the
maximum R\'enyi quantity by mean expected discrepancy, $T-t$ by $T$, and
$\log R_t$ by zero is heuristic. In particular $R_t(\hat\theta_t)\ge E_t$;
substituting $E_t$ into the bound need not preserve its validity. Neither the
constant nor positive detection probability follows from the proposed gate.
The separate simulation check of C1 is a loose diagnostic, not a proof; its
$t=0$ average bound is not tighter than the KL ceiling on the development panel.

Finite-horizon e-value futility is established prior work:
\citet{baas2026curtailment} optimize power and expected sample size using dynamic
programming, explicitly including curtailment. \citet{taga2026horizon} optimize
horizon-aware betting and identify hopeless regions. Our proposition instead
gives a conservative conditional upper bound for a composite universal-inference
detector under adaptive design; we claim no first finite-horizon futility result.
\citet{agrawal2025stopping} study sequential stopping-time lower bounds;
\citet{forre2026typeii} give lower-tail conditions yielding type-II error bounds,
the complementary achievability direction.

\paragraph{Another fresh panel at $T=480$.}
The third registered comparison uses seed 2637601, 24 worlds and 50 campaigns per
world and arm. The calibrated gate detects $468/900=52.00\%$ versus
$463/900=51.44\%$ for class separation, $422/900=46.89\%$ for class EIG,
and $384/900=42.67\%$ for random design. Its class-separation contrast is
$+0.6$ points $[-1.8,2.7]$: no resolved advantage, and no equivalence claim.
Earlier prefixes give 21.78\% versus 20.67\% at $T=120$ and
39.89\% versus 37.89\% at $T=240$; these are descriptive comparisons within
the same campaigns. The frozen gate still fails, at $92/900=10.22\%$.
All five registered criteria pass, including point-threshold competitiveness;
Q5 is an empirical check with Monte Carlo slack, not numerical verification of a
theorem. Null risk premiums are $+0.014\%$ for the calibrated gate and
$+0.365\%$ for class separation. Each arm raises zero alarms in 300 null
campaigns; these observations do not replace the mathematical validity premises.
The $F_t\ge0.2$ gate has identical detection indicators and times on all 900
non-null and 300 null campaigns. There are 468 detected and 432 undetected
non-null campaigns. Its null premium is $+0.026\%$. Equal saved outcomes do
not prove universal behavioral equivalence or identify all intermediate actions.

We retain the original frozen-rule failures and evaluate the calibrated gate as
a revised candidate. Confirmation with the independent-prior pipeline's original
priors and outward-certified detector remains open.

\paragraph{A targeted/random acquisition hedge.}
A registered follow-up splits each falsification step equally between predictive-KL
targeting and a uniformly random design. At the fixed 20\% budget it fails all three
detection gates, while passing its control-cost gate. On the seven random-detectable
Indometh/DNase fitted worlds at $T=30$, detection is $32.57\%$, versus $39.21\%$ for
targeting and $63.00\%$ for fully random designs. On the selected synthetic-right
worlds it is $57.66\%$ versus $73.22\%$ for targeting, missing the allowed three-point
loss. Its worst group-wise detection/random ratio is $0.517$, below targeting's
$0.622$, despite a mean control total-risk premium of only $0.076\%$.
The 30\% budget is descriptive and does not rescue the registered 20\% failures.
These outcome-selected groups and repeated fitted worlds are not external confirmation;
the risk uses prediction after class switching, unlike the fixed-class crossed study.
This acquisition hedge is distinct from mixing detector sequence likelihoods.

\paragraph{Designs aimed at detector growth.}
\label{sec:growth}
We compared predictive-KL targeting with targeting the running maximum-likelihood null
and the exact expected one-step log-growth of the universal-inference statistic
(96 worlds, 200 campaigns, budgets 10\%/20\%, horizons 30/60). At $T=30$, on the registered
random-detectable subgroup and a 20\% budget, the original rule detects in $78.34\%$, the
maximum-likelihood-targeting rule in $76.22\%$, and the growth-targeting rule in $78.13\%$.
The registered five-point improvement criterion fails. The selected growth rule satisfies
the control-cost criterion and at $T=60$ nearly matches random design ($83.56\%$ versus
$83.92\%$), but the original rule already gives $83.49\%$. The experiment does not establish
an improvement from the more exact one-step objective. Selection of the better new rule
and the outcome-selected subgroup remain development analyses.

\section{Language-model experimenters}
\label{sec:llm}

Theorem~\ref{thm:ceiling} applies to any policy, including a prompted language model. We
built a harness in which the LLM only chooses designs. At each step it sees a
natural-language description of the experiment, the working model, the allowed settings
and its own history of settings and outcomes, and names the next setting. The harness
samples the outcome from the planted truth and scores the LLM's data with the same exact
posterior, adequacy e-process and protected target as above. A second prompt adds the
instruction that the working model may be wrong and that measurements should be chosen so
the error would be revealed.

\paragraph{Pilot.} We ran a local 4-billion-parameter model (Qwen3.5-4B, 4-bit, thinking
off) on the four families where detection is possible, using the four misspecified worlds
per family on which random design detected most often in the generality study, plus one
well-specified control per family; $T=30$, 4 campaigns per world and prompt. All 4{,}800
responses parsed. Table~\ref{tab:llm} summarizes the result: the model did not collapse.
Its designs were nearly as spread out as random design and it detected the error as often.
The mechanism is not reassuring, though. Under the plain prompt, 65\% of its steps moved to
the adjacent setting in the list, and some campaigns visited the settings in order. The
model kept its detection power by scanning, not by reasoning about where its model might
fail. The falsification prompt halved the scanning (32\% adjacent moves) and raised detection
by 4.7 points, short of the 10-point effect we pre-registered. Neither prompt raised a false
alarm on the controls. Whether stronger models, which can reason like design optimizers,
concentrate their designs and inherit the ceiling of Theorem~\ref{thm:ceiling} is the
question this pilot leaves open.

\paragraph{Model size.} We repeated the plain-prompt protocol, with identical worlds, seeds
and sampling keys, for 0.8B, 2B and 9B models of the same family
(Table~\ref{tab:llm}). Detection rose monotonically with size, from 32.8\% (0.8B) to 62.5\%
(9B). The smallest model concentrated its designs most (7.1 distinct settings per 30-step
campaign), but by repeating the same setting in 24\% of its steps, not by seeking
informative designs, and it detected least, consistent with Theorem~\ref{thm:ceiling} when
concentration leaves little misfit visible. The 9B model's designs were somewhat more concentrated than the 4B
model's (4.10 versus 4.47 bits) without losing detection. Within this family, then,
capability buys coverage rather than collapse; the pre-registered collapse signals were not
met.

\paragraph{Frontier reasoning models.} We then ran two frontier reasoning models through the
same harness (GPT-6 Astra and GPT-6.1 Sol via the OpenAI API, reasoning effort ``medium''),
as a pre-registered low-budget pilot: plain prompt, the 16 misspecified worlds, 2 campaigns
per world. A hard spend cap stopped Astra after 13 worlds. Both models collapsed
(Table~\ref{tab:llm}). Astra detected the error in 7.7\% of campaigns against 71.3\% for
random design on the same worlds (difference 0.64, 95\% bootstrap interval [0.42, 0.83];
random design detected more often in 12 of 13 worlds, sign test $p=0.002$); Sol detected in
6.2\% against 58.3\% (difference 0.52, [0.31, 0.72]; 14 of 15, $p=0.0005$). Their designs
explain why. Both concentrated on about four settings per 30-step campaign (design entropy
1.60 and 1.84 bits, against 1.40 for greedy EIG and 4.64 for random design), and 61\% of their
measurements sat at the two ends of the range. On all four linear-model worlds their design
distributions were exactly those of greedy EIG: every measurement at one of the two
endpoints, the textbook optimal design for a straight line, on which a straight line can
never be falsified (Corollary~\ref{cor:saturation}). The only detections by either model came in
one Poisson world. A follow-up run completed two of Astra's three missing Poisson worlds; it
never detected there either, giving 6.7\% against 62.1\% over 15 worlds. The models were not careless: on the nonlinear families they reasoned at
length (about 1,300 reasoning tokens per decision) before choosing. They were doing optimal
design well, and on these worlds optimal design leaves almost no misfit visible on its own designs,
which by Theorem~\ref{thm:ceiling} removes the chance to notice a wrong model. Across capability the picture is therefore not monotone: the smallest
models fail by repetition, mid-sized open models keep detection by scanning, and frontier
models lose it by optimizing.

\begin{table}[t]
  \centering
  \small
  \caption{\textbf{Open language models keep detection power; frontier reasoning models from
  two providers lose it, and the blindness monitor restores it.} Pooled over the misspecified worlds in four families, $T=30$. Detection uses the valid
  adequacy test; design entropy is that of the pooled design distribution (uniform over 25
  settings is 4.64 bits). Baselines use 200 campaigns per world; open models 4 campaigns per
  world and prompt. $^\ast$Frontier pilot: 2 campaigns per world; random design detects 0.713
  on the 13 worlds Astra completed. $^\dagger$Second provider: 1--2 campaigns per world (spend
  caps truncated second campaigns), all 16 worlds in both arms.}
  \label{tab:llm}
  \begin{tabular}{@{}lccc@{}}
    \toprule
    Design policy & Detection rate & Design entropy (bits) & Protected risk \\
    \midrule
    Random design & 0.583 & 4.64 & 0.407 \\
    Greedy EIG & 0.001 & 1.40 & 0.421 \\
    Qwen3.5-0.8B, plain prompt & 0.328 & 2.96 & 0.436 \\
    Qwen3.5-2B, plain prompt & 0.453 & 3.92 & 0.423 \\
    Qwen3.5-4B, plain prompt & 0.609 & 4.47 & 0.409 \\
    Qwen3.5-9B, plain prompt & 0.625 & 4.10 & 0.410 \\
    Qwen3.5-4B, falsification prompt & 0.656 & 4.39 & 0.409 \\
    \midrule
    GPT-6.1 Sol, plain prompt$^\ast$ & 0.062 & 1.84 & 0.418 \\
    GPT-6 Astra, plain prompt$^\ast$ (13 worlds) & 0.077 & 1.60 & 0.419 \\
    GPT-6.1 Sol, falsification prompt$^\ast$ & 0.625 & 4.53 & -- \\
    GPT-6.1 Sol, plain prompt + 20\% budget$^\ast$ & 0.469 & 2.34 & -- \\
    GPT-6.1 Sol, plain prompt + blindness monitor$^\ast$ & 0.688 & 3.26 & -- \\
    \midrule
    Gemini 3.8 Flash, plain prompt$^\dagger$ & 0.344 & 2.80 & 0.414 \\
    Gemini 4 Argon, plain prompt$^\dagger$ & 0.094 & 2.16 & 0.415 \\
    Gemini 3.8 Flash + blindness monitor$^\dagger$ & 0.562 & 3.29 & 0.410 \\
    Gemini 4 Argon + blindness monitor$^\dagger$ & 0.750 & 3.57 & 0.409 \\
    \bottomrule
  \end{tabular}
\end{table}

\paragraph{Restoring detection in a frontier model.} We then tested two fixes on GPT-6.1 Sol,
with a harness that stops on any API or quota error and excludes any campaign with a
fallback decision (pre-registered; all 72 campaigns were clean). A one-sentence warning
(``the working model may be wrong somewhere in the allowed range; choose measurements so
that, if it is wrong, your data would reveal it'') raised detection from 6.2\% to 62.5\%, at
or above random design's 58.3\%: Sol spread its measurements almost uniformly (4.53 bits) and
improved in all 10 worlds where the arms differed (sign test $p=0.001$). The falsification
budget, which leaves the prompt unchanged and overrides 20\% of Sol's choices with the
targeted design \eqref{eq:disc}, raised detection to 46.9\% (9 of 10 worlds, $p=0.01$) while
keeping Sol's designs concentrated (2.34 bits). These selected-world results show that
prompt changes and a mechanical budget can restore detection in this tested setting;
they do not establish a general limit or guarantee for frontier experimenters.
The small open model gained only 4.7
points from the same warning. Both fixes were nearly free when the working model was right:
on the four well-specified control worlds (8 campaigns per fix) neither raised a false alarm,
and protected risk changed by $+0.2\%$ (warning) and $+0.6\%$ (budget) relative to Sol's plain
prompt.

\paragraph{Showing the agent its blindness.} The certificate of Section~\ref{sec:certificate} can
be handed to the language model. We added one line to Sol's plain prompt at every step, computed
by the harness from the expansion's posterior: its current ignorance $U_t$ (with the remark that
spreading measurements at random typically ends near 0.02) and the three settings with the
largest per-setting ignorance, with no instruction about what to do (pre-registered; 40
campaigns, all clean). Detection rose from 6.2\% to 68.8\%, above
random design's 58.3\% (gain 0.63, 95\% bootstrap interval [0.31, 0.88]; better in 12 of 13 worlds
where the arms differed, $p=0.002$). Sol spread its designs (3.26 bits per world, against 1.84)
and ended with the lowest ignorance of any arm (0.027, against 0.131 plain, 0.041 with the
warning and 0.038 with the budget). On the well-specified controls it raised no false alarm and
changed protected risk by $+0.7\%$. The score and setting suggestions jointly changed this model's experimental choices.
The intervention includes actionable information and a random-design reference value;
the completed score-only and placebo-suggestion ablation is reported below.
That comparison leaves the score's numerical values confounded with inadequacy
framing and reference guidance unless a constant-score control is included.
The prompt's informal zero-score explanation is stronger than the theorem:
$U_t=0$ does not rule out errors outside the proposed expansion.
(An indexing bug in the run's built-in control analysis was fixed and the
controls re-analysed from the logged campaigns.)

\paragraph{A second provider.} To test whether collapse and its cure are artifacts of one
model family, we repeated the plain and monitor arms with Google's Gemini 4 Argon and the
faster Gemini 3.8 Flash (default dynamic thinking), under the same harness, worlds and
pre-registered criteria: collapse if plain detection is below half of random design's,
restoration if the monitor adds at least 0.20, and safety if the monitor raises no false
alarm on the controls and changes their protected risk by at most 5\%. Each model had a hard
\$8 cap; the jobs were ordered so that the caps truncated only second campaigns, leaving every
world and control with at least one campaign in both arms (138 campaigns, all clean). Argon
collapsed: 9.4\% detection against 58.3\% for random design, with designs nearly as
concentrated as Sol's (2.16 bits). The monitor raised it to 75.0\%
(gain 0.66, 95\% bootstrap interval [0.41, 0.88]; better in 12
of 13 worlds where the arms differed, $p=0.003$), while ignorance fell from 0.119 to 0.030.
Flash, which concentrated less (2.80 bits), collapsed only partly: 34.4\% detection, which
misses the pre-registered collapse threshold of 29\%. The monitor still raised it to 56.2\%
(gain 0.22, [0.00, 0.44]; 7 of 8 worlds, $p=0.07$). Across the four frontier models, detection
falls roughly as design concentration rises (Figure~\ref{fig:capability}), consistent with
Theorem~\ref{thm:ceiling} when concentration lowers the own-design misfit. Neither monitor arm raised a false alarm on the controls.
Flash's control risk changed by $+1.4\%$. Argon failed the pre-registered safety criterion:
pooled over campaigns, its control risk rose by 10.6\%. The spend cap produced unequal campaign counts by control world, so the pooled
contrast mixes monitor effects with differences in world composition. The monitor arm kept both campaigns of the
highest-risk control and the plain arm both of the lowest-risk one, and control by control the
risks agree to within 1.5\% (mean of per-control means $+0.6\%$; this analysis was not
pre-registered).

\paragraph{Separating the monitor prompt from targeted suggestions.}
\label{sec:monitorablation}
A registered ablation removes the suggested settings (score-only prompt) or replaces
them with three uniformly random settings, retaining the true score and reference guidance.
At $T=30$ on the same 16 selected misspecified worlds, score-only detection is
$59.38\%$ for Sol and $56.25\%$ for Argon, meeting the registered twenty-point
improvement over the historical plain arms in both models. Random-suggestion detection
is $62.50\%$ and $56.25\%$. The registered requirement that targeted suggestions add
at least fifteen points passes for Argon and fails for Sol. No new arm raises an alarm
on its eight control campaigns, so both no-alarm gates pass; this small control sample
does not establish general safety or a prediction-cost guarantee.

All 80 Sol campaigns are clean. Argon retains 76 of 80; four campaigns abort after
repeated call failures, with no parse fallbacks among retained campaigns.
In the original ablation, plain and full-monitor references were historical. Matching
all four Argon arms leaves 29 world--seed pairs covering all 16 worlds; the score-only
world-weighted rate becomes $53.13\%$, and the targeted/random-suggestion gap remains
$18.75$ points. Sol's monitor variants have complete 32-pair support; its plain reference
retains per-world aggregate outcomes rather than individual detection indicators.
The declared point criteria are preserved. Descriptive world-bootstrap intervals for
the targeted/random-suggestion gap are $[-15.63,28.13]$ points for Sol and
$[-6.25,43.75]$ for Argon (Figure~\ref{fig:monitorablation}).
Neither resolves a positive setting-specific effect at this sample size.
Allowing arbitrary outcomes for missing Argon campaigns in the planned
32-campaign-per-arm allocation yields a gap from $12.5$ to $25.0$ points, spanning
the registered fifteen-point target. These are sensitivity analyses, not replacement criteria.

\paragraph{Interleaved plain/full reruns.}
A registered follow-up reruns plain and full-monitor prompts with interleaved jobs.
Spend limits leave 44 of 64 planned campaigns for Argon and 35 of 64 for Sol;
available common-world means are $12.50\%$ versus $71.88\%$ on 16 worlds and
$7.69\%$ versus $61.54\%$ on 13 worlds, respectively. Both original available-data
replication/stability criteria pass. No new control campaigns are collected.
The score-only and placebo arms still come from the earlier run.
Matching all four arms on identical world--seed support leaves 21 pairs for
Argon and 15 for Sol. Full-minus-placebo differences become $12.50$ and
$-11.54$ points, with descriptive world-bootstrap intervals $[0,31.25]$ and
$[-42.31,19.23]$ points. Over the full planned 32-campaign-per-arm allocation,
allowing arbitrary missing outcomes gives full-minus-plain bounds
$[9.38,71.88]$ points for Argon and $[-18.75,71.88]$ for Sol.
These bounds do not identify whether the registered forty-point threshold is met;
the original reported passes are retained. Partial interleaving therefore
does not remove missingness or make the four-arm comparison concurrent.

The supported conclusion is that the tested monitor prompt can improve detection without
targeted setting suggestions. It does not identify a causal contribution of the numerical
$U_t$ values: score-only still includes model-inadequacy framing and a random-design
reference, and the ablation lacks a constant-score control. Failure of Sol's pointing gate
does not establish equivalence, and Argon's point-threshold pass does not establish a
statistically resolved or missingness-robust pointing benefit.

\paragraph{A safety layer around selected language models.}
\label{sec:safetylayer}
The harness overrides a proposed design before detection when the
whole-certificate gate $B_t\ge0.1$ or a floor coin with probability
$0.5/\sqrt{t+1}$ calls for a falsification action; otherwise the plain-prompt
language model chooses. These are the original frozen-gate $T=30$ pilots,
not tests of the slightly different $\log(20)/30$ gate.
Registered point criteria require detection gains of at least 40 points over
external plain-arm references, at most five points below monitor references,
and at most one alarm in eight control campaigns with paired risk premium at
most 0.5\%.

Argon detects $26/32=81.25\%$ across 16 selected misspecified worlds.
All 48 planned campaigns complete cleanly, including eight wrapper and eight
plain controls; the paired control premium is $+0.226\%$, with zero wrapper
alarms. The harness takes about 44\% of wrapper steps. External interleaved-rerun
references are 12.5\% plain and 71.9\% monitor; these are not concurrent
randomized wrapper comparisons. All three registered point criteria pass.

Sol completes only 24 of 32 planned non-null campaigns before the spend cap,
with $19/24=79.17\%$ campaign-weighted detection. Because completion counts
vary by world, its equally weighted mean over the 16 represented worlds is
81.25\%. The initial run also completes five wrapper controls, but no paired
plain controls: its recorded W3 failure was unassessable and is retained as
such. A separate registered control-completion run supplies three wrapper and
eight plain controls with the same seeds. Across the eight combined control
pairs, wrapper alarms are zero and the premium is $+0.201\%$; completion W3
passes. W1--W2 pass on available cases against external 7.7\% plain and
61.5\% monitor references. The eight missing non-null campaigns remain missing;
the full planned balanced-campaign detection could range from $19/32=59.38\%$
to $27/32=84.38\%$ without assumptions. The harness takes about 49\% of
initial wrapper steps. These are small, selected pilots, not capability or
provider comparisons.

An API-free control uses the same 16 misspecified worlds, four null controls,
and 200 seeds per arm. Wrapping greedy EIG raises printed detection from
0.1\% to 76.3\%, and wrapping random from 57.3\% to 76.9\%; each wrapped
arm has zero alarms in 800 controls. Paired null risk premiums are $+0.18\%$
and $-0.44\%$. All four registered point criteria pass. A two-seed preview
preceded registration and those seeds remain included. The registered Argon
binomial consistency check passes, but assumes a homogeneous crossing
probability and ignores heterogeneity across paired worlds. The small difference
between the wrapped non-language designers supports robustness on this panel;
it does not establish agent independence, equivalence to the language models,
or detection guarantees for arbitrary agents. The layer directly controls many
acquisitions, so these gains cannot be attributed to improved language-model
reasoning.

\begin{figure}[t]
  \centering
  \begin{minipage}{5.5in}
    \includegraphics[width=\linewidth]{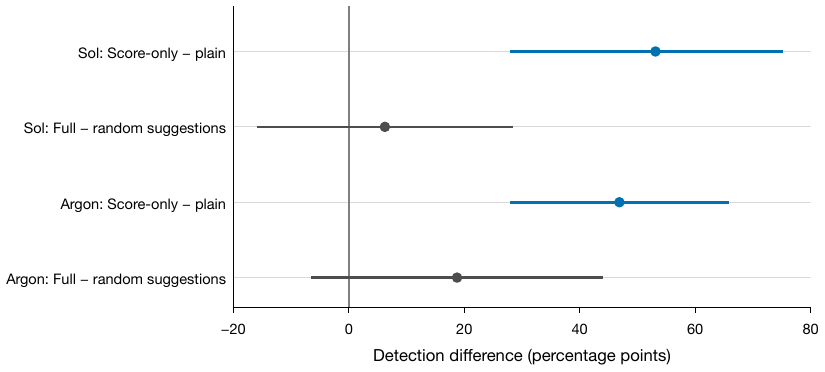}
  \end{minipage}
  \caption{\textbf{Prompt gains are clearer than setting-specific gains in this pilot.}
  Differences across 16 selected worlds at $T=30$; whiskers are descriptive 95\%
  world-bootstrap intervals. The original ablation uses historical reference arms;
  Argon retains one or two campaigns per world after failures and earlier spend caps.
  The score-only prompt also supplies framing and reference guidance.}
  \label{fig:monitorablation}
\end{figure}

\paragraph{Self-reports did not discriminate reliably in the pilot.} After each campaign we asked Sol whether its
working model was adequate for the data. It answered ``no'' in 84\% of misspecified campaigns
but also in 88\% of campaigns where the model was correct, so this self-assessment did not distinguish the two groups; the pre-registered criterion that it discriminate failed.
Gemini 4 Argon erred the other way: under the plain prompt it judged its model adequate in
71\% of the misspecified campaigns it failed to detect.
These prompts did not produce useful adequacy judgments in this pilot; broader
claims about self-assessment require more prompts and independent worlds.

\section{Limitations}
\label{sec:limitations}

The completed empirical worlds are low-dimensional simulators with planted misspecification;
most earlier studies infer on finite grids, while the independent-prior study
profiles a continuous logistic null and retains its negative confirmation decision.
The new crossed study is a deliberately constructed mechanism benchmark, and previously unused
ChickWeight/Puromycin and dose-response/soybean data supply fitted-curve simulations rather than held-out-observation validation;
four of seven generic families stayed essentially undetectable at $T=60$ and $T=200$, and in
the two mechanistic environments the collapse was partial or absent
(Section~\ref{sec:realistic}). The expansion
$\M_1$ is a generic bump class; when the truth lies far from $\M_1$, detection
(Theorem~\ref{thm:ceiling} still applies) and especially the accuracy gain from
expansion will be smaller. Theorem~\ref{thm:achieve} assumes the expansion contains the truth and
that the selected designs accumulate separation against every null parameter; we do not prove that the adaptive
rule \eqref{eq:disc} does. The fixed-truth certificate also requires a finite
expansion containing the truth; randomly drawn off-grid truths need not satisfy
that assumption. The continuous confidence-set extension needs a separately
declared family containing the truth. Its implemented global risk reference
has analytical checks and a small integrated feasibility study, which yields no
intervention savings and does not establish useful performance. Neither calibration justifies
the heuristic $u^\star=\log(1/\alpha)/T$, nor the final gate $b^\star=\log(1/\alpha)/T$. The revised gate passed registered point criteria on two fresh panels and a reused-world horizon-transfer comparison; it is not proved, and the same calibration failed for the earlier ignorance trigger. The frozen absolute gate $b^\star=0.1$ failed on continuous-null panels at $T=240$ and $T=480$. The two-model trigger counterexample
shows that even a correct target choice can cease too early. The
frontier-model evidence is a pilot: four models from two providers, one prompt, 1--2
campaigns per world, and 15 of 16 worlds for GPT-6 Astra. The sampled models show similar effects across two providers; generality across
prompts, selection of worlds, and agent scaffolds is untested. Frontier-model worlds
were chosen for high random-design detection in earlier simulations, and campaign
counts are small and sometimes truncated by spend caps. Capability, architecture,
provider and prompting vary together, preventing a causal capability claim.
Expected Brier risk includes irreducible outcome noise; a small relative premium
in total risk need not be small relative to excess predictive error. Future comparisons
should report both quantities and absolute risk differences. World-level uncertainty
should distinguish variation across worlds from sampling uncertainty within campaigns.
Most older simulator policies expand on rejection, whereas the independent-prior
study and LLM harness score the fixed working model; their prediction-risk
results therefore concern different revision policies. The completed primary
comparison provides no eligible confirmation horizon, and the truth-informed
diagnostic is retrospective rather than an available scientific policy.

\section{Experimental details}
\label{app:details}

\begin{table}[ht]
  \centering
  \footnotesize
  \setlength{\tabcolsep}{4pt}
  \caption{\textbf{The seven design families.} Each has 25 designs and a grid-exact working
  model $\M_0$. ``Deviation scale'' $A$ sets the amplitude of planted deviations and of the
  expansion $\M_1$, in link units (logit, mean, or log rate).}
  \label{tab:families}
  \begin{tabular}{@{}lllll@{}}
    \toprule
    Family & Designs & Working model $\M_0$ (grid) & Outcomes & $A$ \\
    \midrule
    Dose--response & $x\in[-4,8]$ & $\mathrm{logit}\,p = b(x-a)$ ($31\times 20$) & Bernoulli & 3.0 \\
    Linear & $x\in[-1,1]$ & $\mu = \alpha+\beta x$ ($25\times25$) & Gaussian, $\sigma=0.5$ & 1.0 \\
    Michaelis--Menten & $S\in[0.1,100]$, log & $\mu = V_{\max}S/(K_m+S)$ ($25\times25$) & Gaussian, $\sigma=0.1$ & 0.3 \\
    PK decay & $t\in[0.25,12]$ & $\mu = C_0 e^{-kt}$ ($25\times25$) & Gaussian, $\sigma=0.08$ & 0.25 \\
    Poisson rate & $x\in[0,1]$ & $\log\lambda = \alpha+\beta x$ ($25\times25$) & Poisson, 0--15+ & 0.8 \\
    Discounting (2-D) & delay $\times$ ratio, $5\times5$ & $\mathrm{logit}\,p = (R/(1+kD)-1)/\tau$ ($25\times25$) & Bernoulli & 2.0 \\
    Logistic (2-D) & $[-1,1]^2$, $5\times5$ & $\mathrm{logit}\,p = s(u^\top x - c)$ ($16\times9\times4$) & Bernoulli & 2.5 \\
    \bottomrule
  \end{tabular}
\end{table}

\paragraph{Outcomes and inference.} Gaussian outcomes are reported in 16 bins (14 interior
bins between fixed edges plus two open tails), so the likelihood of every grid point is
exact. Poisson counts above 14 share one tail class. Posteriors over $\M_0$ and $\M_1$ start
uniform on their grids and are updated exactly.

\paragraph{Expansion and truths.} $\M_1$ adds $g\,A\,\exp(-\|\tilde x-c\|^2/(2\cdot0.12^2))$ in
link units to every member of $\M_0$, with design coordinates $\tilde x$ scaled to $[0,1]$,
$g\in\{\pm0.5,\pm1\}$ and centres $c\in\{0.1,0.3,0.5,0.7,0.9\}$ (1-D) or
$\{0.15,0.5,0.85\}^2$ (2-D); $g=0$ recovers $\M_0$. A truth starts from a parameter drawn
uniformly from the interior of the $\M_0$ grid. A \emph{bump} adds a Gaussian deviation with
amplitude $\pm U(0.5,1)\,A$, centre $U[0,1]^d$ and width log-uniform on $[0.04,0.3]$. A
\emph{smooth} deviation is a sum of three random cosines with amplitudes $N(0,(A/2)^2)$ and
frequencies $U(0.3,1.5)$. Well-specified controls add nothing.

\paragraph{Policies and metrics.} Greedy EIG maximizes expected information gain about the
$\M_0$ parameter at each step. After detection with expansion on, every policy designs greedily
under $\M_1$. Protected risk is half the expected multiclass Brier score of the posterior
predictive against the truth, averaged uniformly over the 25 designs; for binary outcomes it
equals the usual Brier score. Every (world, policy) cell uses 200 independent seeds; $T=60$,
$\alpha=0.05$.

\paragraph{Reproducibility.} All experiments run from one Python package with a fixed command;
each result is tied to a recorded commit and run. Runs cited in this draft: generality gate
(fixed-class accuracy, Figure~\ref{fig:collapse}b), valid-detector re-test (detection and
budgets, Figures~\ref{fig:collapse}a and~\ref{fig:budget}), and theory check
(Section~\ref{sec:collapse}). A separate run reproduces the original pilot study to five
decimals.

\section{Fixed-truth calibration and limits of adaptive probing}
\label{app:adaptiveaudit}

The following statements concern one fixed, finite, strictly positive expansion.
They do not transfer a finite-grid guarantee to a continuous scientific null or
to truths outside the expansion.

\subsection{A conditional detect-or-resolve statement}
Let $\M_0\subset\M_1$ be finite, $m=|\M_1|$, and let
$P\in\M_1\setminus\M_0$. Use a uniform prior on $\M_1$ and the detector
\eqref{eq:eprocess}. For $g\in\M_1$ write
$S_{g,t}=\sum_{s\le t}d_g(a_s)$, with $d_g$ defined as in
\eqref{eq:cumulative-separation}. Define
$W=\{g:\ p_g(\cdot\mid x)\ne P(\cdot\mid x)\text{ for some }x\text{ with }\mu(x)>0\}$.
If $W$ is empty then $U_t=0$ identically. Otherwise fix $0<u\le\log m$ and set
\[
c_{\rm det}=\log(m/\alpha)+2\log(m/\beta),\qquad
c_{\rm res}=2\log(m/\beta)+2\log(2|W|/u).
\]

\begin{proposition}[Conditional detect or resolve]
\label{prop:detectresolve}
Under every predictable policy there is an event of $P$-probability at least
$1-\beta$ on which, simultaneously in time,
\[
\tau>t\ \Longrightarrow\ R_t<c_{\rm det},\qquad
U_t\ge u\ \Longrightarrow\ \min_{g\in W}S_{g,t}<c_{\rm res}.
\]
Call a design $\Delta$-informative if
$\min_{\theta\in\M_0}d_\theta(a)\ge\Delta$ or
$\min_{g\in W}d_g(a)\ge\Delta$. On the same event, if $\tau>T$ and
$U_T\ge u$, fewer than $(c_{\rm det}+c_{\rm res})/\Delta$ of the first $T$
designs are $\Delta$-informative.
\end{proposition}

\begin{proof}
For each $g\ne P$, the normalized square-root likelihood-ratio martingale from
Theorem~\ref{thm:achieve} and a union bound over at most $m$ members give an event
of probability at least $1-\beta$ on which
\[
\log(L_{g,t}/L_{P,t})<-S_{g,t}+2\log(m/\beta)
\quad\text{for all }g\ne P,t.
\]
The numerator is at least $L_{P,t}/m$, so
$\log E_t>R_t-\log m-2\log(m/\beta)$, proving the detection implication.
Put $\varepsilon=w^1_t(W)$. Members outside $W$ agree with $P$ on the target.
Grouping that common predictive component and then bounding the entropy of the
remaining $|W|$ labels gives, when $\varepsilon>0$,
\[
U_t\le\varepsilon+\varepsilon\log(|W|/\varepsilon)
<2\sqrt{\varepsilon|W|}.
\]
Indeed the common component contributes at most
$-(1-\varepsilon)\log(1-\varepsilon)\le\varepsilon$, and the remaining label
entropy is at most $\varepsilon\log(|W|/\varepsilon)$.
For $v=\varepsilon/|W|\le1$, use
$v\log(1/v)\le(2/e)\sqrt v$ and $\varepsilon\le\sqrt{\varepsilon|W|}$.
If $\varepsilon=0$ then $U_t=0$ directly.
If every $S_{g,t}$ for $g\in W$ were at least $c_{\rm res}$, then
\[
\varepsilon\le\sum_{g\in W}L_{g,t}/L_{P,t}
<|W|(m/\beta)^2e^{-c_{\rm res}}=u^2/(4|W|),
\]
contradicting $U_t\ge u$. Finally, count the designs satisfying each of the two
informativeness conditions. Nonnegative separation gives
$R_T\ge n_{\rm det}\Delta$ and
$\min_{g\in W}S_{g,T}\ge n_{\rm res}\Delta$; their sum bounds the union count.
\end{proof}

Intersect this event at level $\beta/2$ with Proposition~\ref{prop:fixedtruth}
at level $\beta/2$. If every targeted design is $\Delta$-informative, then by
$T=\lceil(c_{\rm det}+c_{\rm res})/\Delta\rceil$, with constants evaluated at
$\beta/2$, some $t\in\{0,\ldots,T\}$ has either $\tau\le t$ or $U_t<u$.
In the latter case the expanded-predictive error is below $2mu/\beta$, while
working-predictive error is below $2m(u+D_t)/\beta$. The timing uses $U_t$ to
choose action $a_{t+1}$; resolution at time zero is allowed. This is an
existence statement, not permanent resolution. The pointwise informativeness
condition is stronger than separation obtained by complementary designs and
has not been established for either greedy acquisition rule in this paper.

\subsection{Predictive cancellation}
\label{app:predictivecancellation}
Let actions be $a,x$, target $\delta_x$, null $\theta=(1/2,1/2)$, and a uniform
expansion $\{\theta,P,P',h\}$, where probability pairs at $(a,x)$ are
$P=(1/2,9/10)$, $P'=(1/2,1/10)$ and $h=(9/10,1/2)$.
Along the auxiliary experiment that always chooses $a$, weights take the form
$(v,v,v,1-3v)$ with $v=1/(3+L_h/L_P)$. The predictive at $x$ is exactly $1/2$,
whereas at $a$ it is $9/10-6v/5>1/2$ for every finite history.
Thus predictive-KL targeting uniquely chooses $a$ in exact arithmetic.
Ville's inequality gives probability at least $3/4$ that $L_h/L_P<4$ for all
time, hence $v>1/7$ and
$U_t>2\operatorname{kl}(9/10\|1/2)/7>0.1$.
Under the always-$a$ experiment, $P$ and $\theta$ induce identical observations;
null validity bounds the probability of any rejection by $\alpha$.
On the intersection of these two events the actual $U$-triggered policy
coincides with the always-$a$ experiment forever. With probability at least
$3/4-\alpha$ it never rejects and its working and expanded target errors both
remain $\operatorname{kl}(9/10\|1/2)$. Finite-precision rounding may change
vanishing score differences into ties; this infinite-horizon argument assumes
exact arithmetic and does not follow from a finite simulation alone.

\subsection{Premature return to the base policy}
\begin{proposition}[A correct target score does not repair the ignorance trigger]
\label{prop:triggerfailure}
There is a strictly positive two-model expansion containing the truth for which
both predictive-KL and posterior-expected-discrepancy targeting choose the useful
design, but the rule with $U$ trigger $0.1$ takes exactly one useful observation
and never rejects at $\alpha=0.05$, on every response path.
\end{proposition}

\begin{proof}
Use actions $a,x$, target $\delta_x$, working law $\theta=(1/2,1/2)$, true law
$P=(1/2,9/10)$ and the uniform expansion $\{\theta,P\}$. Let the base always
choose $a$, a valid deterministic EIG tie rule under the singleton null.
Both target scores are zero at $a$ and positive at $x$ at every finite
nondegenerate posterior. Initially $U_0=0.101749225\ldots>0.1$.
After the first observation at $x$, success gives weight $w(P)=9/14$ and
$U_1=0.097816505\ldots$; failure gives $w(P)=1/6$ and
$U_1=0.052428619\ldots$. Thus both branches return to $a$. Because the models
agree at $a$, all later posterior weights and likelihood ratios remain fixed.
The sequence statistic stays respectively at $7/5$ or $3/5$, below 20, while
working target error remains $\operatorname{kl}(9/10\|1/2)=0.368064207\ldots$.
The same branch inequalities hold at $u=\log(20)/30$. Exact rational weights
and outward 60-digit logarithm enclosures verify the stated strict inequalities;
no sampling is used.
\end{proof}

The stated deterministic base tie rule is essential to the forever claim.
A randomized-tie or forced-coverage base that keeps visiting $x$ can accumulate
evidence after the targeted branch stops. This is not a zero-power statement
for every $U$-triggered policy.

After success in this construction $B_1=0.236612704\ldots>0.1$ despite
$U_1<0.1$, so a whole-$B$ gate continues at that history. After failure,
$B_1=0.061344034\ldots<0.1$; this threshold itself is not a calibrated
fixed-truth guarantee. Targeting, trigger calibration and realized probing
effort therefore require separate comparisons. Neither changing the score
nor using the same numeric threshold on $B$ establishes a universal repair.
In fact the first target failure has true probability $1/10$. On that branch
the $B\ge0.1$ rule also returns permanently to $a$, with statistic $3/5$ and
true target KL error $0.368064207\ldots>0.1$. Thus even with a represented
truth it has probability at least $0.1$ of permanent return above that error
margin. If that return is interpreted as reassurance, its error probability
exceeds a proposed frequentist risk budget $\beta=0.05$.
The same deterministic base assumption is essential. This branch does not
contradict improvement on the separately tested four-model construction or
other empirical worlds; it rules out treating the numeric $B$ trigger as a
general calibrated guarantee. Exact rational branch probability and outward
enclosures verify this deduction without sampling.

\subsection{A conditional forced-targeting safety net}
\label{app:forcedcoverage}
An independent decaying exploration coin addresses permanent cessation, under
a stronger design condition than has been proved for our acquisition objective.
This is a corollary of Theorem~\ref{thm:achieve}, rather than a new inference
principle. Let potential independent coins $F_t$ have probabilities
$\epsilon_t\in[0,1]$, independently of response randomness, and before rejection
choose a targeted design whenever the ordinary gate opens or $F_t=1$.

\begin{proposition}[Conditional eventual detection and extra-probe cost]
\label{prop:forcedcoverage}
Use the finite positive null and represented truth, uniform expansion prior
and proper mixture detector of Theorem~\ref{thm:achieve}. Suppose every forced
target selected before rejection satisfies
$\min_{\theta\in\Theta}d_\theta(a_t)\ge\Delta>0$.
If $\sum_t\epsilon_t=\infty$, then $\Pr(\tau<\infty)=1$.
Writing $S_T=\sum_{t\le T}\epsilon_t$, if
\[
\Delta S_T/2\ge\log(1/\alpha)+\log|\M_1|+2\log(|\Theta|/\beta),
\]
then $\Pr(\tau\le T)\ge1-\beta-\exp(-S_T/8)$.
The expected number of additional forced probes taken while the ordinary
gate is closed is at most $S_T$. For $\epsilon_t=c/\sqrt t$, $0<c\le1$,
this is at most $2c\sqrt T$.
\end{proposition}

\begin{proof}
Let $Z_T=\sum_{t\le T}F_t$, defining potential coins even after rejection.
On $\{\tau>T\}$, at least $Z_T$ observations are informative, hence
$R_T\ge\Delta Z_T$. The restriction to this event matters because targeting
may cease after rejection. On the simultaneous good event of
Theorem~\ref{thm:achieve}, this lower bound forces rejection once sufficiently
large. If the coin probabilities have infinite sum, independent
Borel--Cantelli gives $Z_T\to\infty$ almost surely. Therefore
$\Pr(\tau=\infty)\le\beta$ for every $\beta>0$, which proves eventual rejection.
For finite $T$, the independent Bernoulli Chernoff bound gives
$\Pr(Z_T<S_T/2)\le e^{-S_T/8}$; combine it with the theorem by a union bound.
Finally the additional-probe indicator is bounded by $F_t$; take expectations
and use $\sum_{t\le T}t^{-1/2}\le2\sqrt T$.
\end{proof}

The cost bound concerns additional cold-gate probes, under any truth; it does
not bound all ordinary-gate probes or the total-effort difference between two
policies whose histories diverge. Nor does it calibrate a numeric $B$ threshold.
The separation condition is not established for expected-discrepancy targeting
in general and does not cover complementary designs that are informative only
jointly. Truths outside the finite expansion and numerical posterior underflow
also lie outside this exact-arithmetic guarantee.

\subsection{Cyclic coverage with an alternative that need not contain the truth}
\label{app:cycliccoverage}
The preceding forced-target result assumes each forced target separates the
truth from every null member. A different policy can replace that assumption
with identifiability across the available actions. Forced exploration in
controlled sensing, including schedules producing order-$\sqrt T$ visits to
each control, is established \citep{deshmukh2019controlled}. The following is
a composition with Proposition~\ref{prop:broad}, not a new general testing
principle or a guarantee for the targeted-floor policies already evaluated.

Let $m=|\A|<\infty$, let the outcome alphabet have $K\ge2$ categories, and let
$\Theta$ be a finite nonempty strictly positive null. Assume a stationary,
strictly positive categorical truth $P$ satisfies
\[
\text{for every }\theta\in\Theta,\quad
\text{there exists }a\in\A\text{ with }P_a\ne p_{\theta,a}.
\]
Write $d_\theta(a)$ as in \eqref{eq:cumulative-separation} and define
\[
D_*=\min_{\theta\in\Theta}\sum_{a\in\A}d_\theta(a)>0.
\]
The agent need not know $D_*$ or $P$. Before outcomes are observed, draw
independent potential coins $F_t\sim\mathrm{Bernoulli}(\epsilon_t)$,
independent of response randomness. Before rejection, a forced coin selects
the next action in a fixed full cycle through $\A$, \emph{with priority over
the ordinary targeting gate}. Other choices may follow any predictable rule.
After rejection the override may stop. Use a proper sequence likelihood $Q_t$
with a fixed universal component satisfying $Q_t\ge\lambda L_{\mathrm{sat},t}$,
$0<\lambda\le1$, and reject at
$\tau=\inf\{t:Q_t/\max_\theta L_{\theta,t}\ge1/\alpha\}$.

\begin{corollary}[Coverage sufficient for eventual detection]
\label{cor:cycliccoverage}
Let $S_T=\sum_{t\le T}\epsilon_t$. For $0<\alpha,\beta<1$, if
\begin{align}
\label{eq:cycliccondition}
\left\lfloor\frac{S_T}{2m}\right\rfloor D_*
&\ge \log(1/\alpha)+\log(1/\lambda)
       +2\log(|\Theta|/\beta)\nonumber\\
&\quad +m(K-1)\log(T+K-1),
\end{align}
then $\Pr_P(\tau\le T)\ge1-\beta-\exp(-S_T/8)$.
For $\epsilon_t=c/\sqrt t$, $0<c\le1$, eventual rejection has probability
one. The expected number of completed coverage-route observations before
rejection and through $T$ is at most $S_T\le2c\sqrt T$.
The likelihood test retains level-$\alpha$ validity for each null member.
\end{corollary}

\begin{proof}
Define $Z_T=\sum_{t\le T}F_t$, including potential coins after rejection.
On $\{\tau>T\}$, the full cycle gives at least $\lfloor Z_T/m\rfloor$
observations at every action. Nonnegativity of $d_\theta(a)$ implies
$R_T\ge\lfloor Z_T/m\rfloor D_*$ on this event, even when a single action
has $\min_\theta d_\theta(a)=0$.
Proposition~\ref{prop:broad} and the fixed mixture component give, with
probability at least $1-\beta$, simultaneously at every time,
\[
\log\frac{Q_T}{\max_\theta L_{\theta,T}}
\ge R_T-C_T-2\log(|\Theta|/\beta)+\log\lambda.
\]
The count bound in \eqref{eq:broad-regret} satisfies
$C_T\le m(K-1)\log(T+K-1)$. The independent Bernoulli Chernoff bound is
$\Pr(Z_T<S_T/2)\le\exp(-S_T/8)$. Thus, on the intersection of the two good
events, \eqref{eq:cycliccondition} contradicts $\tau>T$. A union bound proves
the finite-horizon claim. With $\epsilon_t=c/\sqrt t$,
$S_T\sim2c\sqrt T$ dominates $\log T$. For each fixed $\beta>0$ the
condition eventually holds, yielding $\Pr(\tau=\infty)\le\beta$; let
$\beta\downarrow0$. The coverage-route indicator is at most $F_t$, giving
the expected-count bound under any truth. Finally, under a null member,
$Q_t/\max_\theta L_{\theta,t}\le Q_t/L_{\theta,t}$, whose right-hand side
is a nonnegative likelihood-ratio martingale for the same adaptive policy.
Ville's inequality proves validity.
\end{proof}

For example, with two Bernoulli actions, truth $(1/5,1/5)$, and null members
$(1/5,4/5)$ and $(4/5,1/5)$, each action separately has zero minimum
separation. The mismatched affinity is $4/5$, so
$D_*=-2\log(4/5)>0$; cycling supplies joint evidence.
The finite-time bound is conservative and can be vacuous at realistic budgets.
It establishes neither a numerical gate calibration nor superiority over
uniform, balanced or cost-aware allocation. It bounds coverage-route counts,
not all ordinary-gate probes, physical cost without action prices, or the
total-effort difference between policies with different histories.
An unbounded continuation with a diverging coverage budget is essential to
the asymptotic claim; a hard fixed override cap does not satisfy that premise.
The argument does not cover a continuous null by replacing it with a grid,
growing action sets, or nonstationary outcomes. The sufficient universal-coding
argument needs separation to dominate its logarithmic penalty; mere infinite
exploration is not enough to establish that using this bound. This observation
does not prove that slower exploration schedules fail.

\subsection{Compact continuous nulls and numerical likelihood error}
\label{app:compactcoverage}
A finite inference grid does not replace the scientific null. This subsection
extends the coverage calculation to a compact continuous null under an explicit
log-likelihood covering condition. Covering arguments for likelihood confidence
sets and regret-to-confidence conversions are established
\citep{kirschner2025likelihood}; the contribution here is the stated combination
with cumulative separation, universal alternative coding, cyclic coverage and
certified numerical error. It is not a new general inference principle.

Let $\A$ and $\Y$ be fixed finite sets, $m=|\A|\ge1$, $K=|\Y|\ge2$, and
let $P$ be a stationary strictly positive categorical truth: the next response
has conditional law $P(\cdot\mid a_t)$ given the history and chosen action.
Let $\Theta$ be nonempty and
compact, and let $p_\theta(y\mid a)>0$ be continuous in $\theta$ for each
$(a,y)$. Use any common predictable policy and a proper sequence likelihood
$Q_T\ge\lambda L_{\mathrm{sat},T}$ with fixed $0<\lambda\le1$.
Write $f_T(\theta)=\log L_{\theta,T}$,
$f_T^*=\sup_{\theta\in\Theta}f_T(\theta)$, and
$R_T=\inf_{\theta\in\Theta}\sum_{s\le T}d_\theta(a_s)$.
Let $H_j=2^j$, $j=0,1,\ldots$, and let $j(T)=\lceil\log_2 T\rceil$.
For each epoch, suppose a deterministic finite set
$\Theta_j\subset\Theta$, of cardinality $N_j$, satisfies
\begin{equation}
\label{eq:compactcover}
\sup_{\theta\in\Theta}\inf_{\vartheta\in\Theta_j}
\max_{a,y}|\log p_\theta(y\mid a)-\log p_\vartheta(y\mid a)|
\le\kappa_j/H_j.
\end{equation}
The set is a proof device, not a replacement denominator for the test.

\begin{proposition}[Continuous-null cumulative separation]
\label{prop:compactseparation}
For every $0<\beta<1$, with probability at least $1-\beta$, simultaneously
for all positive integers $T$,
\begin{align}
\label{eq:compactseparation}
\log Q_T-f_T^*
&\ge R_T-C_T-\log(1/\lambda)-\kappa_{j(T)}\nonumber\\
&\quad-2\log\!\left(\frac{N_{j(T)}(j(T)+1)(j(T)+2)}{\beta}\right),
\end{align}
where $C_T$ is the coding bound in \eqref{eq:broad-regret}.
The test rejecting when $\log Q_T-f_T^*\ge\log(1/\alpha)$ retains
anytime level $\alpha$ at every $\theta\in\Theta$.
\end{proposition}

\begin{proof}
Set $\beta_j=\beta/[(j+1)(j+2)]$, so $\sum_j\beta_j=\beta$.
For each epoch $j$ and $\vartheta\in\Theta_j$, use the square-root
likelihood-ratio martingale from Theorem~\ref{thm:achieve}. Ville's
inequality and a union bound over the deterministic cover give, except on
an event of probability at most $\beta_j$, simultaneously for every $T$,
\[
\log L_{P,T}-\max_{\vartheta\in\Theta_j}f_T(\vartheta)
\ge\min_{\vartheta\in\Theta_j}\sum_{s\le T}d_\vartheta(a_s)
-2\log(N_j/\beta_j).
\]
The cover is contained in the null, so its minimum separation is at least
$R_T$. At a time with $j=j(T)$ we have $T\le H_j$; hence
condition~\eqref{eq:compactcover} gives
$f_T^*\le\max_{\vartheta\in\Theta_j}f_T(\vartheta)+\kappa_j$.
Union over epochs and combine with the universal coding bound and fixed mixture
weight. The likelihood ratio $Q_T/L_{\theta,T}$ is a nonnegative martingale
under each null member for the same adaptive policy, and dominates
$Q_T/\exp(f_T^*)$. Ville's inequality proves validity. The latter statistic
need not itself be a martingale or have valid conditional multiplicative
increments.
\end{proof}

\paragraph{An explicit logistic cover.}
For $p_\theta(1\mid a)=\sigma(\theta^\top\phi(a))$,
$\Theta=[-B,B]^d$, $B>0$, and
$G\ge\max_a\|\phi(a)\|_1>0$, the log probability has gradient
$(y-p_\theta(1\mid a))\phi(a)$ and is $G$-Lipschitz in the parameter
$\ell_\infty$ norm. A tensor grid with $n_j=\lceil2BGH_j\rceil$ equal
intervals per coordinate is contained in $\Theta$, has
$N_j=(n_j+1)^d$, and places a point within $B/n_j$ in each coordinate.
Thus \eqref{eq:compactcover} holds with $\kappa_j=1/2$.
For the nine-action scientific family $\sigma(a+bx)$ on $[-3,3]^2$ and
$x\in[-1,1]$, one can take $d=2$, $B=3$, $G=2$, and
$N_j=(12H_j+1)^2$. This theoretical cover is not the $9\times9$ or
$17\times17$ inference quadrature used by acquisition.

\paragraph{Separating validity from numerical power.}
Suppose verified numerical bounds provide
$\ell_T\le f_T^*\le U_T$,
$\underline q_T\le\log Q_T\le\overline q_T$, and
$\overline b_T\ge\log(1/\alpha)$.
Set $g_T=U_T-\ell_T$, $w_T=\overline q_T-\underline q_T$, and
$h_T=\overline b_T-\log(1/\alpha)$.
The reported rejection margin obeys, on the same probability event,
\begin{align*}
\underline q_T-U_T-\overline b_T
&\ge R_T-C_T-\log(1/\lambda)-\kappa_{j(T)}\\
&\quad-2\log\!\left(\frac{N_{j(T)}(j(T)+1)(j(T)+2)}{\beta}\right)
-\log(1/\alpha)-(g_T+w_T+h_T).
\end{align*}
Here $U_T-f_T^*\le g_T$, since $\ell_T\le f_T^*$; analogous inequalities
hold for numerator and threshold rounding. A lower threshold enclosure
$\underline b_T\le\log(1/\alpha)$ gives the computable bound
$h_T\le\overline b_T-\underline b_T$. Rejecting only at a nonnegative
reported margin preserves validity. A loose $U_T$ can nevertheless prevent
rejection indefinitely. If an implementation can remain unresolved or suppress
a certified crossing for another reason, that behaviour requires a separate
liveness condition; validity alone does not imply power.

\begin{corollary}[Continuous-null coverage with controlled numerical error]
\label{cor:compactcoverage}
Assume $P\ne p_\theta$ as action-indexed distributions for every
$\theta\in\Theta$. Then
$D_*=\min_{\theta\in\Theta}\sum_a d_\theta(a)>0$.
Use the independent potential coins and full-cycle priority of
Appendix~\ref{app:cycliccoverage}. Suppose all auxiliary acceptance conditions
hold and a nonnegative reported margin triggers rejection at every analyzed
time. On $\{\tau>T\}$, suppose numerical error satisfies the
deterministic envelope $g_T+w_T+h_T\le\eta_T$; exact arithmetic has
$\eta_T=0$. If
\begin{align}
\label{eq:compactcoveragebudget}
\left\lfloor\frac{S_T}{2m}\right\rfloor D_*
&\ge \log(1/\alpha)+\log(1/\lambda)+\kappa_{j(T)}+\eta_T\nonumber\\
&\quad+m(K-1)\log(T+K-1)
+2\log\!\left(\frac{N_{j(T)}(j(T)+1)(j(T)+2)}{\beta}\right),
\end{align}
then $\Pr_P(\tau\le T)\ge1-\beta-\exp(-S_T/8)$.
For the logistic cover, $\epsilon_t=c/\sqrt t$, $0<c\le1$, and
$\eta_T=o(\sqrt T)$, eventual rejection has probability one; expected
coverage-route count through $T$ is at most $2c\sqrt T$.
\end{corollary}

\begin{proof}
Strict positivity, continuity and compactness make
$\sum_a d_\theta(a)$ continuous with a strictly positive minimum: zero would
require equality of both distributions at every action. On $\{\tau>T\}$,
the cycle implies $R_T\ge\lfloor Z_T/m\rfloor D_*$.
Use Proposition~\ref{prop:compactseparation}, the numerical-margin bound,
$C_T\le m(K-1)\log(T+K-1)$ and
$\Pr(Z_T<S_T/2)\le\exp(-S_T/8)$ to contradict $\tau>T$ on both good events.
With the specified logistic cover, all terms except $\eta_T$ are
$O(\log T)$, while $S_T\sim2c\sqrt T$. For every fixed $\beta>0$ the
condition eventually holds. Let $T\to\infty$ and then $\beta\downarrow0$.
The expected-count bound follows from the coverage indicator being at most
$F_t$, independently of the numerical computation.
\end{proof}

The finite-time bound can be extremely conservative. Compactness,
identifiability, a proper universal component, actual cyclic coverage and
numerical liveness remain substantive premises. The theorem does not cover
an unrestricted continuous action set, a noncompact null with zero separation
infimum, nonstationarity, a hard fixed coverage cap or predictive-risk
reassurance. In particular, it does not automatically validate the existing
forced-target floor, the primary benchmark's fixed scheduled allocations, or
a finite-grid detector. The truth-dependent $D_*$ and $R_T$ are analysis
quantities, not computable reassurance certificates for the agent.
Its logarithmic covering penalty is not asserted
optimal. Epoch allocation avoids a union over every integer horizon, but
remains a standard deterministic union-bound construction.

\subsection{Confidence-set inversion beyond prior atoms}
\label{app:continuousriskgate}
The posterior-mass bound in Proposition~\ref{prop:fixedtruth} uses a positive
atom at the true law. Confidence-set inversion avoids that requirement.
Let $\{P_\eta:\eta\in K\}$ be a separately declared compact family with
strictly positive likelihoods and continuous likelihood and target-risk maps.
For any proper predictable response-sequence likelihood $Q_t$, set
\[
\mathcal C_t=\{\eta\in K:L_{\eta,t}\ge\beta Q_t\},\qquad
A_t=\sup_{\eta\in\mathcal C_t}\sum_x\mu(x)
\KL(P_\eta(\cdot\mid x)\|p_t(\cdot\mid x)).
\]
If $\mathcal C_t$ is empty, report expansion insufficiency rather than reassurance;
for the risk gate define $A_t=+\infty$ in that case.

\begin{proposition}[A protected-risk gate from likelihood confidence sets]
\label{prop:continuousriskgate}
For each fixed $P_{\eta^*}$ in the declared family, under any predictable design
policy, with probability at least $1-\beta$ the true target error is at most
$A_t$ simultaneously in time. This does not require the numerator prior to
assign positive point mass, or any support, to $\eta^*$. Consequently returning
to the base only when a certified upper bound on $A_t$ is at most $\varepsilon$
has probability at most $\beta$ of ever reassuring while true protected error
exceeds $\varepsilon$.
\end{proposition}

\begin{proof}
Under the fixed truth $P_{\eta^*}$, $Q_t/L_{\eta^*,t}$ is a nonnegative mean-one
martingale. Ville's inequality bounds the probability that it ever reaches
$1/\beta$ by $\beta$. On the complement,
$L_{\eta^*,t}>\beta Q_t$ for every $t$, so $\eta^*\in\mathcal C_t$ and its
target risk is bounded by the supremum defining $A_t$.
\end{proof}

This is the standard likelihood-ratio confidence-sequence construction
\citep{kirschner2025likelihood} applied to the protected-risk decision, not a new
general inference principle.
It controls target prediction error, not membership in $\M_0$, and assumes the
truth belongs to the separately declared family. It does not require the truth
to be a posterior quadrature point. However, maximizing risk only over that
finite grid does not certify the continuous-family supremum. A valid numerical
gate needs an outer confidence set and a global upper bound on risk. With
outward likelihood intervals, prune a parameter region only if its likelihood
upper bound is below a lower bound on $\beta Q_t$. Bound risk over every
unpruned region; unresolved optimization must not trigger reassurance. The
continuous risk reference is now implemented for finite unions of bounded
two-coefficient logistic families with fixed structural offsets. It retains a
complete binary partition, outward likelihood and risk bounds, and independently
replayable feasible witnesses. Empty sets, exhausted budgets and unresolved
enclosures do not permit reassurance. The reference is now coupled to a proper
sequence numerator and pre-outcome campaign traces; the limited feasibility
evaluation below does not establish an advantageous resource trade-off.
No stopping-time or detection-power guarantee is claimed.

An analytical check separates sample count from design coverage. Let
$a,b\in[-3,3]$, $p_{a,b}(x)=\sigma(a+bx)$, working probabilities $1/2$,
$\beta=0.05$, and collect 100 successes and 100 failures entirely at $x=0$.
Use the proper fixed Bernoulli-$1/2$ sequence numerator $Q=2^{-200}$.
Writing $p_0=\sigma(a)$, the confidence condition is exactly
$[4p_0(1-p_0)]^{100}\ge0.05$. Consequently $|a|\le a_{\max}$ with
$a_{\max}=0.3470284983\ldots$, while $b$ remains unrestricted.
The exact risk suprema for targets $x=0$ and $x=1$ are respectively
$\operatorname{kl}(\sigma(a_{\max})\|1/2)=0.01482998049\ldots$ and
$\operatorname{kl}(\sigma(3+a_{\max})\|1/2)=0.54478917478\ldots$.
The reference certifies the observed target below $\varepsilon=0.02$ and
exhibits a compatible risk exceeding $0.50228$ at the unobserved target.
The latter witness $(a,b)=(0,-3)$ stays compatible for every balanced even
sample size at $x=0$, since its likelihood equals $Q$. These are analytical
histories checked against a separate closed-form calculation, not measured
power or observed truth error. Ten numerical checks and eight literal
certificates validate the reference's stated operations, not campaign-scale
coverage, efficiency or scientific superiority.

\subsection{Integrated risk-gate feasibility}
\label{app:riskgatefeasibility}
A registered short-horizon feasibility study uses six prospectively selected
worlds from an independently generated 24-world development specification:
two nulls, two matched bumps, one near-null departure and one outside-prior
quadratic departure. It runs 42 campaigns at $T=12$: expected discrepancy,
balanced coverage and class information, each with fixed scheduled interventions
or risk-based safe suppression, plus no overrides. Four additional campaigns
check primary/refined quadrature on two worlds and two fixed policies.
Eligible slots number exactly two; suppression does not move a slot.
Risk settings are fixed before collection: $\beta=0.05$, uniform action target,
$\varepsilon=0.1$, 50-digit arithmetic, at most 32 splits and depth 30.
The declared continuous family contains the null and 18 fixed numerical
structural offsets, with $a,b\in[-3,3]$. It is not a guarantee for every generated
departure. Working prediction always remains under $\M_0$.

All 46 histories (552 observations) replay successfully; no collection failure
or unresolved detector decision occurs. Every one of 36 gate evaluations
finds a compatible model whose risk exceeds the margin. Consequently no
intervention is suppressed, and all 18 gated/fixed pairs have identical
actions, outcomes and prediction endpoints. No arm rejects on these short
histories. Gate computation takes 89.0 seconds of 118.7 seconds total collection
time, with 124.7 additional seconds for full replays; paired collection time
is a median 15.3 times that of the corresponding fixed arm in this local run.
Compressed records total 790,287 bytes. These timings are implementation and
short-history measurements, not general complexity bounds. The result establishes
integration feasibility while exposing conservatism and computation without
observed experimental savings. It does not establish power, a preferred policy,
family-wide empirical calibration or a confirmation-ready method.

\section{When an ED gate does and does not imply remaining power}
\label{app:gateachievability}
This analysis follows protocol commit \texttt{df1ce0f}, written after the integration
review and conference draft. It is an analytical contribution with exact
finite-state checks, not another empirical confirmation. The standard
concentration tools are not new: horizon-aware betting already relates drift,
wealth and time-to-go \citep{taga2026horizon}, and lower-tail control is central
to power guarantees for test supermartingales \citep{forre2026typeii}.
Our purpose is to test the logical implication used to motivate the ED gate,
while retaining numerator regret and the remaining budget explicitly.

\subsection{A represented-truth, correctly specified-prior counterexample}
Consider one Bernoulli action, null $p_0=1/2$, expansion
$\{p_-,p_+\}=\{1/100,99/100\}$, and an equal prior. Use that same prior for
both the planning posterior and the detector numerator. Either represented
alternative can be the truth; all outcome probabilities are strictly positive.
With $k$ successes in $t$ observations,
\[
 E_{t,k}=\frac{2^t}{2}\left\{(99/100)^k(1/100)^{t-k}
                    +(1/100)^k(99/100)^{t-k}\right\}.
\]
Both alternatives have the same divergence from the null,
\[
 d=\frac{99}{100}\log\frac{99}{50}
      +\frac1{100}\log\frac1{50}=0.6371456462\ldots.
\]
Consequently $B_t=\mathrm{ED}_t=d$ at every possible history, regardless of
posterior weights. For $\alpha=1/20$ and $T=5$ this exceeds
$\log(20)/5=0.5991464547\ldots$: the gate is always open, every available
experiment is selected, and there is no acquisition bottleneck.
Nevertheless, convexity in $k$ puts the maximum sequence likelihood at $k=0,t$,
so for $1\le t\le5$,
\[
 \max_k E_{t,k}=2^{t-1}\{(99/100)^t+(1/100)^t\}
                  \le2^{t-1}\le16<20.
\]
Detection probability is exactly zero under either represented truth and hence
under the correctly specified expansion prior. At $t=0$, $E=1$. The sharper
maximum at $t=5$ is $15.2158408$. Thus even persistent ED above the proposed
threshold, correct representation, correct prior, and fully informative
acquisition do not by themselves imply any positive power for this detector.
The issue is a finite-horizon evidence margin and mixture coding cost, not
failure to discover an informative setting. At $T=6$ the exact crossing
probability becomes $(99/100)^6+(1/100)^6=0.941480149402$.

\subsection{An exact deadline oracle on the same observations}
\label{app:deadlineoracle}
The preceding zero-power result concerns the specified mixture detector.
It is not an impossibility for all valid tests of these five observations.
Deadline-optimal testing via a Neyman--Pearson rejection event and its null
Doob martingale is established machinery
\citep[Proposition~2]{clerico2026timesensitive}. We apply that construction to
the existing example, allowing an explicit auxiliary test randomization.

Let $C$ be the event that all five outcomes agree, draw an independent
$U\sim\mathrm{Unif}[0,1]$ after the fifth outcome, and reject on
$A=C\cap\{U\le4/5\}$. Then
\[
 \begin{aligned}
 P_0(A)&=\frac45\,\frac2{2^5}=\frac1{20},\\
 P_{p_-}(A)=P_{p_+}(A)
 &=\frac45\left\{(99/100)^5+(1/100)^5\right\}\\
 &=0.76079204.
 \end{aligned}
\]
The null-conditional process $V_t=P_0(A\mid H_t)/\alpha$ is a nonnegative
martingale starting at one. For $1\le t<5$ it equals $2^{t-1}$ on an
all-equal prefix and zero otherwise; $H_5$ includes $U$, so $V_5=20\mathbf1_A$.
It cannot cross earlier, and it controls false rejection under arbitrary
monitoring, including after the deadline if held constant. The auxiliary
randomization is part of this oracle test, not another scientific observation
or a component of the original detector.

This power is sharp for randomized tests against the two specified alternatives.
The equal-mixture likelihood ratio is maximized on $C$, at
$L_{\max}=16\{(99/100)^5+(1/100)^5\}$. Every level-$\alpha$ test $\psi$ has
equal-prior average power at most $\alpha L_{\max}=0.76079204$.
Its worst-case power cannot exceed its average. Our symmetric test attains
this bound under both alternatives, proving minimax optimality for this fixture.
Thus the same informative data support substantial detection with a different
deadline-specific statistic, while the original mixture still has exactly zero
power. This strengthens the distinction between available information and
evidence captured; it does not validate the ED gate constant or supply a
practical continuous-null detector repair.

An exact rational checker covers horizons 1--12, constructs the randomized
likelihood-ratio test, verifies its linear-program dual bound, and checks all
454 null-martingale transitions including terminal randomization. Original
mixture first crossings are separately recomputed. This is an analytical
follow-up to a known example, not a new empirical confirmation.

\paragraph{Conditional increments need a separate check.}
Conditional e-increments are a stronger premise than validity of an entire
profile e-process \citep{sandoval2026multiarmed,lin2026sampleefficient}.
For one Bernoulli action with null parameters $\{1/5,4/5\}$ and an independent
Bernoulli-$1/2$ numerator, $E_1=5/8$. A repeated outcome gives $E_2=25/64$;
a mixed pair gives $E_2=25/16$. Under the true null $p=1/5$, conditional on
$Y_1=1$,
\[
 \begin{aligned}
 \mathbb E_{1/5}[E_2/E_1\mid Y_1=1]
 &=\frac15\frac58+\frac45\frac52\\
 &=\frac{17}{8}>1.
 \end{aligned}
\]
The profile process retains its anytime-valid domination by each fixed-null
likelihood-ratio martingale. Its value ratios cannot simply be inserted as
conditionally valid one-step e-values into another algorithm. A faithful
comparison requires an explicit valid construction for the shared null.

\subsection{A conditional bound that pays the missing costs}
Fix an unrejected history $H_t$, a truth $p_h$, terminal time $T$, and finite
null family $\Theta$ of size $m$. Let $n=T-t$. For each $\theta$ define
\[
 Z_{s,\theta}=\log\frac{p_h(Y_s\mid a_s)}{p_\theta(Y_s\mid a_s)},\qquad
 d_{s,\theta}=\mathbb E_h[Z_{s,\theta}\mid H_{s-1},a_s]
              =\KL(p_h(\cdot\mid a_s)\|p_\theta(\cdot\mid a_s)).
\]
Actions are predictable, including their independent randomization. Suppose
$Z_{s,\theta}$ has conditional range of width at most deterministic
$w_{s,\theta}$ on every continuation, and set
$V_\theta=\sum_{s=t+1}^T w_{s,\theta}^2>0$.
All quantities below are conditional on this fixed history. Assume there is an
event $G$ of probability at least $1-\eta$ on which, simultaneously,
\begin{enumerate}
\item the proper continuation numerator has regret at most $r$:
$\log(Q_T/Q_t)\ge\sum_{s=t+1}^T\log p_h(Y_s\mid a_s)-r$;
\item the selected designs satisfy $\sum_{s=t+1}^T d_{s,\theta}\ge K_\theta$
for every $\theta$;
\item the implemented terminal log statistic is at least
$\log Q_T-\max_\theta\log L_{\theta,T}-\varepsilon$,
and the terminal threshold comparison is evaluated without another suppressing
acceptance rule.
\end{enumerate}
Here $r$, $\varepsilon$ and $K_\theta$ are fixed given $H_t$;
$\varepsilon\ge0$ includes any proved denominator/numerator numerical loss.
The range bounds must hold outside $G$ as well. Failure of one of these premises
must not silently be assigned zero probability.
Write $b=\log(1/\alpha)$,
$R_t(\theta)=Q_t/L_{\theta,t}$ and
\[
 A_\theta=\log R_t(\theta)+K_\theta-r-\varepsilon-b.
\]
\begin{proposition}[Conditional achievability with regret and numerical loss]
\label{prop:conditionalachievability}
If all $A_\theta>0$, the probability of detection by $T$, conditional on $H_t$,
is at least
\[
 \left[1-\eta-\sum_{\theta\in\Theta}
             \exp\{-2A_\theta^2/V_\theta\}\right]_+.
\]
If a margin is nonpositive, replace that parameter's summand by one, yielding
the trivial lower bound. Zero-range parameters can instead be treated by their
deterministic increments directly.
\end{proposition}
\begin{proof}
For a fixed $\theta$, Hoeffding's conditional exponential bound, iterated over
the predictable designs, gives
\[
 P_h\left(\sum_{s=t+1}^T(Z_{s,\theta}-d_{s,\theta})<-a\,\middle|\,H_t\right)
 \le e^{-2a^2/V_\theta}\quad(a>0).
\]
On $G$, the implemented terminal log statistic is at least
\[
 \min_\theta\left\{\log R_t(\theta)+\sum_{s=t+1}^T Z_{s,\theta}\right\}
             -r-\varepsilon.
\]
If no crossing has occurred by $T$, this is below $b$. For some $\theta$ the
centered sum must then be below $-A_\theta$. A union bound over the finite null
and a probability cost $\eta$ for $G^c$ prove the claim. Terminal crossing is
sufficient for crossing by $T$; there is no optional-stopping approximation.
\end{proof}

For an actual mixture containing truth $h$, its detector posterior weight
$\pi_t^{\rm num}(h)$ gives the pathwise conditional regret bound
$r=-\log\pi_t^{\rm num}(h)$, with zero regret-failure probability.
This is the numerator's weight, not an interchangeable planning posterior.
For a continuous scientific null the finite union requires a covering or
other uniform argument, with its own approximation and complexity costs;
substituting the inference grid is not valid. This proposition alone therefore
does not certify the production continuous-null gate.

\subsection{A continuous logistic-null lower bound without a parameter grid}
\label{app:continuousachievability}
The finite-null union above cannot simply be applied to an inference grid.
For the particular Bernoulli logistic family, however, centering removes the
log-partition term and leaves an affine function of the null parameter. The
result below combines this elementary geometry with standard conditional
Hoeffding concentration \citep{howard2018chernoff}. It is a family-specific
sufficient bound, not a new concentration inequality. Lower-tail assumptions
remain essential \citep{forre2026typeii}; no conditional e-increment property
is asserted for ratios of successive profile statistics.

Condition on an unrejected history $H_t$. Let $p_s=p_h(1\mid a_s)\in(0,1)$,
where the actions are predictable, including their independent randomization,
and let $x_s=x(a_s)$. The scientific null is
$p_\theta(1\mid a)=\sigma(a_0+a_1x(a))$ for
$\theta=(a_0,a_1)\in\Theta=[-3,3]^2$.
Define $Z_{s,\theta}$ and $d_{s,\theta}$ as above, and let
$\mathcal V=\{-3,3\}^2$ contain the four vertices. Suppose deterministic
numbers $V_v\ge0$, fixed given $H_t$, satisfy on \emph{every} continuation
\[
 \sum_{s=t+1}^T\{\operatorname{logit}p_s-v_0-v_1x_s\}^2\le V_v,
 \qquad v\in\mathcal V.
\]
These are squared conditional-range bounds for the centered log likelihood,
not observed variances or fitted posterior uncertainty. Suppose an event $G$
has conditional probability at least $1-\eta$ and, on $G$,
\begin{enumerate}
\item the proper numerator satisfies
$\log(Q_T/Q_t)\ge\sum_{s=t+1}^T\log p_h(Y_s\mid a_s)-r$;
\item cumulative separation and the already accumulated evidence satisfy
\[
 \inf_{\theta\in\Theta}\left\{
 \log R_t(\theta)+\sum_{s=t+1}^T d_{s,\theta}\right\}\ge K;
\]
\item the effective terminal log statistic, after conservative threshold
padding, is at least
$\log Q_T-\sup_{\theta\in\Theta}\log L_{\theta,T}-\varepsilon$,
and rejection is evaluated when this effective statistic reaches
$\log(1/\alpha)$; any additional numerical acceptance guard succeeds.
\end{enumerate}
Here $r,K,\varepsilon\ge0$ except that $K$ may have either sign, and these
quantities are fixed given $H_t$; the range bounds hold outside $G$ as well.
In particular, an unresolved production profile-gap guard must be charged to
$G^c$ or otherwise ruled out, rather than silently treated as successful.
The numerical loss $\varepsilon$ includes any conservative threshold padding.

\begin{proposition}[Four-corner continuous-null achievability]
\label{prop:continuousachievability}
Write $A=K-r-\varepsilon-\log(1/\alpha)$. If $A>0$, conditional detection
probability by $T$ is at least
\[
 \left[1-\eta-\sum_{v\in\mathcal V:\,V_v>0}\exp\{-2A^2/V_v\}\right]_+.
\]
For a common envelope $V$, a sufficient margin for probability at least
$1-\eta-\beta$ is
$K\ge r+\varepsilon+\log(1/\alpha)+\sqrt{(V/2)\log(4/\beta)}$,
where $0<\beta<1$. A nonpositive margin gives only the trivial bound.
Zero-range vertices have identically zero centered noise and can be omitted
when $A>0$.
\end{proposition}
\begin{proof}
The Bernoulli likelihood gives the exact centered identity
\[
 Z_{s,\theta}-d_{s,\theta}
 =(Y_s-p_s)\{\operatorname{logit}p_s-a_0-a_1x_s\}.
\]
Thus $N_\theta=\sum_{s=t+1}^T(Z_{s,\theta}-d_{s,\theta})$ is affine in
$\theta$, pathwise. Every interior $N_\theta$ is a convex combination of
the four $N_v$, so $\inf_{\theta\in\Theta}N_\theta=\min_vN_v$.
For each fixed vertex, conditional Hoeffding's lemma and iteration give
$P_h(N_v<-A\mid H_t)\le\exp(-2A^2/V_v)$ for $A>0$.
On $G$, if all $N_v\ge-A$, the implemented terminal statistic is at least
\[
 \inf_{\theta\in\Theta}\left\{\log R_t(\theta)
                +\sum_{s=t+1}^T d_{s,\theta}+N_\theta\right\}
                 -r-\varepsilon
 \ge K-A-r-\varepsilon=\log(1/\alpha).
\]
It rejects by $T$. A union bound over the four noise events and $G^c$ proves
the result; no independence between vertices or between $G$ and the noise is
needed. The common-envelope form follows by bounding the four summands.
\end{proof}

Only the \emph{centered noise} reduces to corners. The cumulative-KL infimum
and likelihood supremum remain over the \emph{whole} continuous box, whose
optimizers can be interior. Replacing the scientific denominator by four
corner likelihoods would invalidate it. For $|x_s|\le1$ and
$p_s\in[\delta,1-\delta]$, $0<\delta<1/2$, one available envelope is
$V=(T-t)\{\log((1-\delta)/\delta)+6\}^2$. This is deliberately conservative.
The result allows arbitrary predictable actions but does not prove that the
ED policy attains $K$ or bounds its numerator regret. It neither validates
the gate constant nor reverses any failed empirical diagnostic.

\paragraph{Declared checks and non-vacuity.}
Exact rational response/action-coin tree checks at horizons 1--5 verify
34,100 centered-likelihood identities on 1,364 leaves with 25 parameter
points, affine corner extrema and normalized exponential-martingale mass.
They use a scaled-logit fixture with rational odds, not new scientific worlds.
They corroborate the algebra; the proof of concentration is the conditional
Hoeffding argument, not these finite checks.

For a separate analytical example declared before evaluation, cycle the nine
dyadic settings $x_j=(j-4)/4$, set truth to $1/4$ at indices $0,1,7,8$ and
$3/4$ elsewhere, use $\alpha=1/20$, and retain the proper partition hedge.
On complete cycles the minimizing null is interior,
$(a_0,a_1)=(\log(19/17),0)$: symmetry makes the slope score zero, the mean
success probability is $19/36$, and convexity of Bernoulli cross entropy
proves global optimality. The per-observation minimum KL is
$d=H(19/36)-H(1/4)=0.12926803125\ldots$, with binary entropy $H$.
For $T$ cycling observations, $K\ge9\lfloor T/9\rfloor d$; any remaining
visit adds nonnegative KL, without assuming that the unbalanced allocation
has the same minimizer. The three truth blocks give the conservative regret
\[
 r\le\log2+\log9+\log28+
       \sum_{B=1}^3\log(N_B+1)+3\log(11/7),
\]
using $\pi/2<11/7$ in Appendix~\ref{app:partitionrepair}. The numerator does
not receive the truth. For the \emph{ideal exact-profile} detector,
$\eta=\varepsilon=0$ and $V=T(\log3+6)^2$.
At the two declared horizons, $T=18$ gives only zero as a lower bound;
$T=10{,}000$ gives a lower bound exceeding $0.9925$, with
$K\ge1292.5510$, $r\le31.6149$ and a failure bound below $0.00749$.
Outward 80-digit arithmetic encloses these expressions. They are mathematical
bounds, not measured powers, a sample-size recommendation, or guarantees for
the production numerical guard. The large sufficient budget illustrates the
bound's looseness. No new horizon or candidate search was performed, and the
registered reused-history failure of this numerator is unchanged.

\subsection{What would make a gate constant sufficient?}
Suppose the protected weighting is a probability measure on available actions,
so ED maximization has score at least $B_s$. Suppose, on $G$, there are at least
$\ell>0$ remaining probes with $B_s\ge c b/T$, and, for their chosen actions,
\[
 \KL(p_h(\cdot\mid a_s)\|p_\theta(\cdot\mid a_s))
       \ge\kappa\,\mathrm{ED}_s(a_s)\quad\text{for every }\theta,
       \qquad\kappa>0.
\]
These are strong truth-specific coverage/calibration assumptions, not consequences
of posterior correctness, an initially open gate, or large mean ED.
Nonnegative KL on other steps gives $K_\theta\ge\kappa\ell c b/T$.
For $V\ge\max_\theta V_\theta$ and $0<\beta<1$, it suffices that
\[
 c\ge\frac{T}{\kappa\ell b}\left[
 b-\min_\theta\log R_t(\theta)+r+\varepsilon
       +\sqrt{\frac V2\log\frac m\beta}\right]
\]
to obtain conditional detection probability at least $1-\eta-\beta$.
A negative right side can be replaced by zero. The chosen $c$ still has to
satisfy the coverage premise; this is a conditional implication, not a circular
method for proving that the gate stays open. Complementary designs can relax
pointwise calibration by directly lower-bounding cumulative $K_\theta$.

At $t=0$, even $\kappa=1$ and $\ell=T$ leave coding, numerical and concentration
penalties beyond $b$. Hence $c=1$ cannot in general guarantee high power.
A late history with little evidence or fewer remaining probes has a larger
required budget. Conversely, favorable accumulated $R_t(\theta)$ can reduce
the needed effort. The original gate uses none of these margins explicitly.
This identifies why its observed successes do not turn its constant into a
theorem. The earlier cumulative R\'enyi sufficiency result remains useful and
requires no bounded log increments; the present conditional KL bound is not
claimed uniformly sharper.

\subsection{Exact checks and their limits}
A rational first-crossing dynamic program retains non-crossed probability mass
at each success count, transfers it under the declared Bernoulli law, and
absorbs mass only when the exact rational statistic reaches 20. All planned
horizons $5,6,12,20,60,120$ are checked under both truths and the null. At
$T=5$, every prefix count is exhaustively below threshold; at $T=6$ the exact
probability above is recovered. At terminal $T=20$, all 21 unrejected count-state
prefixes through $t=5$ have exact conditional power at least the bound. These
checks corroborate arithmetic and indexing, not a proof for arbitrary policies.
No response sequences are sampled. The sufficient bound remains conservative;
its role is an auditable implication with explicit premises, not a fitted
replacement for the observed gate.

\section{Independent audit of the later diagnostic and measured-data studies}
\label{app:oct26audit}
These studies followed the earlier manuscript audit. Their registered failures
remain failures. The initial audit used retained run logs without rerunning
policy campaigns. The subsequent registered correction in
Section~\ref{app:correctedreplay} is a separate empirical study.

\subsection{Four difficult bumps: signal capture on a fixed design}
\label{app:missedbumps}
The post hoc diagnostic selected two matched bumps from each continuous-null
panel: indices 7 and 11 at seed 2637501, $T=240$, and indices 9 and 8 at seed
2637601, $T=480$. The first three have zero observed detection for all four
tested policies; the fourth has 6\% detection under separation and the revised
rule. Thus the description ``every arm misses'' is approximate for the fourth.
The diagnostic uses the same float64 concavity denominator as those panels;
its code's word ``certified'' does not supply outward numerical certification.

For each truth, a floating-point constraint-generation calculation maximizes
\[
D^*=\max_\nu\min_{\theta\in[-3,3]^2}\sum_a\nu(a)
\KL(P_a\|p_{\theta,a}).
\]
A fixed allocation rounds $T\nu^*$ to integers by
largest remainders and permutes its action list before responses are drawn.
All four numerators see the same 100 response sequences per world. The
truth-informed numerator is unavailable to an experimenter. The other three
are a uniform $\M_1$ path mixture, per-action Beta$(1,1)$ prediction, and their
equal path mixture. Table~\ref{tab:bumpsaudit} transcribes the saved log.

\begin{table}[htbp]
\centering\small
\caption{Difficult-bump diagnostics. $T D^*$ is rounded numerical separation,
not achievable power. Detection is out of 100 fixed-design campaigns; the
neighborhood mass uses uniform weights over 38,185 library members.}
\label{tab:bumpsaudit}
\begin{tabular}{lrrrrrrr}
\toprule
Panel/bump & Nearest mean KL & $T D^*$ & $\log\pi_{.005}$ & Oracle & $\M_1$ & Beta & Hedge\\
\midrule
240/1 & .000732 & 4.3 & $-7.99$ & .71 & .03 & .05 & .04\\
240/5 & .000951 & 3.9 & $-7.78$ & .55 & .01 & .01 & .01\\
480/3 & .000160 & 2.9 & $-6.47$ & .41 & .02 & .03 & .03\\
480/2 & .000764 & 5.8 & $-7.15$ & .86 & .07 & .11 & .11\\
\bottomrule
\end{tabular}
\end{table}

The neighborhood contains respectively 13, 16, 59 and 30 members within mean
actionwise KL $.005$. Its negative log mass is a coding-cost proxy, not the
realized sequence regret. Positive nearest-member KL demonstrates close
approximation, rather than exact truth membership in this finite library.
The large oracle/hedge gap establishes numerator loss on these fixed,
truth-informed designs. It does not decompose losses on the policies' actual
adaptive histories or show that every implementable numerator must fail.

The study separately reruns 50 seeds per policy and computes $D$ at their
pooled empirical occupancy. Within each world, list greedy, random,
separation and revised rule in that order. Across the four worlds, $T D$ is
\[
\begin{gathered}
(1.25,1.70,2.58,1.57),\quad (.28,1.71,2.17,.90),\\
(.69,1.23,1.48,1.13),\quad (.02,2.39,3.73,3.71).
\end{gathered}
\]
These are plug-in occupancy estimates, not exact
expected occupancies or oracle replay on saved histories. Separation below
$\log20$ is not a theorem of zero power. The declared numerator rule
(oracle at least $.5$, hedge at most $.2$) triggers in three worlds;
neither rule triggers for bump3. The implemented history rule requires
$T D^*>3\log20$ and all policies below $\log20$; it triggers in none.
The narrative claim of a second formally diagnosed history bottleneck is
therefore unsupported. Acquisition and numerator limitations both remain
plausible, without a complete causal allocation of their effects.

\subsection{A valid three-expert numerator that did not improve the panel}
\label{app:trionegative}
The next study replaces the two-expert hedge with
$Q^{\rm trio}=\tfrac13 Q^{\M_1}+\tfrac13 Q^{\rm Beta}+\tfrac13 Q^{\rm MAP}$.
The last expert predicts $.9p_{\mathrm{MAP}(H_{s-1})}(y\mid a_s)+.1/K$,
using only earlier responses. Each component is a proper predictable path law,
so their mixture is proper and costs at most $\log3$ relative to its best
component. This is a path mixture, not an average of per-step component
predictions with fixed weights. The same floating denominator qualification
applies. Detection also controls later intervention activity, so changes in
this statistic can change subsequent policy histories.

On the reused $T=480$ panel, both hedge and trio detect 52.0\% of 900 non-null
campaigns. One world's rate rises $.06$ to $.08$ and another falls $.66$ to
$.64$; the worst change is $-.02$. Trio has no observed alarms in 300 null
campaigns. Registered C1 (gain at least $.03$) fails; C2 (null alarms at most
$.05$) and C3 (no world loss exceeding $.10$) pass their point criteria.
A two-seed smoke check touched both panels before the main run; these are
reused worlds and seeds, not untouched confirmation units.

At the four fixed designs, the oracle/hedge/trio rates are
$.71/.04/.04$, $.55/.01/.01$, $.41/.03/.01$, and $.86/.11/.15$.
Mean improvement $.005$ fails the declared development threshold $.15$.
The fourth reported arm is plug-in-plus-Beta, with the same rates as trio,
although the description planned plug-in-only. That substitution is a protocol
deviation; plug-in-only performance was not reported. On the reused $T=240$
panel, revised-rule rates round to $.358$ for both numerators; separation
rounds to $.317$ with hedge and $.304$ with trio; no null alarms are recorded.
These negative results reject this candidate improvement. They do not exclude
all prior reweighting, prequential prediction, or GROW-based constructions.
The GROW criterion optimizes worst-case expected log evidence, which differs
from finite-horizon crossing probability \citep{ram2026grow}.

\subsection{Why no universal finite-horizon coding floor follows}
\label{app:codingfloor}
The suggested explanation $T D\gtrsim\log(1/\alpha)+(k/2)\log T$ for every
predictable numerator is not established. Worst-case universal-coding regret,
pointwise almost-everywhere asymptotics, and a finite-horizon lower bound on
testing power have different quantifiers. No verified theorem supplies the
proposed corollary for this adaptive finite library. In particular, the three
bump-shape parameters do not by themselves determine an identifiable regular
dimension under these designs; the baseline parameters are also unknown.

Here is an obstruction to a uniform finite-horizon interpretation.
Let $P_0$ generate i.i.d. Bernoulli observations with success probability
$1/2$ and declared horizon $T$. Let
$A=\{\sum_{s=1}^T Y_s\ge r\}$, and choose its null mass
$\delta=P_0(A)<\alpha$. Define the strictly positive proper path law
\[
 Q=(1-\epsilon)P_0(\cdot\mid A)+\epsilon P_0,\qquad 0<\epsilon<1.
\]
Its prefix likelihood ratio is the nonnegative $P_0$ martingale
\[
 E_s=\epsilon+(1-\epsilon)P_0(A\mid H_s)/\delta.
\]
Conditional factorization gives a predictable numerator without access to the
alternative truth. At $T$, $E_T=\epsilon+(1-\epsilon)/\delta$ on $A$ and
$\epsilon$ otherwise. Ville's inequality controls all prefix crossings;
terminal power alone is at least $P_p(A)$ whenever the first value exceeds
$1/\alpha$.

For $T=10000$, $r=5099$, $\epsilon=.1$, $\alpha=.05$, numerical binomial
summation gives $\delta=.02441647$, $E_T=36.96036$ on $A$, and
$P_{.515}(A)=.84860251$. Yet
\[
 T\KL(\operatorname{Ber}(.515)\|\operatorname{Ber}(.5))=4.50067524
 <\log20+\tfrac12\log10000=7.60090246.
\]
These numerical values use two checked floating binomial-sum formulations,
not interval arithmetic; the martingale construction and validity are exact.
This is a horizon-specific construction, not a proposed production detector.
For local alternatives $p_T=1/2+1.5/\sqrt T$, $T\KL\to4.5$ and the terminal
power stays nontrivial while the proposed logarithmic penalty diverges.
This does not contradict an almost-everywhere asymptotic regret statement
for a fixed universal predictor and fixed parameters. It prevents importing
such a statement as a universal finite-budget impossibility claim.

\subsection{A larger replay of measured drug-combination wells}
\label{app:largereplay}
\textbf{Disposition: withheld as supporting evidence pending corrected
preprocessing and a registered rerun.} The audit identified contamination
of negative-control means by solvent annotations on active-drug wells.
The original result is preserved below as an experiment with a data-preparation
error; its registered G1 failure remains a failure.

The later replay uses the raw combination screen of \citet{jaaks2022combinations},
archived as Figshare dataset 19141916, rather than an independent prospective
screen. The committed input is named \texttt{gdsc2\_units300.csv.gz}; this
filename does not make it a GDSC2 monotherapy validation dataset. Replicate-count
eligibility requires two anchor concentrations times seven library doses,
plus seven library-alone and two anchor-alone settings, each with at least
three recorded outcomes. A seed-2026 label-blind sample selects 300 units
(cell line, anchor, library drug) from 10,427 eligible units. Labels were not
used for selection, but all recorded outcomes are subsequently used to fit
labels, grid centers and noise levels before campaigns.

The preprocessing intends to divide intensity minus its plate's mean blank by
mean negative control minus mean blank, following the original screen's
normalization formula. However, its selector combines all \texttt{DMSO}
annotation rows with \texttt{NC} rows. Many DMSO rows describe solvent in
active-drug wells, including positive controls: they are not untreated wells.
In plate14512, 374 of 384 DMSO annotation rows share positions and identical
intensities with active single-agent or positive-control annotations.
Their mean intensity is21,887.01, versus25,846.82 for NC-tagged wells;
the blank mean is272.39. The resulting denominator is materially contaminated.
These exact well matches establish the defect without relying on a statistical
label or model fit. Selecting physical control wells by validated plate roles
must precede a new normalization; a corrected replay has not been run.

The script joins anchor/library annotations by plate and well
position and pools single-agent outcomes across plates. This audit verified
the original archive MD5 against the publisher-linked dataset, reconstructed
the eligible pool and exact selected identities, and matched every selected
recorded viability to the cached normalized-well table. A bounded-memory raw
check covers eight deterministic sampled plates, not the whole archive.
Preprocessing scripts were recovered from the author's cache; they were not
part of the frozen run's source snapshot. Their present hashes and audit scope
are recorded separately. Exact reconstruction of the flawed values is a
provenance check, not validation of that normalization. The eight-plate audit
also records unpaired library annotations discarded by the inner join.
Their well roles require explicit disposition rather than silent imputation.

A subsequent archive-wide provenance check streams all 3,106 plates and
reconstructs all 368,661 historical recorded values exactly. It identifies
98 selected combination wells and 112 selected single-agent wells with multiple
active annotations, affecting units73,114,120. Eight other units use the same
compound as anchor and library. The original 300-unit correction therefore
fails its physical-role data gate. A separately registered cohort excludes
those eleven whole units using metadata before any new labels or policy
outcomes: 289 distinct-drug units,25 cell lines and355,353 recorded observations.
Its corrected data export passes reconstruction, control-role, coordinate and
23-action checks. That data-only review supplied no replacement performance
claim; the subsequent registered campaign is reported in
Section~\ref{app:correctedreplay}. The original failed criterion and withheld
300-unit result remain unchanged.

The 23-action, $T=30$ replay samples outcomes with replacement. There are
\emph{31} categories from 30 edges, correcting the description's ``30 bins.''
The null is a four-parameter Bliss-style dose-response grid of 1,925 members;
the expansion adds two potency shifts, giving 22,925 members including the
null. Grid centers and Gaussian noise use all observations, not held-out
calibration data. The Gaussian-bin model has positive probability in every
category; sparse empirical replay distributions have zero categories. Thus
the empirical truths are outside the finite grid null, including units labelled
``adequate.'' Their alarm rates are not estimates of Type I error under true
null membership.

Labels use an approximate nonlinear nested-model $F$ comparison (four versus
six parameters): $p<.01$ yields 168 ``misspecified'' units, $p>.2$ yields 108
``adequate'' units, and 24 are ambiguous. The first group contains 151 labelled
synergistic and 17 antagonistic shifts. Non-rejection does not establish model
adequacy; nonlinearity, shared controls, pooled singles and plate effects limit
this label interpretation. These are fit-based categories, not biological
ground truth or validated synergy discoveries.

\begin{table}[htbp]
\centering\small
\caption{Historical large-replay summaries from flawed normalization;
withheld as supporting evidence. There are 50 campaigns/unit/arm. Detection
averages the 168 low-$p$ units; alarms and relative prediction-risk premiums
average the 108 high-$p$ units. Values retain the run log's rounding.}
\label{tab:largereplayaudit}
\begin{tabular}{lrrr}
\toprule
Arm & Low-$p$ detection & High-$p$ alarm & High-$p$ risk premium\\
\midrule
Greedy & .040 & .0011 & $0.00\%$\\
Random & .078 & .0000 & $-1.65\%$\\
Class EIG & .273 & .0206 & $-.89\%$\\
Separation & .264 & .0196 & $-.57\%$\\
Revised rule & .247 & .0006 & $-.21\%$\\
\bottomrule
\end{tabular}
\end{table}

The logged unit-bootstrap contrasts for the revised rule are $+.169$
$[.131,.209]$ against random, $-.026$ $[-.046,-.007]$ against EIG, and
$-.017$ $[-.041,.004]$ against separation. The corresponding unit
win/tie/loss counts are $114/41/13$, $39/57/72$, and $52/54/62$.
G1 collapse fails; G2 restoration, G3 the $.05$ point margin, and G4 the
high-$p$ alarm/premium point thresholds pass. G3 is not a noninferiority
inference. The rounded G1 gap does not justify an exact $.001$ margin claim.
EIG has higher measured detection, so this replay does not establish method
superiority for the proposed rule.

Crucially, the frozen driver saves only rounded aggregate and cell-line
summaries, not per-unit or campaign records. This audit reconciles the
transcription and implemented criteria, but cannot independently recompute
those bootstrap intervals or fit-label counts from retained results. Units
also share cell lines, drugs and single-agent pools; the unit bootstrap does
not account for that clustering. A follow-up must retain records and assess
cell-line and shared-pool dependence. The very small final high-$p$ alarm count
does not support a general safety advantage. With the normalization defect,
this run currently supplies no accepted extension of the real-data evidence.
Corrected well roles, retained unit results, clustered uncertainty and independent
calibration are required before reassessing its conclusions.

\subsection{Registered correction: a larger descriptive replay, with a mixed result}
\label{app:correctedreplay}
This registered correction retains the previously selected 289-unit cohort and
physical normalization above; it is not untouched-unit or prospective validation.
All units remain: 127 low-$p$ fit-labelled, 85 high-$p$, and 77 ambiguous.
The low-$p$ group spans 20 cell lines, with 75 fitted synergistic and 52
antagonistic directions. These approximate nonlinear-$F$ labels are not biological
ground truth. Historical 300-unit labels were 168/108/24; missing per-unit records
prevent identifying individual label transitions or causally separating
normalization and cohort changes.

Each unit has 50 response seeds per arm at $T=30$, with five unchanged arms,
31 Gaussian-bin categories and the fitted 1,925-member null/22,925-member
expansion. The revised gate is $\log(20)/30$, not exactly $.1$. Sampling is
with replacement from recorded outcomes. All 72,250 campaigns retain indicators,
zero-based first-crossing indices, terminal protected Brier/2 risk and action
counts summing to 30. Model/noise fitting uses the full data. Prediction switches
to the expansion after rejection, so risk differences include that revision.

\begin{table}[htbp]
\centering\small
\caption{Corrected replay. Detection among 127 low-$p$ units (6,350 campaigns/arm);
high-$p$ alarms and relative risk premium among 85 high-$p$ units (4,250/arm).
The last column is a \emph{separate} known-grid-null simulation (5,780/arm),
not an empirical adequacy label. Means weight units equally.}
\label{tab:correctedreplay}
\begin{tabular}{lrrrr}
\toprule
Arm & \shortstack{Detection\\(\%)} & \shortstack{High-$p$ alarm\\(\%)} & \shortstack{Risk premium\\(\%)} & \shortstack{Simulated-null\\alarms}\\
\midrule
Greedy & 2.38 & .094 & 0 & 0\\
Random & 3.42 & .212 & $-.787$ & 0\\
Class EIG & 17.37 & 1.153 & $+.105$ & 1\\
Separation & 16.02 & .353 & $+.028$ & 0\\
Revised rule & 14.16 & .494 & $+.017$ & 0\\
\bottomrule
\end{tabular}
\end{table}

G1 \textbf{fails again}: $2.38>3.42/2$. G2 passes its ten-point random-gain
criterion (observed 10.74); G3 passes its five-point comparator margin (class-EIG
deficit 3.21); G4 passes its high-$p$ alarm/risk thresholds. These point passes
establish neither inferential noninferiority, a population ten-point gain,
nor real-data Type I control.

\begin{table}[htbp]
\centering\small
\caption{Revised-rule minus comparator, percentage points. Registered paired
bootstrap sensitivity uses 5,000 draws; cell resampling retains all units in each
drawn cell line. All intervals are descriptive and conditional on fitted labels.}
\label{tab:correctedclusters}
\begin{tabular}{lrrr}
\toprule
Comparator & \shortstack{Unit point and\\IID interval} & \shortstack{Unit-weighted\\cell interval} & \shortstack{Equal-cell point\\and cell interval}\\
\midrule
Random & $10.74\ [7.09,14.83]$ & $[6.46,16.24]$ & $16.01\ [8.01,25.73]$\\
Class EIG & $-3.21\ [-5.39,-1.34]$ & $[-5.70,-.68]$ & $-1.67\ [-4.24,.93]$\\
Separation & $-1.86\ [-4.61,.85]$ & $[-5.07,1.12]$ & $.85\ [-2.65,5.23]$\\
\bottomrule
\end{tabular}
\end{table}

The unit-weighted cell interval puts the revised rule below class EIG; its
lower endpoint also extends beyond the allowed five-point point margin.
The equal-cell estimand gives a less resolved contrast and does not replace the
registered equal-unit primary. The random-gain interval includes effects below
ten points. The result supports recovery over weak random sampling, with a
stronger classical comparator and unresolved population margins. Cell resampling
does not capture every shared-drug or shared-well dependence.

A pre-launch amendment selects one fitted null member per unit before responses,
then simulates 20 campaigns/arm with distinct fixed seeds shared across arms.
All 28,900 campaigns are retained; chosen laws are positive normalized 23-by-31
arrays. Revised-rule alarms are 0/5,780, class EIG 1/5,780, all other arms zero;
none exceeds the report-only Hoeffding/union warning level 6.996\%.
Conditioning on fitted finite grids and independent response streams does not
make empirical high-$p$ units true nulls, validate a continuous Gaussian null
using a grid denominator, or certify floating arithmetic.

Independent reanalysis reproduces labels from $F$/degrees of freedom, seed lists,
endpoint means, point criteria and IID/cell intervals. Closed-form RSS agrees
within $1.43\times10^{-14}$, reconstructed null probabilities within
$6.94\times10^{-16}$, and the corrected file hash matches pre-launch export.
These deterministic checks neither refit nor resimulate policies. Counts/times
alone cannot reconstruct adaptive histories or verify each first crossing.
Earlier normalization/role failures and original G1 failure remain recorded.

\subsection{Quantifying rounded null-simulation kernels}
\label{app:roundingtransfer}
An exact continuous-null detector does not make a stored simulation kernel an
exact member of that null. We separately audit this representation issue for
the running original-prior, horizon-480 development study; this is an input
audit, with no new response histories or power claim. Its registration,
eligibility rules and source remain fixed.

\paragraph{Finite-horizon transfer.}
Let $P$ and $\widetilde P$ use the same history-dependent randomized policy,
numerical updates and rejection map. Suppose their conditional response
kernels, on every common history and chosen action at step $s$, differ in total
variation by at most a deterministic $\epsilon_s$. Couple policy randomness
identically and response kernels maximally until the first disagreement.
Conditional on agreement so far, the next disagreement probability is at most
$\epsilon_s$. Therefore, for any trajectory event $A$, including rejection at
any time through $H$,
\[
 |\widetilde P(A)-P(A)|
 \le 1-\prod_{s=1}^H(1-\epsilon_s)
 \le \sum_{s=1}^H\epsilon_s.
\]
Here total variation is $\sup_A|P(A)-\widetilde P(A)|$. The argument allows
discontinuous action ties, adaptive stopping and post-rejection policy changes;
it does not condition on a future adaptive action sequence. If exact-null
rejection probability is at most $\alpha$, the rounded law consequently has
bound $\min\{1,\alpha+\sum_s\epsilon_s\}$. This elementary coupling transfer
is not a new general testing theorem, nor an unbounded-horizon guarantee.

\paragraph{Saved-input check.}
We checked all 49 frozen source-manifest entries and all 24 saved world hashes,
then numerically interpreted only the six unperturbed null worlds. Coefficients,
coordinates and stored probabilities were exported as their exact binary64
values. The coefficients lie in $[-3,3]^2$, and the nine coordinates are exact
multiples of $1/4$. Outward arithmetic at 80 decimal digits encloses each
$\operatorname{sigmoid}(a+bx)$ and bounds its distance from the stored
probability. An independent exact-rational, 128-term exponential Taylor bound,
with a geometric remainder, verifies all 54 enclosures and kernel-error bounds.
No truths were regenerated or policies rerun. Hash consistency is not a defense
against malicious re-signing or a new audit of campaign outcomes.

For each of these six nulls, $480\max_x\epsilon(x)<7\times10^{-14}$.
Thus rounding contributes an additive allowance below that value to the
finite-horizon false-alarm bound, under the common-policy conditional-response
model. It does not supply an observed false-alarm rate. We also checked a
finite uniform grid explicitly: for independent uniform
$U\in\{0,\ldots,2^{53}-1\}/2^{53}$ and strict $U<p$, the success probability
is $\lceil 2^{53}p\rceil/2^{53}$. Using these exact grid probabilities still
gives an allowance below $7\times10^{-14}$ for all six nulls. Independence of
these ideal grid draws and policy randomness is an assumption; this audit does
not certify that fixed PCG64 seeds produce independent random bits. The stored
kernels remain approximations, the result is limited to these null inputs and
$H=480$, and no other detector's float arithmetic or unresolved decisions are
validated by this calculation.

\subsection{A constructive repair of a numerator bottleneck}
\label{app:partitionrepair}
The truth-numerator diagnostic does not give an implementable replacement.
Here an exact constructed example shows that sharing observations across
neighboring settings can repair a coding bottleneck, using a proper numerator
that receives no truth argument. This is analytical development, not an
empirical improvement on the primary panel or a new confirmation. Bayesian
partition mixtures and their dynamic programs are established
\citep{solari2026segmentation}; contextual multiscale sequential audits already
use mixtures over tree prunings with explicit coding penalties
\citep{ullah2026multiscale}. General sequential probability assignment with
side information has a substantial regret literature
\citep{jia2025sequential}. We claim the controlled construction and comparison,
not the invention of these principles.

\paragraph{A fixed proper local-sharing likelihood.}
Order the $m$ fixed actions before observing outcomes. For each contiguous
partition $\mathcal P$ into $k$ blocks put
\[
 w(\mathcal P)=\frac{1}{m\binom{m-1}{k-1}},\qquad
 K(s,f)=\frac{(2s)!(2f)!}{4^{s+f}s!f!(s+f)!}.
\]
There are $\binom{m-1}{k-1}$ such partitions, so the weights sum to one.
Within a block, $K$ is the Beta$(1/2,1/2)$ ordered-sequence likelihood, without
binomial multiplicities. With pooled block counts $(s_B,f_B)$, define
\[
 Q_t^{\rm part}=\sum_{\mathcal P}w(\mathcal P)
                    \prod_{B\in\mathcal P}K(s_B,f_B).
\]
Each response belongs to one block per component. Given the current action,
its next-response probabilities are posterior-weighted Beta predictives;
they sum to one even when the action depends on previous outcomes. Thus
$Q_t^{\rm part}$ is a proper predictable sequence law. This mixes complete
path likelihoods, rather than repeatedly resetting fixed per-step weights or
multiplying overlapping cells. The dynamic program sums block evidences by
last block and block count in $O(m^3)$ arithmetic operations; it does not
post-select the best partition.

For any truth constant within a partition $\mathcal P$ of $k$ blocks,
the arcsine prior density is at least $2/\pi$ times the uniform density.
The uniform Beta sequence law has regret at most $\log(N_B+1)$ against any
fixed block probability, since a binomial event has probability at most one.
Consequently, pathwise,
\[
 \log Q_t^{\rm part}\ge\log L_{P,t}-\log m
 -\log\binom{m-1}{k-1}
 -\sum_{B\in\mathcal P}\log(N_B+1)-k\log(\pi/2).
\]
This conservative bound makes the cost of sharing explicit. Keeping the
all-singleton partition also gives
$Q_t^{\rm part}\ge m^{-1}(2/\pi)^m Q_t^{\rm flat}$ for the existing independent
per-action Beta$(1,1)$ law. This is a log-likelihood comparison, not pathwise
dominance, preserved crossing times or a power guarantee.

Let $S_t$ be the original stored structural marginal with its weights normalized
exactly. Both $Q_t^{\rm old}=(S_t+Q_t^{\rm flat})/2$ and
$Q_t^{\rm new}=(S_t+Q_t^{\rm part})/2$ are proper. Dividing either by the
\emph{same continuous} null profile gives anytime level-$\alpha$ tests under
predictable actions, by fixed-null likelihood-ratio domination.

\begin{proposition}[Computer-assisted continuous-null coding witness]
\label{prop:partitionrepair}
Use nine settings $x_j=(j-4)/4$, $j=0,\ldots,8$, a fixed twice-repeated cycle,
$T=18$, $\alpha=1/20$, and the scientific null
$\sigma(a+bx)$ on $[-3,3]^2$. Keep the original 9-by-9 inference prior and
structural library. There is a stationary strictly positive Bernoulli truth for
which the original hedge's detection probability is at most
$1-(.99)^{18}<.16549$, while the new hedge's is at least
$(.99)^{18}>.83451$, with identical observation allocation.
\end{proposition}
\begin{proof}
Set the success probabilities to $.01$ at indices $0,1,7,8$ and $.99$ at the
other five settings. Let $C$ be the event that both observations at each setting
take its more likely outcome. Then $P(C)=(.99)^{18}$.
On $C$, each setting has at most two identical responses at every prefix.
For the flat numerator, a first visit has likelihood ratio one against the
constant-$1/2$ null, and a second visit multiplies it by $4/3$. Hence every
prefix ratio is at most $(4/3)^9<20$. The original structural-plus-flat hedge
requires its separate check: exact rational evaluation of the stored normalized
prior gives
\[
 \max_{0\le t\le18}2^tQ_t^{\rm old}<7.320<20.
\]
The continuous null contains the constant-$1/2$ member, so its profile is at
least $2^{-t}$; the original test cannot reject anywhere on $C$.

At $t=18$, symmetry gives zero slope score. The intercept score vanishes at
$a=\log(5/4)$, $b=0$, an interior feasible point. Concavity of the logistic
log likelihood proves the exact global profile
$L_0^*=(5/9)^{10}(4/9)^8$. Independent exact partition enumeration and the
dynamic program agree on
\[
 Q_{18}^{\rm part}=\frac{2329271797}{7215545057280},\qquad
 \frac{Q_{18}^{\rm new}}{L_0^*}
 \ge\frac{Q_{18}^{\rm part}}{2L_0^*}>37.85>20.
\]
The new test rejects by $18$ on $C$. Its power is therefore at least $P(C)$;
the old test's power is at most $P(C^c)$. These are bounds, not exact full
crossing probabilities. Neither numerator receives the truth probabilities.
\end{proof}

\paragraph{Checks and limits.}
Five deterministic fixture methods check exact normalization on adaptive
response trees, dynamic-program agreement with a separate polynomial-integration
and partition enumeration, predictive telescoping, empty histories and unsafe
inputs. The stored-prior comparison uses exact binary64 values as rationals;
its approved prior hash begins \texttt{31ec062d213c}, with the full identity
in the machine-readable proof record. The finite rational computation is part
of the computer-assisted proof, rather than a symbolic bound for arbitrary priors;
all 19 checked prefix marginals also lie within their earlier outward Decimal
enclosures. The separately declared 19-step continuation also passes a
terminal crossing check with an outward continuous-profile certificate; the
18-step example uses no optimizer or logarithmic arithmetic. Both are reported.
The actual-prior check and strictly positive corollary were disclosed follow-ups
after the first sharing-only calculation, not blind empirical confirmations.

This deliberately aligned, ordered, piecewise-constant construction shows
possibility, not prevalence or superiority on smooth/off-library worlds.
Partition coding has its own costs, and the numerator may lose useful crossings
elsewhere. It does not validate the ED gate constant, infer a practical power
guarantee from mean discrepancy, change any registered failure, or establish a
new benchmark result. Fresh registered evaluation and resource accounting are
required before adopting this candidate in the production rule.

\paragraph{Registered reused-history diagnostic: failure.}
After fixing this candidate, we registered a separate diagnostic on all original
block-0, nine-action, $T=120$ discrepancy-guided and cycling histories:
24 worlds, 16 repeats and two history arms, giving 768 histories and 92,160
prefixes. These worlds, outcomes and previous negative summaries were already
known. We changed only the numerator score, retaining saved structural
marginals, outward continuous-profile bounds, threshold enclosures and the
original $10^{-7}$ profile-gap guard. No actions, responses, prior fits or
denominators were regenerated. The partition component was evaluated exactly;
the new hedge used 80-digit outward arithmetic.

All histories completed without numeric failures, all original first crossings
reproduced, and possible/certain crossing counts agreed with no guard-blocked
possible prefixes. A separate retained-record audit reconciled every prefix
flag and first event, source/input identities, count summaries and bootstrap
arithmetic; it did not independently recompute every partition likelihood or
profile certificate. Non-null counts were:
\begin{center}
\begin{tabular}{lrrr}
\toprule
Fixed history arm & Original hedge & Partition alone & New hedge\\
\midrule
Discrepancy-guided & $13/288$ & $11/288$ & $10/288$\\
Cycling & $30/288$ & $20/288$ & $20/288$\\
\bottomrule
\end{tabular}
\end{center}
The new hedge gained no events and lost three and ten, respectively.
Its new-minus-original contrasts were $-1.04$ percentage points
(descriptive paired-world percentile interval $[-2.78,0]$) and $-3.47$ points
($[-8.33,0]$), using the registered 5,000 resamples of 18 whole non-null worlds,
with all 16 repeats retained. These are reused-world development intervals,
without multiplicity-adjusted confirmation or population coverage claims.
All three numerators had zero alarms among each arm's 96 null histories;
this descriptive observation is not a calibration proof.

The primary requirement of at least five percentage points of improvement
\emph{failed}. The requirement that no non-null world lose more than ten points
also \emph{failed}: the largest loss was six of 16 histories in one cycling
world. Integrity and the null point-fraction criterion passed, so the overall
registered diagnostic failed. The secondary partition-only result cannot
replace that failure. Scoring took approximately 114 seconds wall time,
excluding previous collection, acquisition and profile computation; no
end-to-end resource advantage is established. We retain the original hedge
and do not promote this candidate. The constructed proposition proves
possibility in its aligned example, while this diagnostic demonstrates its
limited transfer to the previously studied worlds. Fixed-history rescoring
does not establish behavior of a policy whose future actions depend on a
changed detector, or performance on fresh scientific units.

\subsection{Audit scope and assisted research workflow}
Here ``independent audit'' means a separate code and proof-checking pass with
access to completed results, including language-model-assisted review and
deterministic checks, not blinded external replication or human peer review.
Language models also assisted coding and drafting. Versioned source,
recoverable results, proofs and retained failures support the claims;
assistant agreement is not evidence. Human authors remain responsible for
the submission's content and disclosures.

\end{document}